\documentclass[acmtog]{acmart}

\usepackage{bm}
\usepackage{amsfonts}
\usepackage{booktabs}
\usepackage{multirow}
\usepackage{tikz}
\usepackage{pgfplots}
\usepackage{algorithmicx}
\usepackage{algorithm}
\usepackage{algpseudocode}
\usepackage{enumitem}
\usepackage{afterpage}

\pgfplotsset{compat=1.11}
\usetikzlibrary{quotes, angles, shadings, intersections}
\usetikzlibrary{external}
\pgfplotsset{compat=newest}
\usepgfplotslibrary{fillbetween}
\usetikzlibrary{shadows}
\definecolor{red-base}{rgb}{0.925,0.05,0.04}
\definecolor{red-base-dark}{rgb}{0.7,0.2,0.3}
\definecolor{blue-base}{rgb}{0.08,0.52,0.8} 
\definecolor{aqua-base}{rgb}{0.34,0.87,1.0}
\definecolor{green-base}{rgb}{0.1,0.6,0.33}
\definecolor{green-base-shiny}{rgb}{0.17,1.0,0.54}
\definecolor{green-base-dark}{rgb}{0.09,0.5,0.27}
\definecolor{lime-base}{rgb}{0.34,1.0,0.19}
\definecolor{yellow-base}{rgb}{1.0,0.84,0.06}
\definecolor{orange-base}{rgb}{1.0,0.49,0.08}
\definecolor{pink-base}{rgb}{0.98,0.27,1.0}
\definecolor{gray-base}{rgb}{0.88,0.85,0.77}
\definecolor{gray-base-thick}{rgb}{0.78,0.75,0.68}
\definecolor{gray-base-heavy}{rgb}{0.63,0.61,0.55}
\definecolor{gray-base-sombre}{rgb}{0.38,0.36,0.33}
\definecolor{gray-base-dark}{rgb}{0.15,0.25,0.13}

\usepackage{wrapfig}
\usepackage{subfigure}
\usepackage{array}
\usepackage{xspace}
\newcommand{\PreserveBackslash}[1]{\let\temp=\\#1\let\\=\temp}
\newcolumntype{C}[1]{>{\PreserveBackslash\centering}p{#1}}
\newcolumntype{R}[1]{>{\PreserveBackslash\raggedleft}p{#1}}
\newcolumntype{L}[1]{>{\PreserveBackslash\raggedright}p{#1}}
\newcommand{\MethodName}{BEDRF\xspace}

\newcommand{\RedComment}[1]{{\color{red} \emph{#1}}}

\newcommand{\comments}[1]{}

\renewcommand{\RedComment}[1]{} 

\title{BEDRF: Bidirectional Edge Diffraction Response Function for Interactive Sound Propagation}

\author{Chunxiao Cao}
\email{ccx4graphics@gmail.com}
\affiliation{%
  \institution{State Key Laboratory of CAD\&CG, Zhejiang University}
  \city{Hangzhou}
  \country{China}
  \postcode{310058}
}
\author{Zili An}
\email{22021151@zju.edu.cn}
\affiliation{%
  \institution{State Key Laboratory of CAD\&CG, Zhejiang University}
}

\author{Zhong Ren}
\email{renzhong@zju.edu.cn}
\affiliation{%
  \institution{State Key Laboratory of CAD\&CG, Zhejiang University}
}

\author{Dinesh Manocha}
\email{dmanocha@umd.edu}
\affiliation{%
  \institution{Department of Computer Science, University of Maryland}
  \streetaddress{8125 Paint Branch Drive}
  \city{College Park}
  \state{Maryland}
  \country{USA}
  \postcode{20742}
}

\author{Kun Zhou}
\email{kunzhou@zju.edu.cn}
\affiliation{%
 \institution{State Key Laboratory of CAD\&CG, Zhejiang University}
}

\begin{document}



\begin{abstract}
We introduce {\em{bidirectional edge diffraction response function (BEDRF)}}, a new approach to model wave diffraction around edges with path tracing. The diffraction part of the wave is expressed \RedComment{the local expression is the point} as an integration on path space, and the wave-edge interaction is expressed using only the localized information around points on the edge similar to a bidirectional scattering distribution function (BSDF) for visual rendering. For an infinite single wedge, our model generates the same result as the analytic solution \RedComment{equivalence with BTM is expected, but not guaranteed mathematically in this paper}. Our approach can be easily integrated into interactive geometric sound propagation algorithms that use path tracing to compute specular and diffuse reflections. Our resulting propagation algorithm can approximate complex wave propagation phenomena involving high-order diffraction, and is able to handle dynamic, deformable objects and moving sources and listeners. We highlight the performance of our approach in different scenarios to generate smooth auralization.
\end{abstract}

\maketitle

\section{Introduction}

Interactive sound simulation and rendering is becoming increasingly popular in games and virtual environments. Accurate simulation of sound phenomena can significantly improve the plausibility of a VR environment as well as the immersive experience for the user. The most accurate methods for sound simulation are based on wave-based methods \cite{Pind2020Wave} \RedComment{not a survey, but chapter 2 is close}. However, a physically accurate simulation of sound propagation, using a numerical wave equation solver, is computationally-intensive. Despite various efforts to increase the efficiency of the wave equation solvers \cite{Raghuvanshi2009EfficientAA, Savioja2010REALTIME3F,Mehra2012AnEG,Allen2015AerophonesIF}, they are not fast enough for interactive applications, especially when dealing with complex scenes with moving objects.

For interactive simulation, a large family of algorithms, based on the theory of geometric acoustics (GA), has been developed for sound propagation. Geometric optics (GO), the counterpart of GA in the field of computer graphics, has been the foundation of photorealistic rendering. GO expresses light transport with an integral called the transport equation or the rendering equation \cite{kajiya1986the}. This equation has been extensively studied, and many efficient integrators have been developed and used in various applications \cite{Veach1995BidirectionalEF, Veach1997MetropolisLT,Keller1997InstantR,Jensen2001RealisticIS}.

Many researchers have borrowed techniques from GO and have extended them to the field of acoustics. However, the accuracy of geometric approximation depends on the wavelength. One of the reasons for the success of GO in light propagation simulation is that the wavelength of visible light in a vacuum ranges from $380\text{nm}$ to $740\text{nm}$, a very small value compared to the scale of common objects. In comparison, the wavelength of audible sound waves in the air ranges from $0.017\text{m}$ to $17\text{m}$, which makes wave-based sound phenomena easily perceptible by human ears. Among these phenomena, a key issue is diffraction or occlusion effects, which tend to be important for receivers located inside the shadow region of an obstacle. Due to the discontinuity of the visibility function, the ray-based propagation model will generate ``shadows''. Without simulating the diffraction effect, the user would perceive a sudden change of sound quality when entering a shadow zone, which contradicts our aural experiences in the physical world. It has been shown that inaccurate modeling of diffraction effects can result in loss of realism in virtual environments~\cite{Rungta2016Psychoacoustic}.

Since diffraction is a low-frequency effect, it is especially difficult to simulate for interactive GA-based algorithms based on ray or path tracing. Many techniques have been proposed to approximate diffraction effects with methods based on the unified theory of diffraction (UTD, \cite{tsingos2001modeling,micah2012guided, schissler2014high-order}), Biot-Tolstoy-Medwin model (BTM, \cite{calamia2007fast, Calamia2009, Antani2010}), diffraction kernels \cite{rungta2018diffraction}, learning methods \cite{Tang2021learning}, the Heisenberg uncertainty principle \cite{Stephenson0improved}, and volumetric diffraction and transmission \cite{Pisha2020VDaT}. However, their accuracy can vary based on the environment, moving objects, and the location of the source or the receiver relative to the diffraction edges.

\noindent {\bf Main Results:} In this paper, we present a new method to compute the diffraction effects in an interactive GA framework, with the following contribution:

\begin{itemize}
    \item We present \MethodName, a localized edge-diffraction representation around convex wedges. Our novel representation relies only on the incident and outgoing wave direction and the local boundary condition (wedge angle and material). This is similar to a bidirectional scattering distribution function (BSDF) \cite{Asmail1991BidirectionalSD} for visual rendering. Our new representation is derived from the exact solution for planar incident waves. We also present an importance sampling strategy for this representation, which is used for interactive propagation.
    \item A novel, efficient edge diffraction solver based on Monte Carlo path tracing. We present a rendering equation for diffraction and an integrator that solves this rendering equation. Our integrator supports the simulation of diffraction effects via the BEDRF representation. With the concept of ``meta-path space'', our modified intersection detector allows paths to interact with edges in the scene, while being compatible with traditional path tracing techniques like multiple importance sampling (MIS). We compared the accuracy of our solver with state-of-the-art diffraction algorithms based on UTD and BTM.
    \item Several specially designed techniques that improve the accuracy and performance of our diffraction path tracer. The effect and performance of these techniques are demonstrated with simulation results in actual scenes.
\end{itemize}
We combine our approach with an interactive geometric propagation approach that uses path tracing for specular and diffuse reflections. We highlight its benefits in different benchmarks in terms of impulse response (IR) computation and auralization.

\section{Related Works}

\subsection{Sound Propagation and Geometric Acoustics}

The existing algorithms for sound propagation simulation can be divided into two categories \cite{SoundSurvey}. The first category is ``wave-based'' algorithms that compute the propagation result with a numerical wave equation solver. The solver may be based on the finite-element method (FEM) \cite{Thompson2006review}, the boundary-element method (BEM) \cite{Kirkup2019boundary} or the finite-difference time domain method (FDTD) \cite{Hamilton2021Tutorial}. These methods are accurate, but usually very slow for complex acoustic scenes. Moreover, their complexity can increase as a fourth power of the simulation frequency. The second category is GA-based algorithms, which can be accurate at high frequencies, have been shown to be effective in large, dynamic scenes, and are used for interactive applications.

While earlier algorithms like the image source method \cite{Vorlander1989simulation} use specific transport equations, many current GA algorithms are based on the acoustic rendering equation in \cite{siltanen2007the}. Since this equation is very similar to the rendering equation for visual rendering, many acoustic solvers take inspiration from existing visual rendering algorithms like Monte Carlo path tracing \cite{lentz2007virtual,micah2012guided,schissler2014high-order,cao2016interactive,Schissler2016interactive} or photon mapping \cite{Bertram2005phonon, Jenkin07acousticalmodeling}.

\subsection{Diffraction Algorithms in Geometric Acoustics}

Many techniques have been proposed to approximate wave diffraction within the framework of GA. A direct solution is to use a hybrid framework that combines the numeric and GA solution \cite{yeh2013wave-ray,mehra2013wave-based,rungta2018diffraction}. The wave propagation around scene objects is precomputed with a numerical solver, and the long-distance propagation in empty areas is modeled with GA. Some of these methods are limited to static scenes, and others only consider the local scattering field around the objects.

Another approach is to analyze the diffraction wave around local geometry of the objects, and combine the local solutions into a global one. As scene geometries are usually represented with triangle meshes, the edge-diffraction models, which give the diffraction solution around the intersecting edge of two half-planes forming a wedge, can be useful. Most edge-diffraction models can only compute the solution for convex wedges, as solutions for non-convex wedges are complicated due to the the presence of interreflection between two half-planes and are unsuitable for real-life applications. The most popular models in sound propagation are the uniform geometrical theory of diffraction (UTD) and the Biot-Tolstoy-Medwin model (BTM).

UTD \cite{UTD1,UTD2} is widely used for simulating diffraction effects of electromagnetic waves. There have also been some successful applications in simulating interactive sound propagation\cite{tsingos2001modeling, micah2012guided, schissler2014high-order, Schissler2021FastDP}. The UTD model gives the solution in various conditions, such as spherical or planar incident wave and flat or curved surface. However, it also suffers from many issues, including the fact that its solution is non-exact, especially for finite-length edges, Moreover, its solution is expressed in the frequency domain instead of the time domain. The last point has a particular impact on the resulting quality, as the solvers in actual applications usually simulate wave propagation on 3-8 frequency ranges, and are unable to process phase information.

BTM \cite{svensson1999an} was developed by combining the Biot-Tolstoy solution\cite{biot1957formulation} with Medwin's assumption\cite{medwin1982impulse}. The Biot-Tolstoy solution gives the exact description of the propagation of a spherical wave around a wedge, and Medwin's assumption expresses such a solution with ``secondary sources'' on the edge of the wedge, which allows BTM to deal with finite-edge diffraction. BTM requires integration operations on geometry edges, which makes it difficult to combine with path tracing (integrating on the path space). In current auralizers using BTM \cite{calamia2007fast, Calamia2009, Antani2010}, these integrations are computed separately and are used for non-interactive applications. Since path tracing also involves numerical integration, we reformulate the integration on the path space, which allows us to use the integrator for rendering equation to solve the diffraction problem. We use this idea to develop the notion of BEDRF and use our approach for interactive propagation.

There are also some other solutions for GA diffraction. A very promising approximative model for edge diffraction is the directive line source model (DLSM) \cite{Menounou2000DLSM}. An improved version of DLSM \cite{Menounou2017DLSM2} gives the exact solution for plane wave diffraction around a half-plane and works well in many other simple cases. \cite{Stephenson0improved, Stephenson2010energetic} compute convex edge diffraction with the Heisenberg uncertainty principle and model edge diffraction with a diffraction angle probability density function (DAPDF). This method is conceptually simple, but lacks a solid physical basis and is non-exact. \cite{tsingos2007instant} combines the Kirchhoff approximation and GPU rasterization to compute the first-order diffraction wave. This method is valid for arbitary geometric shapes, but difficult to extend to high-order diffraction. The volumetric diffraction and transmission (VDaT) method \cite{Pisha2020VDaT} is an approximation of BTM, which uses a large number of probe rays and simplifies all occluding geometries into a combination of planar rings. VDaT is unable to express high-order diffraction, and the accuracy of this approach depends on the similarity of the original and simplified representations.

\section{Background}

The notation used in our paper mostly come from \cite{siltanen2007the} and \cite{keller1951diffraction}, with a few modification. We highlight the symbols used in this paper  in \autoref{tbl:symbols}. We also use the ``arrow notations'' to simplify path-tracing-related expressions.

\begin{table}[h]
  \begin{tabular}{ll}
    \toprule
    Symbol & Explanation\\
    \midrule
    $\mathbb{R}$ & set of real numbers \\
    $||\mathbf{x}||_2$ & 2-norm (Euclidian length) of vector $x$ \\
    $\nabla x$ & gradient of $x$ \\
    $E[x]$ & expectation of random variable $x$ \\
    $\mathcal{F}(f)$ & Fourier transform of function $f$ \\
    $L(\mathbf{x},\mathbf{v}_i,t)$ & sound wave at $\mathbf{x}$, \\ 
    & received from the direction $\mathbf{v}_i$ at time $t$ \\
    $L_0(\mathbf{x},\mathbf{v}_o,t)$ & initial wave from source $\mathbf{x}$, \\
    & to the direction $\mathbf{v}_o$ at time $t$ \\
    $M(\mathbf{x},\mathbf{x}',t)$ & media term \\
    $V(\mathbf{x},\mathbf{x}')$ & visibility between point $\mathbf{x}$ and $\mathbf{x}'$, \\
    & 0 when invisible, 1 when visible \\
    $\rho(\mathbf{x},\mathbf{v}_i,\mathbf{v}_o)$ & BSDF (\autoref{section:PT}) / BEDRF (\autoref{section:BEDRF}) at $\mathbf{x}$, \\
    & with incident direction $\mathbf{v}_i$ \\
    & and outgoing direction $\mathbf{v}_o$ \\
    $c$ & sound speed \\
    $\varphi_i, \varphi_o, \theta_i, \theta_o$ & local coordinate angles, see \autoref{fig-coord} \\
    \midrule
    Arrow Notation & Expanded Form\\
    \midrule
    $L(\mathbf{x}'\rightarrow \mathbf{x},t)$ & $L(\mathbf{x},\frac{\mathbf{x}-\mathbf{x}'}{||\mathbf{x}-\mathbf{x}'||_2},t)$ \\
    $L_0(\mathbf{x}''\rightarrow \mathbf{x}',t)$ & $L_0(\mathbf{x}'',\frac{\mathbf{x}'-\mathbf{x}''}{||\mathbf{x}'-\mathbf{x}''||_2},t)$ \\
    $\rho(\mathbf{x}''\rightarrow\mathbf{x}'\rightarrow\mathbf{x})$ & $\rho(\mathbf{x}',\frac{\mathbf{x}'-\mathbf{x}''}{||\mathbf{x}'-\mathbf{x}''||_2},\frac{\mathbf{x}-\mathbf{x}'}{||\mathbf{x}-\mathbf{x}'||_2})$ \\
    \bottomrule
  \end{tabular}
  \caption{Common symbols used in this paper.}
  \label{tbl:symbols}
\end{table}

\subsection{Geometric Acoustics}

The theory of geometric acoustics is based on the Eikonal approximation, which assumes that the solution for the wave equation $f$ takes the following form:
\begin{equation}
    f(\mathbf{x},t)=e^{A(\mathbf{x})+i(\frac{1}{\lambda}B(\mathbf{x})-ct)},
    \label{WKB}
\end{equation}
where $\lambda$ is the wave length and $A(\mathbf{x})$ and $B(\mathbf{x})$ express the amplitude and phase delay at different positions. Combined with the wave equation in homogenous media
\begin{equation}
    \frac{\partial^2f}{\partial t^2}=c^2\nabla^2f,
    \label{WaveEq}
\end{equation}
we have the following equation:
\begin{equation}
  ||\nabla B||_2^2-\frac1{\lambda^2} = \nabla^2 A + ||\nabla A||^2_2.
  \label{WKB2}
\end{equation}

If $\lambda$ is sufficiently small, the right side of the equation can be ignored, and we have
\begin{equation}
  ||\nabla B||_2^2\approx \frac{1}{\lambda^2},
  \label{eikonal}
\end{equation}
which indicates that $\nabla B$ is of constant length, and every trajectory generated by this vector field is a straight line. In this way, the propagation of a sound wave can be modeled with wave packets of different amplitudes and frequencies, travelling on paths (sometimes called ``rays'') with direction $\nabla B$ at a constant speed. With this approximation, the original wave propagation result expressed as a solution of a differential equation becomes an integration on the path space.

\subsection{Rendering Equation} 
\label{section:PT}

The purpose of a geometric wave propagation algorithm is to solve the rendering equation, Since we are dealing with sound waves, we use the time-dependent rendering equation proposed by \cite{siltanen2007the}. A reformulated version is presented below:

\begin{equation}
\begin{aligned}
  L(\mathbf{x}'\rightarrow \mathbf{x},t) & = L_0(\mathbf{x}'\rightarrow \mathbf{x},t) \\
          & + \int_\Omega\big(L(\mathbf{x}'\rightarrow \mathbf{x}'',t)V(\mathbf{x}'',\mathbf{x})\rho(\mathbf{x}'\rightarrow\mathbf{x}''\rightarrow\mathbf{x})\big. \\
          & \qquad \big.* M(\mathbf{x}'',\mathbf{x}, t)\big)dA_{\mathbf{x}''},
\end{aligned}
\label{eq:RenderEquation}
\end{equation}

In \autoref{eq:RenderEquation}. $\Omega$ represents the scene geometry. $\rho(\mathbf{x}'',\mathbf{v}_i,\mathbf{v}_o)$ is the bidirectional scattering distribution function (BSDF) describing the interaction between the wave and the scene at $\mathbf{x}''$. The media term $M(\mathbf{x}'',\mathbf{x}, t)$ represents the effects of the medium on the wave during the propagation from $\mathbf{x}''$ to $\mathbf{x}$, including time delay, propagation decay and energy absorption. Its form is usually given like this:
\begin{equation}
  M(\mathbf{x}'',\mathbf{x}, t)=\frac{e^{-\alpha r}\delta(t-\frac{r}{c})}{r^2}, r=||\mathbf{x}-\mathbf{x}''||_2,
  \label{eq:media}
\end{equation}
where $\alpha$ is the absorption factor of the media and $\delta(t)$ is the Dirac delta function. The media term is applied on the left side of the wave with a convolution (denoted by the asterisk $*$). $A$ is the area measure defined on $\Omega$, and $dA_{\mathbf{x}''}$ in \autoref{eq:RenderEquation} indicates that we integrate on the measure $A$ with respect to the variable $\mathbf{x}''$. In our algorithm, $A$ is extended to measure both the area of the geometry and the length of the convex edges on it. The media term is also modified to suit our amplitude-based framework, which will be discussed in \autoref{section:NewPT}.

The transport equation is easier to understand in its operator form:

\begin{equation}
  L = L_0 + TL,
  \label{eq:OpFormTransport}
\end{equation}
where
\begin{equation}
  TL = \int_\Omega (LV\rho * M)dA_{\mathbf{x}''}.
  \label{eq:OpTransport}
\end{equation}

The operator $T$ has an intuitive explanation as an unoccluded propagation of wave from a point of the scene $\Omega$ to another. We can see that the rendering equation is a formal expression of the fact that the total wave equals the source wave plus the wave induced by wave-geometry interaction.

Expanding \autoref{eq:OpFormTransport} into a Neumann series \cite{kreyszig1978introductory}, we have
\begin{equation}
  L = \sum_{i=0}^\infty T^iL_0.
  \label{eq:NeumannExpansion}
\end{equation}
This is the expression used in actual algorithms like path tracing.

\subsubsection{Amplitude-Based Rendering Equation}
\label{section:NewPT}

The media term in \autoref{eq:media} decays in an inverse-square way relative to $r$, which matches the inverse-square law for intensity. However, for an amplitude-based rendering equation, we expect its solution to behave like one for the wave equation. The general solution of spherical waves is given below:
\begin{equation}
  f(\mathbf{x},t)=\sum_{l,m}Y_{lm}J_l(r)(a_{lm}\delta(t-\frac{r}{c})+b_{lm}\delta(t+\frac{r}{c}))
\end{equation}
where $Y_{lm}$ is the $m$-th spherical harmonic function of degree $l$, $J_l$ is the spherical Bessel function of degree $l$, and $a_{lm}$ and $b_{lm}$ are constants. When $r$ is sufficiently large, we have $|J_l| \approx 1/r$ for every $l$. Thus we need to change the denominator $r^2$ in \autoref{eq:media} to $r$.



\subsection{Monte Carlo Path Tracing}




A common algorithm that solves the integration in \autoref{eq:OpTransport} is Monte Carlo integration, which is a probabilistic method using the fact that the mathematical expectation of a random variable is an integration on the probability space:
\begin{equation}
  E\left[\frac{f(X)}{p(X)}\right] = \int_\Omega f(x)\frac{d\mu_1}{d\mu_2}d\mu_2 = \int_\Omega f(x)d\mu_1,
  \label{eq:IntegralEstimator}
\end{equation}
where $p(X)$ is the Radon-Nikodym derivative \cite{royden1988real} $d\mu_2/d\mu_1$ between the two measures $\mu_1$ and $\mu_2$ on the space $\Omega$. $\mu_1$ is usually a natural measure defined on $\Omega$, like $A$ in \autoref{eq:RenderEquation}. $\mu_2$ is the sampling probability measure that depends on the algorithm.

To calculate $T^iL_0$ in \autoref{eq:NeumannExpansion}, we need to take samples from $\Omega^i$. Each sample is a path that connects the source and the listener over $i$ interactions with the scene geometry. Each segment of the path corresponds to the media term $M$ in \autoref{eq:RenderEquation}, and each vertex corresponds to the BSDF $\rho$. In path tracing, the generation process of each segment of the path is mutually independent. Therefore. the $f/p$ in \autoref{eq:IntegralEstimator} could be expanded into the following form:
\begin{equation}
  \frac{f}{p} = \frac{L_0(\mathbf{x}_0\rightarrow\mathbf{x}_1)*M(\mathbf{x}_0,\mathbf{x}_1)\cdot\rho(\mathbf{x}_0\rightarrow\mathbf{x}_1\rightarrow\mathbf{x}_2)*M(\mathbf{x}_1,\mathbf{x}_2)\cdots}{p(\mathbf{x}_0\rightarrow\mathbf{x}_1)\cdot p(\mathbf{x}_0,\mathbf{x}_1)\cdot p(\mathbf{x}_0\rightarrow\mathbf{x}_1\rightarrow\mathbf{x}_2)\cdot p(\mathbf{x}_1,\mathbf{x}_2)\cdots}
  \label{eq:IntegralEstimatorExpanded}
\end{equation}
where $\mathbf{x}_i$ is the $i$-th path node and $p$ is the probability for each path-building event. In the following sections, we will discuss the details of each term in \autoref{eq:IntegralEstimatorExpanded}.

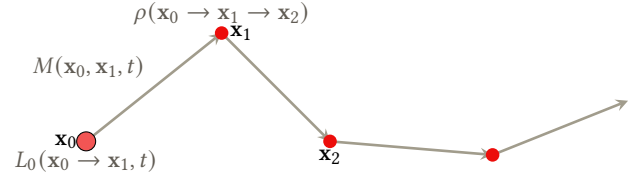
\begin{figure}[ht]
  \centering
  \begin{tikzpicture}[scale=1.8]
    \coordinate (node1) at (-2,-0.3);
    \coordinate (node2) at (-1,0.5);
    \coordinate (node3) at (-0.2,-0.3);
    \coordinate (node4) at (1,-0.4);
    \coordinate (node5) at (2,0);

    \definecolor{bounce4}{rgb}{0.1,0.3,0.1}
    \definecolor{bounce3}{rgb}{0.1,0.5,0.2}
    \definecolor{bounce2}{rgb}{0.2,0.7,0.5}
    \definecolor{bounce1}{rgb}{0.2,0.9,0.7}
    \definecolor{bounce0}{rgb}{0.4,1.0,0.9}

    \draw[-stealth, gray-base-heavy, line width=1pt](node1)--(node2) node[pos=0.5,gray-base-sombre,above left] {$M(\mathbf{x}_0,\mathbf{x}_1,t)$};
    \draw[-stealth, gray-base-heavy, line width=1pt](node2)--(node3);
    \draw[-stealth, gray-base-heavy, line width=1pt](node3)--(node4);
    \draw[-stealth, gray-base-heavy, line width=1pt](node4)--(node5);

    \draw [fill = red-base!70](node1)circle(0.07) node[black,left] {$\mathbf{x}_0$}  node[gray-base-sombre,below] {$L_0(\mathbf{x}_0\rightarrow\mathbf{x}_1,t)$};
    \fill [red-base](node2)circle(0.05) node[black,right] {$\mathbf{x}_1$}  node[gray-base-sombre,above] {$\rho(\mathbf{x}_0\rightarrow\mathbf{x}_1\rightarrow\mathbf{x}_2)$};
    \fill [red-base](node3)circle(0.05) node[black,below] {$\mathbf{x}_2$};
    \fill [red-base](node4)circle(0.05);
  \end{tikzpicture}
\caption{For a path propagating through the scene, the path segments correspond to the media term of the transport equation, and the vertices correspond to the BSDF. The events of path propagation and object interaction are independent.}
\label{fig:RenderEquationExplanation}
\end{figure}

\subsection{Wedge Coordinate}

As mentioned above, the BSDF $\rho(\mathbf{x}''\rightarrow \mathbf{x}'\rightarrow \mathbf{x})$ represents the interaction between wave and scene geometry. Here we'll write this function into a simpler form that is independent from the local coordinate frame: $\rho(\mathbf{x}''\rightarrow \mathbf{x}'\rightarrow \mathbf{x})=\rho_\mathbf{x}(\mathbf{v}_i,\mathbf{v}_o)$. $\mathbf{v}_i=\mathbf{T}^{-1}\frac{\mathbf{x}'-\mathbf{x}}{\left||\mathbf{x}'-\mathbf{x}\right||_2}$, $\mathbf{v}_o=\mathbf{T}^{-1}\frac{\mathbf{x}''-\mathbf{x}'}{\left||\mathbf{x}''-\mathbf{x}'\right||_2}$. $\mathbf{T}$ denotes the local coordinate frame we used at position $\mathbf{x}$. We can see that $\mathbf{v}_i$ and $\mathbf{v}_o$ represents the normalized incident and outgoing wave direction vector.

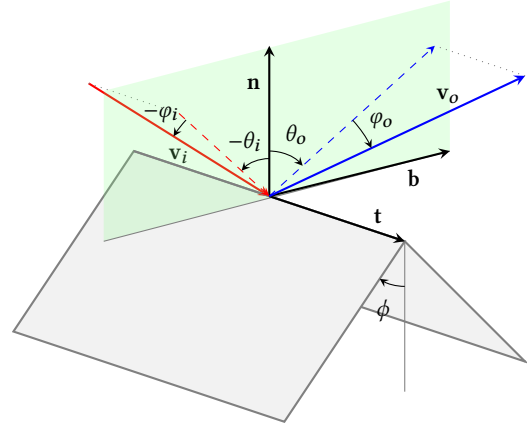
\begin{figure}
  \centering
  \begin{tikzpicture}
    \coordinate (pf1) at (-5.2,-1.2);
    \coordinate (pf2) at (-3.6,1.2);
    \coordinate (pf3) at (-0.5818,-0.8727);
    \coordinate (pf4) at (-1.6,-2.4);
    \coordinate (pf5) at (0,0);
    \coordinate (pf6) at (1.6,-1.6);
    \coordinate (top) at (-1.8,2.6);
    \coordinate (middle) at (-1.8,0.6);
    \coordinate (bottom) at (0,-2);
    \coordinate (incident) at (-4.2,2.1);
    \coordinate (incident_proj) at (-3.0,1.7);
    \coordinate (outgoing) at (1.6,2.2);
    \coordinate (outgoing_proj) at (0.4,2.6);
    \coordinate (plane1) at (-4,0);
    \coordinate (plane2) at (0.6,1.2);
    \coordinate (plane3) at (-4,2);
    \coordinate (plane4) at (0.6,3.2);
  
    \fill[gray!10] (pf1) -- (pf2) -- (pf5) -- (pf4) -- cycle;
    \fill[gray!10] (pf3) -- (pf6) -- (pf5) -- (pf2) -- cycle;
    \draw[thick, gray] (pf1) -- (pf2) -- (pf5) -- (pf4) -- cycle;
    \draw[red, thick, -stealth] (incident) -- (middle) node[pos=0.5,black,below] (vec_vi) {$\mathbf{v}_i$};
    \draw[gray, dotted] (incident) -- (incident_proj);
  
    \fill[green!50, opacity=0.2] (plane1) -- (plane2) -- (plane4) -- (plane3) -- cycle;
    \draw[gray] (plane1) -- (plane2);

    \draw[thick, gray] (pf3) -- (pf6) -- (pf5) -- (pf2);
    \draw[gray] (pf5) -- (bottom);
    \draw[thick, black, -stealth] (middle) -- (top) node[pos=0.75,left] (vec_n) {$\mathbf{n}$};
    \draw[thick, black, -stealth] (middle) -- (pf5) node[pos=0.8,above] (vec_t) {$\mathbf{t}$};
    \draw[thick, black, -stealth] (middle) -- (plane2) node[pos=0.8,below] (vec_b) {$\mathbf{b}$};
    
    \draw[red, -stealth, dashed] (incident_proj) -- (middle);
    \draw[blue, thick, -stealth] (middle) -- (outgoing) node[pos=0.7,black,above] (vec_vo) {$\mathbf{v}_o$};
    \draw[blue, -stealth, dashed] (middle) -- (outgoing_proj);
    \draw[gray, dotted] (outgoing) -- (outgoing_proj);
    
    \pic ["$\phi$", draw, stealth-, angle eccentricity=1.6, angle radius=0.6cm]{angle=pf4--pf5--bottom};
    \pic ["$-\varphi_i$", draw, -stealth, angle eccentricity=1.2, angle radius=1.5cm]{angle=incident_proj--middle--incident};
    \pic ["$-\theta_i$", draw, -stealth, angle eccentricity=1.6, angle radius=0.5cm]{angle=top--middle--incident_proj};
    \pic ["$\theta_o$", draw, stealth-, angle eccentricity=1.5, angle radius=0.6cm]{angle=outgoing_proj--middle--top};
    \pic ["$\varphi_o$", draw, stealth-, angle eccentricity=1.2, angle radius=1.5cm]{angle=outgoing--middle--outgoing_proj};
  
  \end{tikzpicture}
  \caption{Symbols of the local wedge coordinate system used in this paper. The ranges for the angles are $(\phi-\pi, \pi-\phi)$ for $\theta_i$ and $\theta_o$, $(-\pi/2, \pi/2)$ for $\varphi_i$ and $\varphi_o$, and $(0,\pi/2)$ for $\phi$. The dashed vectors are on the normal plane determined by $\mathbf{n}$ and $\mathbf{b}$. }
  \label{fig-coord}
\end{figure}

On the surface of the geometry, we only need to take wave reflection and refraction effect into consideration. Since the scene geometry is described with triangle meshes, the surface is locally flat almost everywhere, and the only natural direction given by the geometry is the normal direction. However, for edges of the triangle mesh, we can derive a natural local coordinate frame from the geometry.

The coordinate system used in this paper is illustrated in \autoref{fig-coord}. Similar to previous local frames, we name the three vectors in the local coordinate frame $\mathbf{T}$ normal vector $\mathbf{n}$, tangent vector $\mathbf{t}$ and bitangent vector $\mathbf{b}$. The expressions of incident and outgoing vectors in the local coordinate frame are given below:

\begin{equation}
  \begin{aligned}
  \mathbf{v}_i & = & -\cos\theta_i\cos\varphi_i\mathbf{n}-\sin\theta_i\cos\varphi_i\mathbf{b}-\sin\varphi_i\mathbf{t}, \\
  \mathbf{v}_o & = & \cos\theta_o\cos\varphi_o\mathbf{n}+\sin\theta_o\cos\varphi_o\mathbf{b}+\sin\varphi_o\mathbf{t}. \\
  \end{aligned}
\end{equation}

These expressions allow us to express BEDRF with angles: \par\noindent $\rho_\mathbf{x}(\varphi_i, \theta_i, \varphi_o, \theta_o)$. The reciprocity of the wave equation's solution \cite{Case1993StructuralAA} demands that the BSDF satisfies the constraint $\rho_\mathbf{x}(\mathbf{v}_i, \mathbf{v}_o) = \rho_\mathbf{x}(-\mathbf{v}_o, -\mathbf{v}_i)$. Replacing the vector parameters with angles, the constraint becomes $\rho_\mathbf{x}(\varphi_i, \theta_i, \varphi_o, \theta_o) = \rho_\mathbf{x}(\varphi_o, \theta_o, \varphi_i, \theta_i)$.

\section{bidirectional edge diffraction response function (BEDRF)}
\label{section:BEDRF}

Our new diffraction representation is derived from \cite{keller1951diffraction}, which give the analytic solution of a plane wave reflected and diffracted by a wedge consisting of two half-planes with Dirichlet or Neumann boundary condition.

We introduce the following symbols:

\begin{equation}
  r=\frac{1-\sin\varphi_i\sin\varphi_o-|\sin\varphi_i-\sin\varphi_o|}{\cos\varphi_i\cos\varphi_o}
  \label{eq:r}
\end{equation}

\begin{equation}
  \kappa=\frac{\pi}{2\pi-2\phi}
\end{equation}

\begin{equation}
  f(\theta_0,\theta_1,\theta,r)=\frac1\pi\arctan\left(\frac{\sinh(\kappa r)\sin\left(\kappa\frac{\theta_0-\theta_1}{2}\right)}{\cos\kappa(\theta-\frac{\theta_0+\theta_1}{2})-\cosh(\kappa r)\cos\left(\kappa\frac{\theta_0-\theta_1}{2}\right)}\right)
  \label{eq:f}
\end{equation}

\begin{equation}
  \begin{array}{rcl}
    \omega_{lu} & = & \theta_i-\pi \\
    \omega_{lr} & = & -\theta_i-\pi+2\phi \\
    \omega_{ru} & = & \theta_i+\pi \\
    \omega_{rr} & = & -\theta_i+\pi-2\phi \\
  \end{array}
\end{equation}

For the Dirichlet boundary condition, the BEDRF is given below:

\begin{equation}
    \rho=\frac12\left(\frac{df}{dr}(\omega_{lu},\omega_{ru},\theta_o, \ln r)+\frac{df}{dr}(\omega_{lr},\omega_{rr},\theta_o, \ln r)\right) 
    \label{EQ:BSDF-DIRICHLET}
\end{equation}


For the Neumann boundary condition, the BEDRF becomes

\begin{equation}
    \rho=\frac12\left(\frac{df}{dr}(\omega_{lu},\omega_{ru},\theta_o, \ln r)-\frac{df}{dr}(\omega_{lr},\omega_{rr},\theta_o, \ln r)\right) 
    \label{eq:bsdf-neumann}
\end{equation}


The expression of $\frac{df}{dr}$ is given in the supplementary material, along with the detailed derivation process and the proof for reciprocity.

\begin{figure*}
  \centering
  \includegraphics[width=0.45\linewidth]{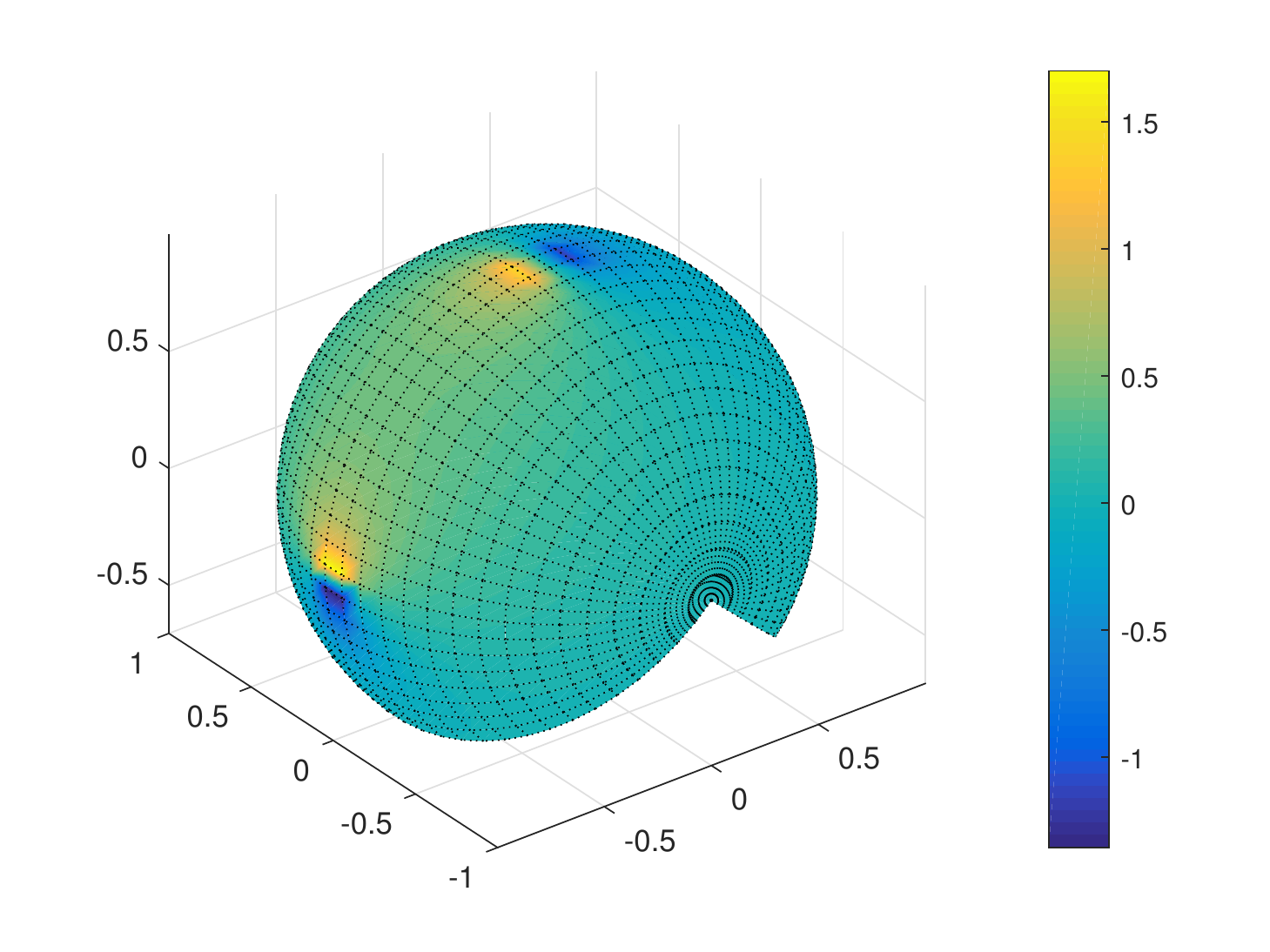}
  \includegraphics[width=0.45\linewidth]{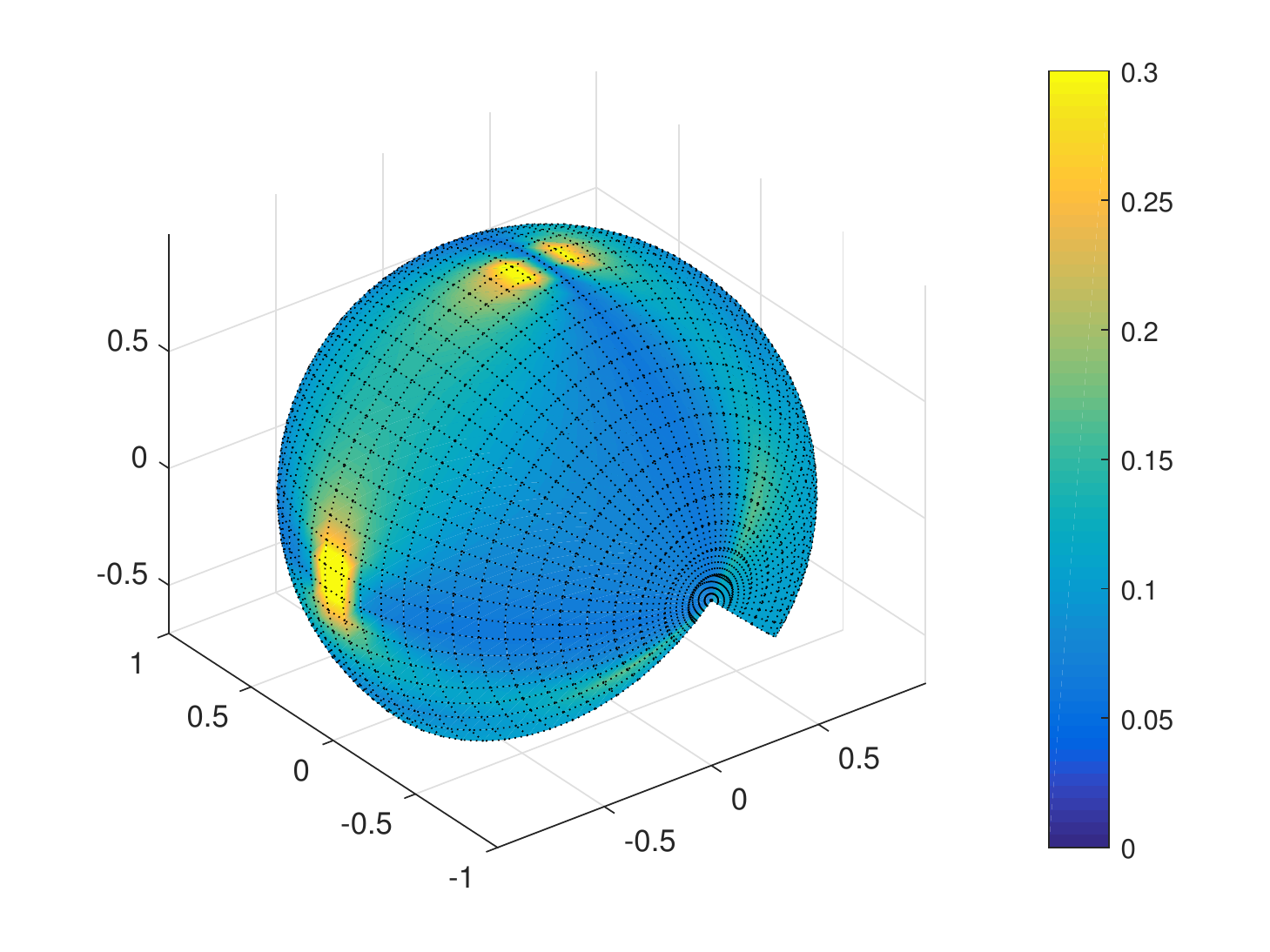}
  \caption{Visualization of BEDRF (left) given by \autoref{EQ:BSDF-DIRICHLET} and the sampler PDF (right) for the Dirichlet boundary condition, with $\phi=\pi/4$, $\varphi_i=0$ and $\theta_i=\pi/2$. The figure shows that the sampler PDF concentrates on the intensive part of BEDRF. \RedComment{ The most prominent part of the BEDRF is around the incident and reflection direction. Not really something useful to know.}}
  \label{fig-bsdf}
\end{figure*}

\autoref{fig-bsdf} visualizes our BEDRF in a special case. Unlike traditional BSDF, our function is not non-negative. The function also has two singularities on the incident, outgoing and/or reflection direction (the natural outgoing direction may not exist in the domain), and the value of \MethodName varies a lot around these singularities. The variation of this BSDF requires us to design a specialized sampler for high quality sampling.

\subsection{Importance Sampling}

In Monte Carlo integration, a good sampler can significantly lower the result variance and reduce the numerical error. When we design an importance sampler for function $f$, we are looking for a random variable with a probability density function (PDF) similar to $f$. In the ideal case, the PDF of the sampler should be $kf$, where $k$ is an constant. Such a sampler can be constructed for an explicitly given and integrable $f:\mathbb{R}\mapsto\mathbb{R}$ using the inverse transform sampling scheme \cite{Ross1987IntroductionTP}: Let $F(x)=\int_{-\infty}^{x}f(y)dy/\int_{-\infty}^{+\infty}f(y)dy$, $S$ be a random variable that distributes uniformly on $[0,1]$, then $F^{-1}(S)$ has the cumulative distribution function (CDF) $F$, and PDF $dF/dx = kf$, where $k=1/\int_{-\infty}^{+\infty}f(x)dx$.

The PDF of a random variable is always non-negative. However, unlike normal BSDFs, our BSDF has both positive and negative parts, which rules out the possibility of finding an ideal sampler. Nevertheless, we can design a function that is "similar" to $F^{-1}$ by using a few of its features to construct our sampler.

Our sampling process proceeds in the following manner. With the predetermined incident ray angle $\theta_i$ and $\varphi_i$, we sample $\theta_o$ first and $\varphi_o$ second. The PDF of our sampler can be written as
\begin{equation}
  p(\theta_o, \varphi_o) = \frac{p_\theta(\theta_o)p_\varphi(\theta_o,\varphi_o)}{\cos\varphi_o},
\end{equation}
where $p_\theta$ and $p_\varphi$ are PDFs for $\theta_o$ and $\varphi_o$ satisfying the following equations:

\begin{equation}
  \int_{\phi-\pi}^{\pi-\phi}p_\theta(\theta_o)d\theta_o=1
\end{equation}

\begin{equation}
  \int_{-\pi/2}^{\pi/2}p_\varphi(\theta_o,\varphi_o)d\varphi_o=1
\end{equation}

Under these conditions, we have $\int p(\theta_o, \varphi_o)\cos\varphi_od\theta_od\varphi_o=1$. For $\theta_o$, we simply use function $p_\theta(\theta_o) = \frac{1}{2(\pi-\phi)}$ as our sampler. For $\varphi_o$, we use the following function:

\begin{equation}
  \label{eq:final-sampler}
  p_\varphi = a+(1-a)\frac{\kappa r^\kappa}{2\cos\varphi_o}\frac{dP_c(\theta_a,\theta_b)}{dx}(r^k), a\in [0,1].
\end{equation}

The expression of $a$, $P_c$, $\theta_a$ and $\theta_b$ are complex and one can find their definitions in the supplementary material of our paper. The shape of the resultant PDF is very similar to BEDRF and can neutralize its singularities during sampling.

\section{Geometric Propagation using Path Tracing}
\label{section:path-tracing}

We use the bidirectional path tracing (BDPT) algorithm from \cite{cao2016interactive} as the backbone of our new integrator. To integrate the diffraction phenomenon into path tracing, we need to perform integration on edges of the scene objects. However, in traditional path tracing, the probability of hitting a point on any edge is exactly zero. Thus we need to make several modifications to the original BDPT algorithm while ensuring that the path tracing method is unbiased.

\subsection{Meta-Path Space}

To intersect a geometry with zero hit probability, a natural solution is to intersect the ray with a ``proxy geometry'' of nonzero hit probability, and map the intersection to a point on the original geometry. If we denote the intersection on the proxy geometry with $\mathbf{x}'$, and the mapping function with $f_m$, then the pair $(\mathbf{x}',\mathbf{x}=f_m(\mathbf{x}'))$ can be regarded as an generalized form of intersection.

We call the series
\begin{equation}
  \mathbf{x}^*_1\to\mathbf{x}^*_2\to\cdots\to\mathbf{x}^*_n, \mathbf{x}^*_i\in\Omega^*
\end{equation}
a \textit{meta-path} on the geometry $\Omega$ if $\Omega^*$ is a measure space and there is a projection $\Pi:\Omega^*\to\Omega$ that preserves measurability. All possible meta-paths containing $n$ nodes form a measure space $\Omega^{*n}$, called the $n$-th \textit{meta-path space}.

An example of the meta-path space is the ``single-point augmented path space'' described above, where every node of the meta-path is a point pair $(\mathbf{x}',\mathbf{x})$. The first point is on the proxy geometry, and the second point is on the real one. The projection function for such space is $\Pi((\mathbf{x}',\mathbf{x}))=\mathbf{x}$. There could be other forms of meta-path nodes that may contain more points or other types of information.

In traditional path tracing, the intersection probability gives a probability measure $\mu$ on the path space, allowing us to do the Monte Carlo integration. For non-intersectable geometries like infinitely thin edges, this measure no longer exists. However, since the proxy geometry is intersectable, we can find a similar probability measure $\mu^*$ defined on some certain meta-path space. Now, it can be proved that the canonical projection $\Pi$ induces a probability measure $\Pi(\mu^*)$ on the original path space. In addition, every integral on the original path space can be converted to an integral on $\mu^*$. Thus the Monte Carlo integral can be done on the meta-path space. All the sampling techniques, including MIS and metropolis sampling, are also feasible on $\mu^*$.   More information about measure projection is given in~\cite{Tarantola08measure} . We use this theoretical basis to work out the necessary details of our method.

\begin{figure}
  \centering
  \begin{tikzpicture}
    \coordinate (source) at (-4.8, 2.5);
    \coordinate (project) at (-4.5, 0.2855);
    \coordinate (plane-l1) at (-5.75,-0.4);
    \coordinate (plane-l2) at (-2.8,-0.5);
    \coordinate (plane-r1) at (-4.4,0.9);
    \coordinate (plane-r2) at (-4,0);
    \coordinate (plane-r3) at (-4.2,0.45);
    \coordinate (plane-r4) at (-5.6,-0.3177);
    \coordinate (plane-r5) at (-5.1,-0.0435);
    \coordinate (plane-u4) at (-5.3750,0.4750);
    \coordinate (plane-u5) at (-5.0140,0.6860);
    \coordinate (plane-a1) at (-5.6,0.7);
    \coordinate (plane-a2) at (-5.3,0.4);
    \coordinate (plane-a3) at (-4.9,0.8);

    \fill[gray!10] (plane-l1) -- (plane-r1) -- (plane-l2) -- (plane-r2) -- cycle;
    \fill[gray!10] (plane-a1) -- (plane-a2) -- (plane-a3) -- cycle;
    \draw[thick, gray] (plane-l1) -- (plane-r1) -- (plane-r2) -- cycle;
    \draw[thick, gray] (plane-r1) -- (plane-l2) -- (plane-r2);
    \draw[thick, gray] (plane-a1) -- (plane-a2) -- (plane-a3) -- cycle;
    \draw[thick, black] (plane-r1) -- (plane-r2);
    \fill[green!60, opacity=0.2] (source) -- (plane-l1) -- (plane-r4) -- (plane-u4) -- (plane-u5) -- (plane-r5) -- (plane-r3) -- (plane-l2) -- cycle;
    \draw[blue-base, thick] (plane-l1) -- (plane-r4);
    \draw[gray] (plane-r4) -- (plane-r5);
    \draw[blue-base, thick] (plane-r5) -- (plane-r3) -- (plane-l2);
    \draw[black, thick, -stealth] (source) -- (project);
    \draw[gray-base-sombre, thick, -stealth] (source) -- (plane-r3);
    \draw[gray!50] (source) -- (plane-l1);
    \draw[gray!50] (source) -- (plane-l2);
    \draw[gray!50] (source) -- (plane-r4);
    \draw[gray!50] (source) -- (plane-r5);
    \draw[gray!50] (plane-u4) -- (plane-u5);
    \fill[red-base](plane-r1)circle(0.05) node[black,above right] {$\mathbf{a}$};
    \fill[red-base](plane-r2)circle(0.05) node[black,below] {$\mathbf{b}$};
    \fill[red-base](plane-l1)circle(0.05) node[black,below] {$\mathbf{c}_1$};
    \fill[red-base](plane-l2)circle(0.05) node[black,below] {$\mathbf{c}_2$};
    \fill[red-base](project)circle(0.05) node[black,above left] {$\mathbf{x}'$};
    \fill[red-base](plane-r3)circle(0.05) node[black,right] {$\mathbf{x}$};
    \fill[red-base](source)circle(0.05) node[black,above] {$\mathbf{x}_s$};

  \end{tikzpicture}
  \caption{An illustration of our edge detection algorithm. We emit a ray from the source position $\mathbf{x}_s$ and intersect with the scene geometry at $\mathbf{x}' \in \triangle\mathbf{abc}_1$. Suppose that $\mathbf{ab}$ is a convex edge. We extend $\mathbf{c}_1\mathbf{x}'$ and intersect $\mathbf{ab}$ at $\mathbf{x}$, which we take as the real intersection point. After intersection, we will also calculate the proxy set $S_\mathbf{x}$ of $\mathbf{x}$, which is composed of segments on $\mathbf{xc}_1$ and $\mathbf{xc}_2$ visible from $\mathbf{x}_s$ (blue lines).}
  \label{fig-proxy-mesh}
\end{figure}
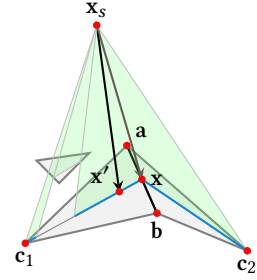

\subsection{Integration on Edges}
\label{subsection:integration-on-edges}

It is not difficult to realize that, since we are using triangle meshes for scene geometry, the geometry itself can be used as a natural proxy geometry for all its convex edges. Compared with other possible choices for proxy meshes, using the original geometry requires almost no precalculation, and allows our algorithm to support dynamic, deformable objects with no extra overhead.

When detecting edges, we first emit a random ray from the source $\mathbf{x}_s$ and perform an intersection test with the scene geometry $S$. We name the intersection point $\mathbf{x}' \in \triangle\mathbf{abc}$ as the ``pseudo-intersection point''. If $\triangle\mathbf{abc}$ has convex edges, we will select one of them and map $\mathbf{x}'$ onto it. The selection probability of each convex edge is equal and we note it with $p_{\triangle\mathbf{abc}}$. After selection, we connect $\mathbf{x}'$ with the opposite vertex of the selected edge, extend the segment, and intersect the selected edge at $\mathbf{x}$, which we called the ``real intersection point''. $\mathbf{x}$ and $\mathbf{x}'$ form a node $(\mathbf{x}',\mathbf{x})$ in our meta path space. Note that $\mathbf{x}$ is not necessarily visible from $\mathbf{x}_s$. If $\mathbf{x}$ is invisible, then we consider this intersection as a failure.

Now, we need to convert the integration on the edge to an integration on its neighbor triangles. Take the case in \autoref{fig-proxy-mesh} as an example. The conversion can be expressed like this:

\begin{equation}
\int_{x\in\mathbf{ab}}fd\mu=\int_{x\in\triangle\mathbf{abc}_1\cup\triangle\mathbf{abc}_2}f\frac{d\Pi^{-1}(\mu)}{d\mu^*}d\mu^*.
\end{equation}
Note that one convex edge has two neighbor triangles, and we need to consider them both. We first calculate all possible pseudo-intersections on the neighboring triangles that possibly project to $\mathbf{x}$. The set of these pseudo-intersections is called the ``proxy set'' of $\mathbf{x}$, noted as $S_\mathbf{x}$. We intersect $\triangle \mathbf{x}_s\mathbf{x}\mathbf{c}_1$ and $\triangle \mathbf{x}_s\mathbf{x}\mathbf{c}_2$ with $S$, and $S_\mathbf{x}$ is the unoccluded parts of the triangles intersecting with $S$.

Afterwards, we map the set $S_\mathbf{x}$ to $[0,1]$ in the following manner:

\begin{equation}
  \begin{aligned}
    q(\mathbf{x''}) & =\frac{||\mathbf{x''}-\mathbf{c}_1||_2}{||\mathbf{x}-\mathbf{c}_1||_2}, \mathbf{x}''\in S_\mathbf{x}\cap \mathbf{xc}_1, \\
    q(\mathbf{x''}) & =\frac{||\mathbf{x''}-\mathbf{c}_2||_2}{||\mathbf{x}-\mathbf{c}_2||_2}, \mathbf{x}''\in S_\mathbf{x}\cap \mathbf{xc}_2. \\
  \end{aligned}
\end{equation}
Then the expression of $d\Pi^{-1}(\mu)/d\mu^*$ can be given as
\begin{equation}
  \begin{aligned}
    \frac{d\mu^*}{d\Pi^{-1}(\mu)}(\mathbf{x}') &=\frac{p_{\triangle\mathbf{abc}_1}S\triangle\mathbf{abc}_1}{||\mathbf{b}-\mathbf{a}||_2}\int_{q(S_\mathbf{x}\cap \mathbf{xc}_1)}xdx, \\
    & +\frac{p_{\triangle\mathbf{abc}_2}S\triangle\mathbf{abc}_2}{||\mathbf{b}-\mathbf{a}||_2}\int_{q(S_\mathbf{x}\cap \mathbf{xc}_2)}xdx,
  \end{aligned}
\end{equation}
where $S\triangle\mathbf{abc}$ is the area of $\triangle\mathbf{abc}$ and $dx$ is the ordinary differential on $[0,1]$. The proof of this equation is given in the supplementary material.

\subsection{MIS Issue for Meta-Paths}

\RedComment{Need an introduction to MIS. \cite{Veach1995OptimallyCS}.}

In traditional bidirectional path tracing, it is possible to generate the same path from the source to the listener from multiple sample strategies. For example, a path with 4 nodes could be generated by strategy $(0,3)$, $(1,2)$, $(2,1)$ or $(3,0)$, where the two numbers are forward and backward path depths. The generation probability of a path for different strategies allows us to use the MIS technique. However, there is some additional complexity for our meta-paths.

Consider a meta path generated with the algorithm in the previous subsection:

\begin{equation}
  (\mathbf{x}'_0,\mathbf{x}_0)\to(\mathbf{x}'_1,\mathbf{x}_1)\to\cdots\to(\mathbf{x}'_n,\mathbf{x}_n)
\end{equation}
If $V$ is the visibility function, we know from the generation process that $V(\mathbf{x}_i,\mathbf{x}_{i+1})=1$, $V(\mathbf{x}_i,\mathbf{x}'_{i+1})=1$. Now, MIS requires that the inverse of the path

\begin{equation}
  (\mathbf{x}'_0,\mathbf{x}_0)\leftarrow\cdots\leftarrow(\mathbf{x}'_{n-1},\mathbf{x}_{n-1})\leftarrow(\mathbf{x}'_n,\mathbf{x}_n)
\end{equation}
can be generated by the backward path tracer and its generation probability computable. However, this requires an additional condition $V(\mathbf{x}'_i,\mathbf{x}_{i+1})=1$. Since our edge tracing algorithm is not symmetric, this is not guaranteed by the generation process.

To solve this problem, we use a twice-augmented path space in our renderer, where the path is expressed as

\begin{equation}
  (\mathbf{x}''_0,\mathbf{x}'_0,\mathbf{x}_0)\to(\mathbf{x}''_1,\mathbf{x}'_1,\mathbf{x}_1)\to\cdots\to(\mathbf{x}''_n,\mathbf{x}'_n,\mathbf{x}_n).
\end{equation}
Both $\mathbf{x}'_i$ and $\mathbf{x}''_i$ projects to $\mathbf{x}$. We require that the path satisfies one more condition $V(\mathbf{x}''_i,\mathbf{x}_{i+1})=1$. In the generation process, this is achieved by choosing a random $\mathbf{x}''_i$ from the proxy segment visible from $\mathbf{x}_{i+1}$ (in simpler words, ``reverse tracing''). In this way, the three visible conditions are symmetric under the inverse transform. We can then use MIS without any problems.

\subsection{Inverse Russian Roulette}

We have seen from \autoref{subsection:integration-on-edges} that our edge intersection algorithm may fail when the real intersection point is occluded. This is not an uncommon case in actual application. Hence it is harder to generate path segments for diffraction. To exploit every valid intersection at maximum, we use a technique similar to the Russian roulette in \cite{Guibas1997RobustMC}, but reversing the process: instead of discarding low-intensity path nodes, we divide high-intensity nodes into multiple shares to generate more path segments.

After each bounce of path tracing, we will duplicate successful intersections and replace the failed intersections sharing the same starting node. When an intersection is duplicated $n$ times, we will multiply the generation probability of the original intersection and all its duplicates by $n+1$. If there are multiple successful intersections, the number of duplications of each intersection is largely proportional to its intensity, which is $f/p$ in \autoref{eq:IntegralEstimatorExpanded}. In this way, the intensity of duplicated intersections would be close to each other. All the intersections generated in this way forms a forward and backward tree, where intensive nodes have more successor branches than other nodes. In this way, we are able to generate much more valid path nodes without significantly increasing the number of intersection tests. 

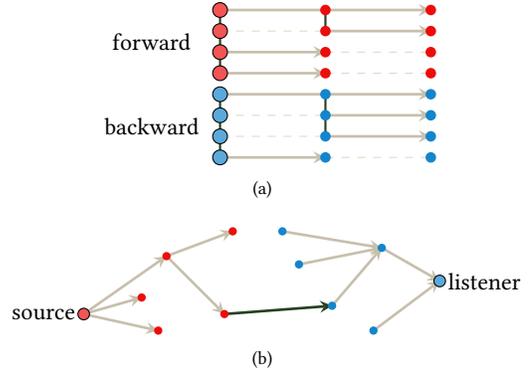
\begin{figure}[ht]
  \centering
  \subfigure[]{
    \begin{tikzpicture}[scale=1.4]
    \coordinate (node1-1) at (-2,0);
    \coordinate (node1-2) at (-1,0);
    \coordinate (node1-3) at (0,0);
    \coordinate (node1-4) at (1,0);
    \coordinate (node1-5) at (2,0);
    \coordinate (node2-1) at (-2,-0.2);
    \coordinate (node2-2) at (-1,-0.2);
    \coordinate (node2-3) at (0,-0.2);
    \coordinate (node2-4) at (1,-0.2);
    \coordinate (node2-5) at (2,-0.2);
    \coordinate (node3-1) at (-2,-0.4);
    \coordinate (node3-2) at (-1,-0.4);
    \coordinate (node3-3) at (0,-0.4);
    \coordinate (node3-4) at (1,-0.4);
    \coordinate (node3-5) at (2,-0.4);
    \coordinate (node4-1) at (-2,-0.6);
    \coordinate (node4-2) at (-1,-0.6);
    \coordinate (node4-3) at (0,-0.6);
    \coordinate (node4-4) at (1,-0.6);
    \coordinate (node4-5) at (2,-0.6);

    \coordinate (node5-1) at (-2,-0.8);
    \coordinate (node5-2) at (-1,-0.8);
    \coordinate (node5-3) at (0,-0.8);
    \coordinate (node5-4) at (1,-0.8);
    \coordinate (node5-5) at (2,-0.8);
    \coordinate (node6-1) at (-2,-1.0);
    \coordinate (node6-2) at (-1,-1.0);
    \coordinate (node6-3) at (0,-1.0);
    \coordinate (node6-4) at (1,-1.0);
    \coordinate (node6-5) at (2,-1.0);
    \coordinate (node7-1) at (-2,-1.2);
    \coordinate (node7-2) at (-1,-1.2);
    \coordinate (node7-3) at (0,-1.2);
    \coordinate (node7-4) at (1,-1.2);
    \coordinate (node7-5) at (2,-1.2);
    \coordinate (node8-1) at (-2,-1.4);
    \coordinate (node8-2) at (-1,-1.4);
    \coordinate (node8-3) at (0,-1.4);
    \coordinate (node8-4) at (1,-1.4);
    \coordinate (node8-5) at (2,-1.4);

      \definecolor{bounce4}{rgb}{0.1,0.3,0.1}
      \definecolor{bounce3}{rgb}{0.1,0.5,0.2}
      \definecolor{bounce2}{rgb}{0.2,0.7,0.5}
      \definecolor{bounce1}{rgb}{0.2,0.9,0.7}
      \definecolor{bounce0}{rgb}{0.4,1.0,0.9}

      \draw[gray-base-dark, thick](node1-1)--(node4-1);
      \draw[gray-base-dark, thick](node1-2)--(node2-2);
      \draw[gray-base-dark, thick](node5-1)--(node8-1);
      \draw[gray-base-dark, thick](node5-2)--(node7-2);
      \draw[gray-base, dashed](node2-1)--(node2-2);
      \draw[gray-base, dashed](node3-2)--(node3-3);
      \draw[gray-base, dashed](node4-2)--(node4-3);
      \draw[gray-base, dashed](node6-1)--(node6-2);
      \draw[gray-base, dashed](node7-1)--(node7-2);
      \draw[gray-base, dashed](node8-2)--(node8-3);
      \draw[-stealth, gray-base-thick, line width=1pt](node1-1)--(node1-2)--(node1-3);
      \draw[-stealth, gray-base-thick, line width=1pt](node2-2)--(node2-3);
      \draw[-stealth, gray-base-thick, line width=1pt](node3-1)--(node3-2);
      \draw[-stealth, gray-base-thick, line width=1pt](node4-1)--(node4-2);
      \draw[-stealth, gray-base-thick, line width=1pt](node5-1)--(node5-2)--(node5-3);
      \draw[-stealth, gray-base-thick, line width=1pt](node6-2)--(node6-3);
      \draw[-stealth, gray-base-thick, line width=1pt](node7-2)--(node7-3);
      \draw[-stealth, gray-base-thick, line width=1pt](node8-1)--(node8-2);

      \draw [fill = red-base!70](-2.65,-0.3) node[black] {forward};
      \draw [fill = red-base!70](-2.65,-1.1) node[black] {backward};

      \foreach \i in {1,...,4} {
        \draw [fill = red-base!70](node\i-1)circle(0.07);
        \fill [red-base](node\i-2)circle(0.05);
        \fill [red-base](node\i-3)circle(0.05);
      }

      \foreach \i in {5,...,8} {
        \draw [fill = blue-base!70](node\i-1)circle(0.07);
        \fill [blue-base](node\i-2)circle(0.05);
        \fill [blue-base](node\i-3)circle(0.05);
      }
      
    \end{tikzpicture}
  } \\
  \subfigure[]{
    \begin{tikzpicture}[scale=1.1]
      \coordinate (node1) at (-2,-0.3);
      \coordinate (node2) at (-1,0.4);
      \coordinate (node2-2) at (-1.3,-0.1);
      \coordinate (node2-3) at (-1.1,-0.5);
      \coordinate (node3) at (-0.3,-0.3);
      \coordinate (node3-2) at (-0.2,0.7);
      \coordinate (node5) at (1.0,-0.2);
      \coordinate (node5-2) at (0.6,0.3);
      \coordinate (node5-3) at (0.4,0.7);
      \coordinate (node6) at (1.6,0.5);
      \coordinate (node6-2) at (1.5,-0.5);
      \coordinate (node7) at (2.3,0.1);
  
      \definecolor{bounce4}{rgb}{0.1,0.3,0.1}
      \definecolor{bounce3}{rgb}{0.1,0.5,0.2}
      \definecolor{bounce2}{rgb}{0.2,0.7,0.5}
      \definecolor{bounce1}{rgb}{0.2,0.9,0.7}
      \definecolor{bounce0}{rgb}{0.4,1.0,0.9}
  
      \draw[-stealth, gray-base-thick, line width=1pt](node1)--(node2);
      \draw[-stealth, gray-base-thick, line width=1pt](node1)--(node2-2);
      \draw[-stealth, gray-base-thick, line width=1pt](node1)--(node2-3);
      \draw[-stealth, gray-base-thick, line width=1pt](node2)--(node3);
      \draw[-stealth, gray-base-thick, line width=1pt](node2)--(node3-2);
      \draw[-stealth, gray-base-dark, line width=1pt](node3)--(node5);
      \draw[-stealth, gray-base-thick, line width=1pt](node5)--(node6);
      \draw[-stealth, gray-base-thick, line width=1pt](node5-2)--(node6);
      \draw[-stealth, gray-base-thick, line width=1pt](node5-3)--(node6);
      \draw[-stealth, gray-base-thick, line width=1pt](node6)--(node7);
      \draw[-stealth, gray-base-thick, line width=1pt](node6-2)--(node7);
  
      \draw [fill = red-base!70](node1)circle(0.07) node[black,left] {source};
      \fill [red-base](node2)circle(0.05);
      \fill [red-base](node2-2)circle(0.05);
      \fill [red-base](node2-3)circle(0.05);
      \fill [red-base](node3)circle(0.05);
      \fill [red-base](node3-2)circle(0.05);
      \fill [blue-base](node5)circle(0.05);
      \fill [blue-base](node5-2)circle(0.05);
      \fill [blue-base](node5-3)circle(0.05);
      \fill [blue-base](node6)circle(0.05);
      \fill [blue-base](node6-2)circle(0.05);
      \draw [fill = blue-base!70](node7)circle(0.07) node[black,right] {listener};
    \end{tikzpicture}
  }
\caption{Illustration of inverse Russian roulette. We duplicate successful intersections and replace failed intersections with the same starting node with duplicated intersections. In Fig.\. (a), we use dashed lines and solid arrows to represent failed/successful intersection, and vertical black lines to represent node duplication. When connecting forward and backward paths, we are actually connecting nodes between a forward and a backward tree, which is shown in Fig.\.(b).}
\label{fig:PathReuse}
\end{figure}

\subsection{Outlier Suppression}

We know from \autoref{subsection:integration-on-edges} that multiple pseudo-intersections can be mapped to the same real intersection point. Different pseudo-intersection positions lead to different sample generating probabilities, and occasionally it could be very low. When the direction of some certain path segment is also near the singularities of the BEDRF, the sample intensity would be very high and produces an outlier in the result. The position of the pseudo-intersection is determined by the outgoing direction of the previous BEDRF sampler and is difficult to control. But it is not hard to identify these samples and suppress their intensity.

Here we note the starting position of a ray segment as $\mathbf{x}_s$, the pseudo-intersection as $\mathbf{x}'$, the incident angle at the pseudo-intersection as $\theta$, the real intersection as $\mathbf{x}$, and the outgoing probability of the sampler at $\mathbf{x}_s$ on the direction $\mathbf{v}$ as $p(\mathbf{v})$. It is not hard to calculate

\begin{equation}
  C_o=\frac{p(\mathbf{x}-\mathbf{x}_s)}{p(\mathbf{x}'-\mathbf{x}_s)}\cdot\frac{||\mathbf{x}'-\mathbf{x}_s||_2^2}{||\mathbf{x}-\mathbf{x}_s||_2^2\cos\theta}.
\end{equation}
A large $C_o$ value indicates that the sample probability on the direction $\mathbf{x}'-\mathbf{x}_s$ is much smaller than that on the direction $\mathbf{x}-\mathbf{x}_s$, thus will likely generate an outlier. The $C_o$ value of a sample (a full path) is the product of all $C_o$ values of its non-connection path segments.

To suppress the outliers, we allow the user to set a maximum threshold for $C_o$, which we note as $C^*_o$ here. We then multiply the intensity of every sample with a logistic suppression coefficient $C_s$. The expression of $C_s$ is given below:

\begin{equation}
  \label{eq:logistic-suppressor}
  C_s = \frac{C_o^*(1-e^{-2C_o/C_o^*})}{C_o(1+e^{-2C_o/C_o^*})}
\end{equation}

Notice that when we use outlier suppression, the result of our path tracer is no longer unbiased. However, the suppression coefficient will only affect samples with high $C_o$ and low probability, and it will not affect the intensity of the simulation result too much when the scene geometry is simple and the sample quality is high. This is also shown in \autoref{FIG:TECHNIQUES} in the result section.

\section{Implementation and Results}
In this section, we describe our implementation and highlight its performance on different benchmarks. Some of the results are presented in the supplementary video of this paper.

\subsection{Implementation and Performance}

\begin{table*}[htbp]
    \centering
    \begin{tabular}{cccc}
    \toprule
    model & \textbf{Boxes} & \textbf{Roomset} & \textbf{Sponza} \\
    \midrule
    & \includegraphics[width=0.15\linewidth]{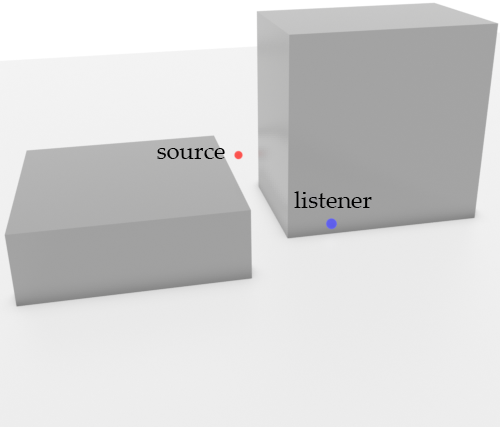} & \includegraphics[width=0.28\linewidth]{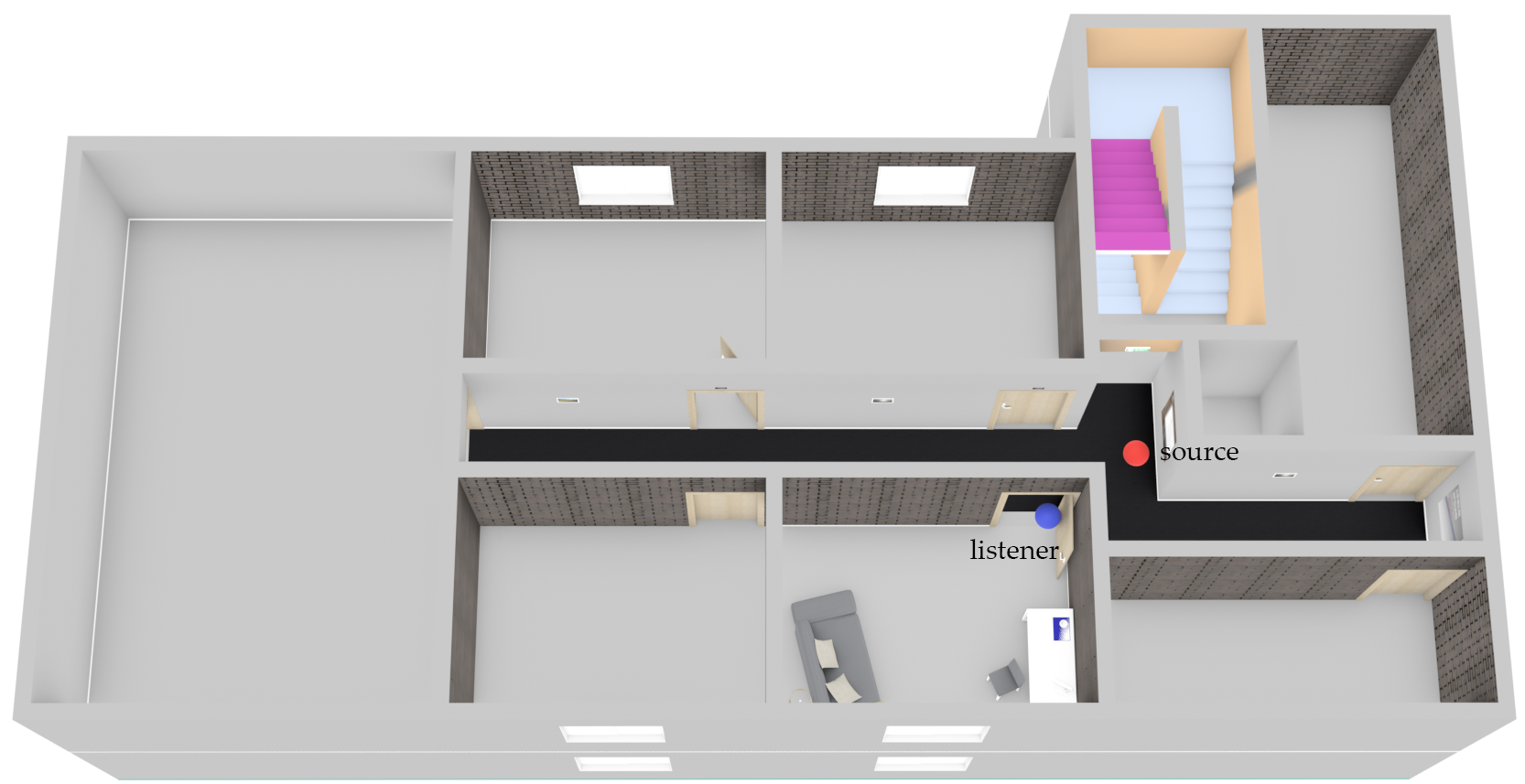} & \includegraphics[width=0.26\linewidth]{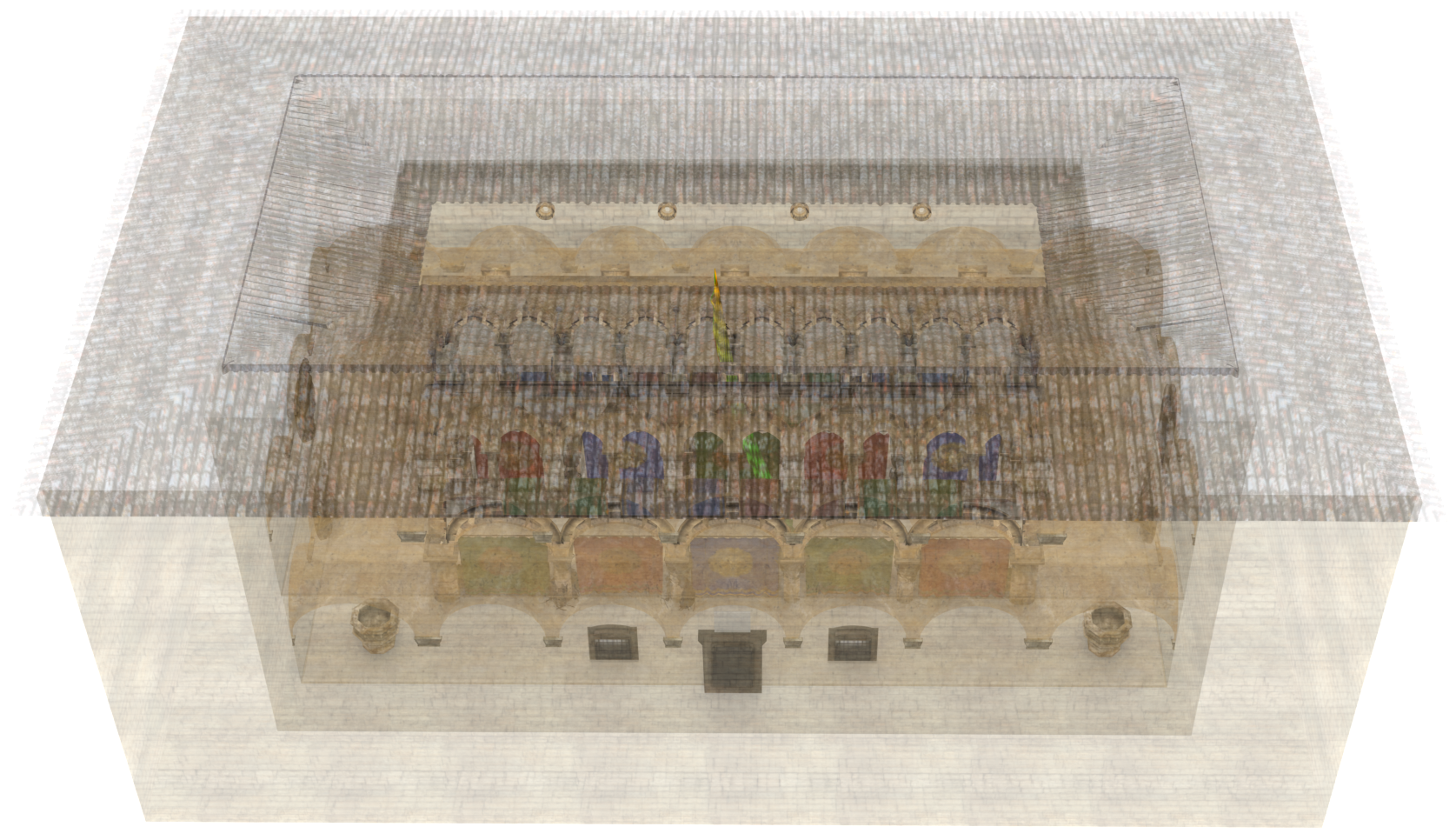} \\
    \hline
    vertices & 20 & 11087 & 153635 \\
    triangles & 26 & 22974 & 279133 \\
    diffraction edges & 16 & 17184 & 220056 \\
    other features & & complex occlusion & \\
    \hline
    time cost per frame & 34ms & 140ms & 205ms \\
    \bottomrule
    \end{tabular}
    \caption{Information of benchmark models used for auralization and performance of our diffraction path tracer in these models.}
    \label{tbl:models}
\end{table*}

\begin{figure*}[htbp]
  \centering
  \begin{tabular}{cc|cc}
    \toprule
    \multicolumn{2}{c|}{\textbf{Boxes}} & \multicolumn{2}{c}{\textbf{Roomset}} \\
    \midrule
    \multicolumn{4}{c}{Plain BDPT} \\
    \hline
    IR SNR: -3.1618 dB & time cost: 19ms & IR SNR: -4.5834 dB & time cost: 60 ms \\
    \hline
    \multicolumn{2}{c|}{\begin{tikzpicture}
      \begin{axis}[
        width = 0.4\textwidth,
        height = 0.23\textwidth,
        axis y line = left,
        axis x line = bottom,
        xlabel      = {Time/ms},
        ylabel      = {pressure/Pa},
        xmin = 30, xmax = 65,
        ymin = -500, ymax = 2500,
        legend cell align = left,
        legend pos = north east,
        legend transposed = true,
        legend style = {draw=none, fill=none},
        grid = major,
        /pgf/number format/.cd,
        use comma,
        1000 sep={}
      ]
        \addplot[gray-base, unbounded coords=jump] table [x expr=\thisrow{time}, y=variance, col sep=comma] {img/IR_BB.txt};
        \addlegendentry{stdvar};
        \addplot[blue-base, unbounded coords=jump] table [x expr=\thisrow{time}, y=pressure, col sep=comma] {img/IR_BB.txt};
        \addlegendentry{IR};
      \end{axis}
    \end{tikzpicture}} & \multicolumn{2}{c}{\begin{tikzpicture}
      \begin{axis}[
        width = 0.4\textwidth,
        height = 0.23\textwidth,
        axis y line = left,
        axis x line = bottom,
        xlabel      = {Time/ms},
        ylabel      = {pressure/Pa},
        xmin = 0, xmax = 65,
        ymin = -10000, ymax = 25000,
        legend cell align = left,
        legend pos = north east,
        legend transposed = true,
        legend style = {draw=none, fill=none},
        grid = major,
        /pgf/number format/.cd,
        use comma,
        1000 sep={}
      ]
        \addplot[gray-base, unbounded coords=jump] table [x expr=\thisrow{time}, y=variance, col sep=comma] {img/IR_RS.txt};
        \addlegendentry{stdvar};
        \addplot[blue-base, unbounded coords=jump] table [x expr=\thisrow{time}, y=pressure, col sep=comma] {img/IR_RS.txt};
        \addlegendentry{IR};
      \end{axis}
    \end{tikzpicture}} \\
    \hline
    \multicolumn{4}{c}{BDPT+MIS} \\
    \hline
    IR SNR: -2.4164 dB & time cost: 22ms & IR SNR: -4.4580 dB & time cost: 73 ms \\
    \hline
    \multicolumn{2}{c|}{\begin{tikzpicture}
      \begin{axis}[
        width = 0.4\textwidth,
        height = 0.23\textwidth,
        axis y line = left,
        axis x line = bottom,
        xlabel      = {Time/ms},
        ylabel      = {pressure/Pa},
        xmin = 30, xmax = 65,
        ymin = -500, ymax = 2500,
        legend cell align = left,
        legend pos = north east,
        legend transposed = true,
        legend style = {draw=none, fill=none},
        grid = major,
        /pgf/number format/.cd,
        use comma,
        1000 sep={}
      ]
        \addplot[gray-base, unbounded coords=jump] table [x expr=\thisrow{time}, y=variance, col sep=comma] {img/IR_BB_MIS.txt};
        \addlegendentry{stdvar};
        \addplot[blue-base, unbounded coords=jump] table [x expr=\thisrow{time}, y=pressure, col sep=comma] {img/IR_BB_MIS.txt};
        \addlegendentry{IR};
      \end{axis}
    \end{tikzpicture}} & \multicolumn{2}{c}{\begin{tikzpicture}
      \begin{axis}[
        width = 0.4\textwidth,
        height = 0.23\textwidth,
        axis y line = left,
        axis x line = bottom,
        xlabel      = {Time/ms},
        ylabel      = {pressure/Pa},
        xmin = 0, xmax = 65,
        ymin = -10000, ymax = 25000,
        legend cell align = left,
        legend pos = north east,
        legend transposed = true,
        legend style = {draw=none, fill=none},
        grid = major,
        /pgf/number format/.cd,
        use comma,
        1000 sep={}
      ]
        \addplot[gray-base, unbounded coords=jump] table [x expr=\thisrow{time}, y=variance, col sep=comma] {img/IR_RS_MIS.txt};
        \addlegendentry{stdvar};
        \addplot[blue-base, unbounded coords=jump] table [x expr=\thisrow{time}, y=pressure, col sep=comma] {img/IR_RS_MIS.txt};
        \addlegendentry{IR};
      \end{axis}
    \end{tikzpicture}} \\
    \hline
    \multicolumn{4}{c}{BDPT+MIS+inverse Russian roulette} \\
    \hline
    IR SNR: 0.0630 dB & time cost: 34ms & IR SNR: -7.6249 dB & time cost: 138 ms \\
    \hline
    \multicolumn{2}{c|}{\begin{tikzpicture}
      \begin{axis}[
        width = 0.4\textwidth,
        height = 0.23\textwidth,
        axis y line = left,
        axis x line = bottom,
        xlabel      = {Time/ms},
        ylabel      = {pressure/Pa},
        xmin = 30, xmax = 65,
        ymin = -500, ymax = 2500,
        legend cell align = left,
        legend pos = north east,
        legend transposed = true,
        legend style = {draw=none, fill=none},
        grid = major,
        /pgf/number format/.cd,
        use comma,
        1000 sep={}
      ]
        \addplot[gray-base, unbounded coords=jump] table [x expr=\thisrow{time}, y=variance, col sep=comma] {img/IR_BB_MIS_IRR.txt};
        \addlegendentry{stdvar};
        \addplot[blue-base, unbounded coords=jump] table [x expr=\thisrow{time}, y=pressure, col sep=comma] {img/IR_BB_MIS_IRR.txt};
        \addlegendentry{IR};
      \end{axis}
    \end{tikzpicture}} & \multicolumn{2}{c}{\begin{tikzpicture}
      \begin{axis}[
        width = 0.4\textwidth,
        height = 0.23\textwidth,
        axis y line = left,
        axis x line = bottom,
        xlabel      = {Time/ms},
        ylabel      = {pressure/Pa},
        xmin = 0, xmax = 65,
        ymin = -10000, ymax = 25000,
        restrict y to domain*=-10000:30000,
        legend cell align = left,
        legend pos = north east,
        legend transposed = true,
        legend style = {draw=none, fill=none},
        grid = major,
        /pgf/number format/.cd,
        use comma,
        1000 sep={}
      ]
        \addplot[gray-base, unbounded coords=jump] table [x expr=\thisrow{time}, y=variance, col sep=comma] {img/IR_RS_MIS_IRR.txt};
        \addlegendentry{stdvar};
        \addplot[blue-base, unbounded coords=jump] table [x expr=\thisrow{time}, y=pressure, col sep=comma] {img/IR_RS_MIS_IRR.txt};
        \addlegendentry{IR};
      \end{axis}
    \end{tikzpicture}} \\
    \hline
    \multicolumn{4}{c}{BDPT+MIS+inverse Russian roulette, outlier suppressed} \\
    \hline
    IR SNR: 0.1271 dB & time cost: 34ms & IR SNR: -0.5212 dB & time cost: 140 ms \\
    \hline
    \multicolumn{2}{c|}{\begin{tikzpicture}
      \begin{axis}[
        width = 0.4\textwidth,
        height = 0.23\textwidth,
        axis y line = left,
        axis x line = bottom,
        xlabel      = {Time/ms},
        ylabel      = {pressure/Pa},
        xmin = 30, xmax = 65,
        ymin = -500, ymax = 2500,
        legend cell align = left,
        legend pos = north east,
        legend transposed = true,
        legend style = {draw=none, fill=none},
        grid = major,
        /pgf/number format/.cd,
        use comma,
        1000 sep={}
      ]
        \addplot[gray-base, unbounded coords=jump] table [x expr=\thisrow{time}, y=variance, col sep=comma] {img/IR_BB_MIS_IRR_OS.txt};
        \addlegendentry{stdvar};
        \addplot[blue-base, unbounded coords=jump] table [x expr=\thisrow{time}, y=pressure, col sep=comma] {img/IR_BB_MIS_IRR_OS.txt};
        \addlegendentry{IR};
      \end{axis}
    \end{tikzpicture}} & \multicolumn{2}{c}{\begin{tikzpicture}
      \begin{axis}[
        width = 0.4\textwidth,
        height = 0.23\textwidth,
        axis y line = left,
        axis x line = bottom,
        xlabel      = {Time/ms},
        ylabel      = {pressure/Pa},
        xmin = 0, xmax = 65,
        ymin = -10000, ymax = 25000,
        legend cell align = left,
        legend pos = north east,
        legend transposed = true,
        legend style = {draw=none, fill=none},
        grid = major,
        /pgf/number format/.cd,
        use comma,
        1000 sep={}
      ]
        \addplot[gray-base, unbounded coords=jump] table [x expr=\thisrow{time}, y=variance, col sep=comma] {img/IR_RS_MIS_IRR_OS.txt};
        \addlegendentry{stdvar};
        \addplot[blue-base, unbounded coords=jump] table [x expr=\thisrow{time}, y=pressure, col sep=comma] {img/IR_RS_MIS_IRR_OS.txt};
        \addlegendentry{IR};
      \end{axis}
    \end{tikzpicture}} \\
    \bottomrule
  \end{tabular}
  \caption{IRs simulated in different scenes with different rendering technique combinations. We highlight the performance on two challenging benchmarks for geometric acoustics in terms of diffraction effects. The average IR SNR and the computation time cost of a single IR is given for each case. We show the relative benefits of MIS, inverse Russian roulette, and outlier suppressed for each of these complex models. }
  \label{FIG:TECHNIQUES}
\end{figure*}
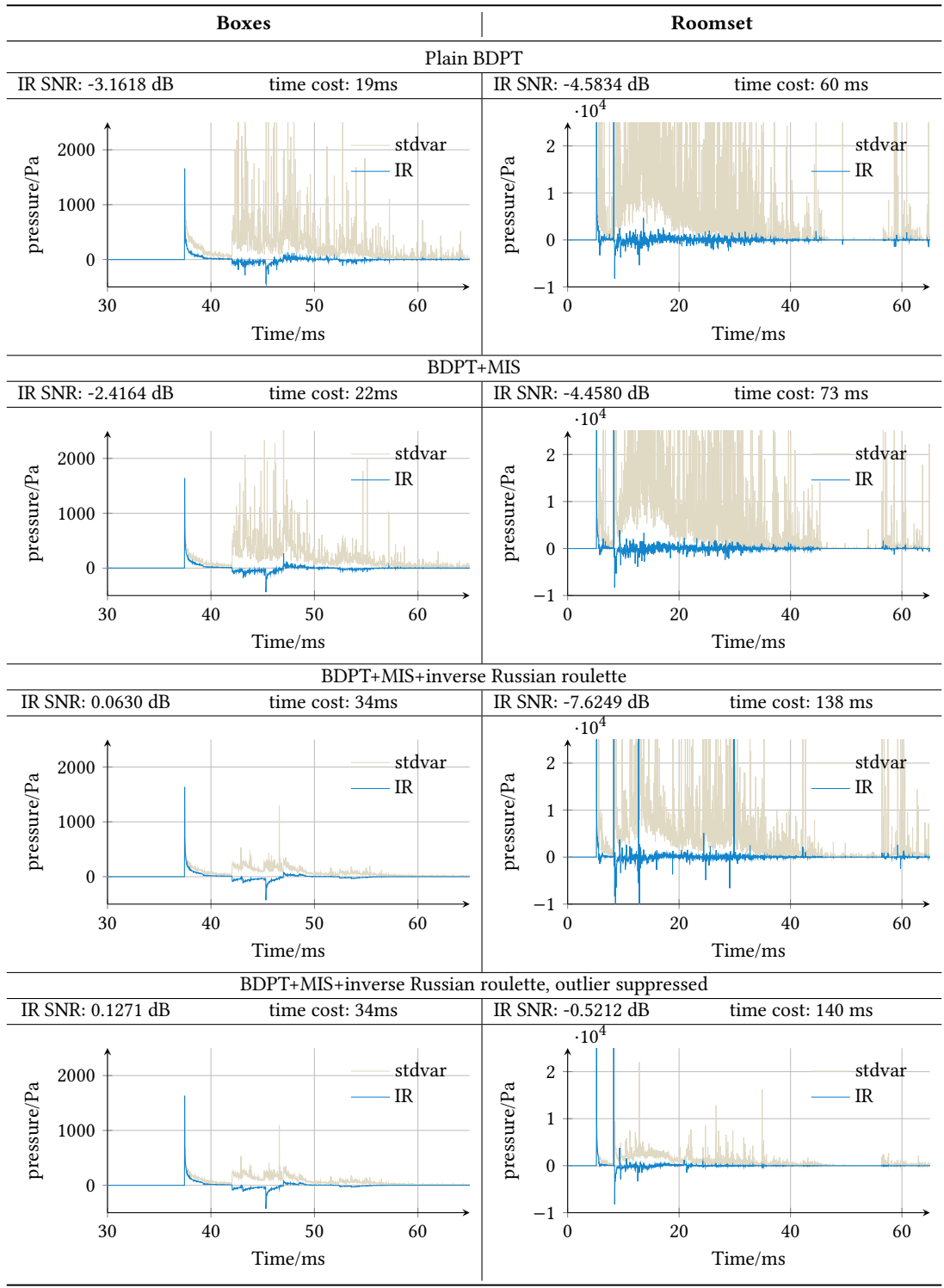

Our path tracer runs on a commodity PC with an Intel i7 2.60Ghz CPU and 12GB memory, and executes on a single thread. To evaluate the result quality under a certain render configuration, we compute 1000 impulse responses (IRs) with 12,000 samples. After this, we calculate the expectation of all the result IRs, and consider the difference between each calculated IR and the expectation IR as its error. We evaluate the render quality by computing the signal-to-noise ratio (SNR) for IR on its spectral domain. Given an IR function $f$ and its numeric error $f_e$, its IR SNR at frequency $\omega$ is calculated in the following way:
\begin{equation}
  \text{SNR}(f)(\omega)=10\log_{10}\frac{|\mathcal{F}(f)(\omega)|}{|\mathcal{F}(f_e)(\omega)|}
\end{equation}
Where $\mathcal{F}(f)$ is the Fourier spectrum of $f$. In \autoref{FIG:TECHNIQUES}, the IR SNR given is the average SNR of all computed IRs in the audible frequency range (20--20,000 Hz). When we use outlier suppression during quality evaluation, we set the threshold $C^*_o$ in \autoref{eq:logistic-suppressor} to 100.

To demonstrate the influence and computation efficiency of various path tracing techniques, we calculate the diffraction result in different scenes with our path tracer using different render configurations. We test our path tracer with the box scene in \autoref{FIG:DIFFRACTION-IR} and a more complex scene called ``Roomset'', which is an indoor scene with several small rooms connected with narrow corridors. The complex occlusion relationships between faces and edges in this scene is challenging for interactive geometric acoustic methods.

\autoref{FIG:TECHNIQUES} shows the IR and its standard variance produced by our path tracer under different rendering configurations, along with the average IR SNR and the average time cost of IR calculation. We notice in \autoref{FIG:TECHNIQUES} that when we use more techniques in \autoref{section:path-tracing} to improve our approach (e.g., MIS, inverse Russian Roulette), the accuracy of the resulting IR improves. The only exception is the inverse Russian roulette (IRR) in the Roomset scene. However, if we compare the variance curve with the one without IRR, we can see that IRR actually lowers the variance curve in most places, and the IR SNR decrease is caused by a few outliers. After outlier suppression, the IR SNR improves considerably.
We can also see from \autoref{FIG:TECHNIQUES} that MIS and outlier suppression have little impact on the simulation efficiency, while IRR almost doubled the simulation time cost. Considering that doubling the path tracing samples will only increase the IR SNR by approximately 1.5 dB, IRR clearly has an advantage in terms of improving the accuracy.

We also demonstrated the auralization result of our path tracer in our video. We used the aforementioned box scene, Roomset and a more complex model ``Crytek Sponza'' with more than 14k triangles for demonstration. We use 36k reflection samples and 12k diffraction samples (if diffraction is present) for audio generation in each frame. In the video, we put a single source in the scene and move the audio receiver around. During the motion, the source is often totally occluded from the receiver by the scene geometry, and higher order diffraction plays a major role in terms of generating smooth auralization results. The time cost of our diffraction path tracer is between 30--205 ms per frame. Information about auralization models and results can be found in \autoref{tbl:models}. Further details are presented in the video.

\subsection{Comparisons}

\begin{figure*}[htbp]
  \centering
  \begin{tabular}{cc}
    \begin{tikzpicture}[scale=0.75]
      \usetikzlibrary{arrows.meta}

      \coordinate (pf1) at (-5.2,-1.2);
      \coordinate (pf2) at (-3.6,1.2);
      \coordinate (pf3) at (-0.5818,-0.8727);
      \coordinate (pf4) at (-1.6,-2.4);
      \coordinate (pf5) at (0,0);
      \coordinate (pf6) at (1.6,-1.6);
      \coordinate (middle) at (-1.8,0.6);
      \coordinate (incident) at (-3.1856,-0.7464);
      \coordinate (outgoing) at (-0.9,1.3);
      \coordinate (outgoing_proj) at (-1.8,1.6);
      \coordinate (plane1) at (-3.4,0.2);
      \coordinate (plane2) at (-0.2,1);
      \coordinate (plane3) at (-3.4,2.2);
      \coordinate (plane4) at (-0.2,3);
      \coordinate (plane5) at (-3.4,-1.8);
      \coordinate (plane6) at (-0.2,0.0667);

      \fill[gray!10] (pf1) -- (pf2) -- (pf5) -- (pf4) -- cycle;
      \fill[gray!10] (pf3) -- (pf6) -- (pf5) -- (pf2) -- cycle;
      \draw[thick, gray] (pf1) -- (pf2) -- (pf5) -- (pf4) -- cycle;
      \draw[gray-base-sombre, |{Stealth}-{Stealth}| ] (incident) -- (middle) node[black,midway,sloped,below] {2m};
      \draw [fill = red-base!70](incident)circle(0.07) node[black,left] {source};

      \fill[green!50, opacity=0.2] (plane5) -- (middle) -- (plane6) -- (plane4) -- (plane3) -- cycle;
      \draw[gray] (plane1) -- (plane2);

      \draw[thick, gray] (pf3) -- (pf6) -- (pf5) -- (pf2);
      
      \draw[gray-base-sombre, |{Stealth}-{Stealth}| ] (middle) -- (outgoing);
      \draw[gray-base-sombre, dashed] (middle) -- (outgoing_proj) node[black,midway,left] {1m};
      \draw[gray, dashed] (outgoing) -- (outgoing_proj) node[black,midway,above] {1m};
      \draw [fill = blue-base!70](outgoing)circle(0.07) node[black,right] {listener};

      \draw[gray-base-sombre, |{Stealth}-{Stealth}| ] (-5.22,-1.25) -- (-1.62,-2.45) node[black,midway,sloped,below] {2m};

      \pic ["$\frac\pi 2$", draw, stealth-, angle eccentricity=2, angle radius=0.3cm]{angle=pf4--pf5--pf6};
      \pic ["$\frac\pi 6$", draw, stealth-, angle eccentricity=1.8, angle radius=0.5cm]{angle=plane1--middle--incident};
      \pic ["$\frac\pi 4$", draw, stealth-, angle eccentricity=2, angle radius=0.3cm]{angle=outgoing--middle--outgoing_proj};
    \end{tikzpicture} & \begin{tikzpicture}
      \begin{axis}[
        width = 0.45\textwidth,
        height = 0.25\textwidth,
        axis y line = left,
        axis x line = bottom,
        xlabel      = {Time/ms},
        ylabel      = {pressure/Pa},
        xmin = 0, xmax = 20,
        ymin = -400, ymax = 1600,
        legend cell align = left,
        legend transposed = true,
        legend style = {draw=none, fill=none, at={(1.01,1.01)},anchor=north east},
        grid = major,
        /pgf/number format/.cd,
        use comma,
        1000 sep={}
      ]
        \addplot[red-base, thick, dashed, unbounded coords=jump] table [x expr=\thisrow{time} * 1000, y=pressure_BTM, col sep=comma] {img/IR_single_edge.txt};
        \addlegendentry{BTM (ground truth)};
        \addplot[green-base, thick, dotted, line cap=round, unbounded coords=jump] table [x expr=\thisrow{time} * 1000, y = pressure_UTD, col sep=comma] {img/IR_UTD.txt};
        \addlegendentry{UTD (full quality)};
        \addplot[yellow-base, unbounded coords=jump] table [x expr=\thisrow{time} * 1000, y = pressure_UTD, col sep=comma] {img/IR_UTD_8Band.txt};
        \addlegendentry{UTD (8 bands)};
        \addplot[blue-base, unbounded coords=jump] table [x expr=\thisrow{time} * 1000, y=pressure_ours, col sep=comma] {img/IR_single_edge.txt};
        \addlegendentry{Ours};
      \end{axis}
    \end{tikzpicture} \\
    \begin{tikzpicture}[scale=0.75]
      \usetikzlibrary{arrows.meta}

      \coordinate (pf1) at (-3.8,-0.8);
      \coordinate (pf2) at (-1.8,1.2);
      \coordinate (pf3) at (-0.2,-0.2);
      \coordinate (pf4) at (-2,-2);
      \coordinate (pf5) at (0,0);
      \coordinate (pf6) at (1,-1);
      \coordinate (pf7) at (2,0);
      \coordinate (pf8) at (4,-2);
      \coordinate (pf9) at (1.6,-0.4);
      \coordinate (pf10) at (0.2,1.2);
      \coordinate (pf11) at (-0.6,0.4);

      \coordinate (incident) at (-2.9,-0.4);
      \coordinate (left) at (-2.9,0.6);
      \coordinate (middle) at (-0.9,0.6);
      \coordinate (middle2) at (1.1,0.6);
      \coordinate (right) at (3.1,0.6);
      \coordinate (outgoing) at (3.1,-0.4);
    
      \fill[gray!10] (pf1) -- (pf2) -- (pf11) -- (pf10) -- (pf7) -- (pf8) -- (pf9) -- (pf6) -- (pf3) -- (pf4) -- cycle;
      \draw[thick, gray] (pf1) -- (pf2) -- (pf5) -- (pf4) -- cycle;
      \draw[thick, gray] (pf3) -- (pf6) -- (pf5) -- (pf2);
      \draw[thick, gray] (pf6) -- (pf7) -- (pf8) -- (pf9);
      \draw[thick, gray] (pf11) -- (pf10) -- (pf7) -- (pf6) -- (pf5) -- cycle;

      \draw[gray-base-sombre, |{Stealth}-{Stealth}|] (left) -- (middle)  node[pos=0.5,black,above] {2m};
      \draw[gray-base-sombre, |{Stealth}-{Stealth}|] (middle) -- (middle2)  node[pos=0.5,black,above] {2m};
      \draw[gray-base-sombre, |{Stealth}-{Stealth}|] (middle2) -- (right)  node[pos=0.5,black,above] {2m};
      \draw[gray-base-sombre, |{Stealth}-{Stealth}|] (incident) -- (left)  node[pos=0.5,black,left] {1m};
      \draw[gray-base-sombre, |{Stealth}-{Stealth}|] (right) -- (outgoing)  node[pos=0.5,black,right] {1m};
      \draw[gray-base-sombre] (incident) -- (middle);
      \draw [fill = red-base!70](incident)circle(0.07) node[black,right] {source};
      \draw[gray-base-sombre] (middle2) -- (outgoing);
      \draw [fill = blue-base!70](outgoing)circle(0.07) node[black,right] {listener};

      \draw[gray-base-sombre, |{Stealth}-{Stealth}| ] (-3.84,-0.85) -- (-2.04,-2.05) node[black,midway,sloped,below] {2m};

      \pic ["$\frac\pi 2$", draw, stealth-, angle eccentricity=2, angle radius=0.3cm]{angle=pf4--pf5--pf6};
      \pic ["$\frac\pi 2$", draw, stealth-, angle eccentricity=2, angle radius=0.3cm]{angle=pf6--pf7--pf8};
    \end{tikzpicture} & \begin{tikzpicture}
      \begin{axis}[
        width = 0.45\textwidth,
        height = 0.25\textwidth,
        axis y line = left,
        axis x line = bottom,
        xlabel      = {Time/ms},
        ylabel      = {pressure/Pa},
        xmin = 0, xmax = 30,
        ymin = -20, ymax = 60,
        legend cell align = left,
        legend transposed = true,
        legend style = {draw=none, fill=none, at={(1.01,1.01)},anchor=north east},
        grid = major,
        /pgf/number format/.cd,
        use comma,
        1000 sep={}
      ]
        \addplot[red-base, thick, dashed, unbounded coords=jump] table [x expr=\thisrow{time} * 1000, y=pressure_BTM, col sep=comma] {img/IR_double_edge.txt};
        \addlegendentry{BTM (ground truth)};
        \addplot[blue-base, unbounded coords=jump] table [x expr=\thisrow{time} * 1000, y expr=\thisrow{pressure_ours} * -1, col sep=comma] {img/IR_double_edge.txt};
        \addlegendentry{Ours};
      \end{axis}
    \end{tikzpicture} \\
    \includegraphics[width=0.18\linewidth]{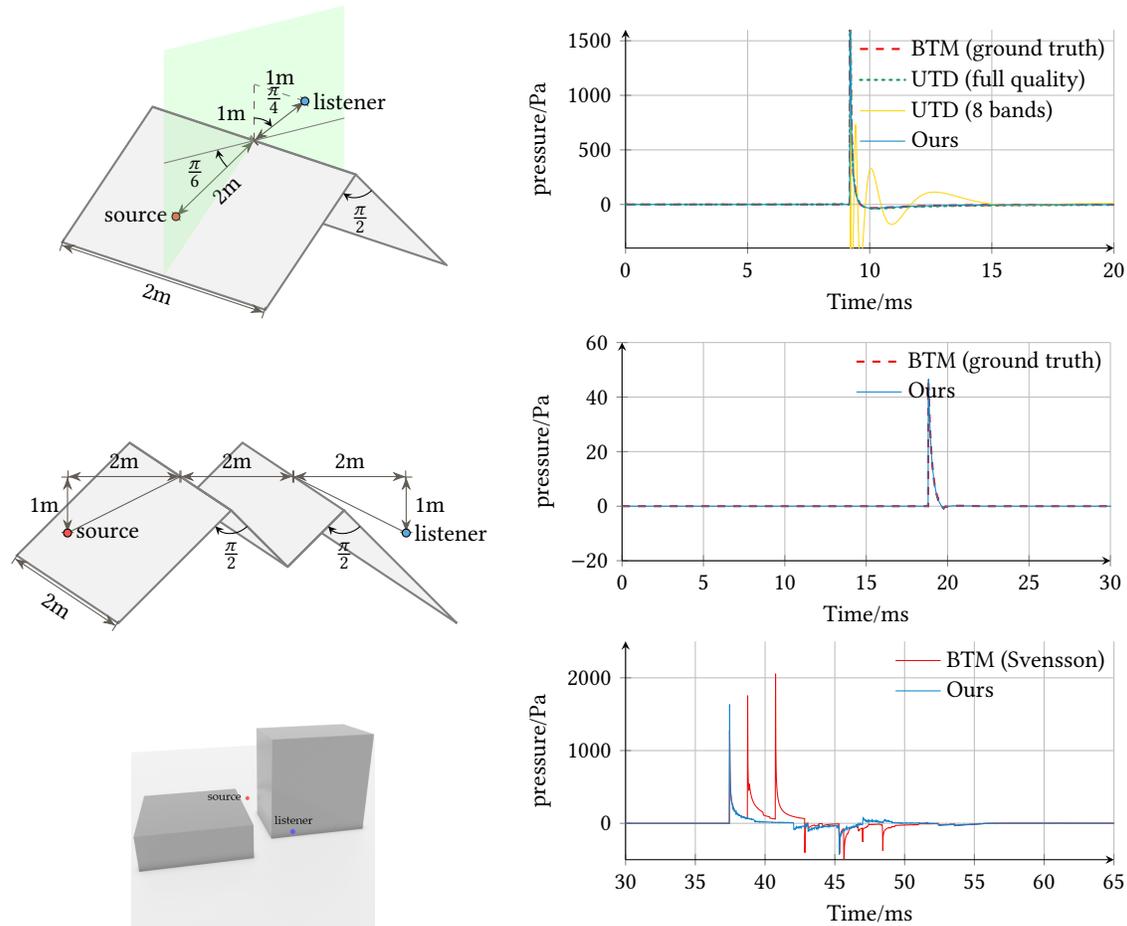} & \begin{tikzpicture}
      \begin{axis}[
        width = 0.45\textwidth,
        height = 0.25\textwidth,
        axis y line = left,
        axis x line = bottom,
        xlabel      = {Time/ms},
        ylabel      = {pressure/Pa},
        xmin = 30, xmax = 65,
        ymin = -500, ymax = 2500,
        legend cell align = left,
        legend transposed = true,
        legend style = {draw=none, fill=none, at={(1.01,1.01)},anchor=north east},
        grid = major,
        /pgf/number format/.cd,
        use comma,
        1000 sep={}
      ]
        \addplot[red-base, unbounded coords=jump] table [x expr=\thisrow{time}, y=pressure_BTM, col sep=comma] {img/IR_BoxBuildings.txt};
        \addlegendentry{BTM (Svensson)};
        \addplot[blue-base, unbounded coords=jump] table [x expr=\thisrow{time}, y=pressure_ours, col sep=comma] {img/IR_BoxBuildings.txt};
        \addlegendentry{Ours};
      \end{axis}
    \end{tikzpicture}
  \end{tabular}
  \caption{IRs of diffraction wave in several simple scenes. The diffraction IRs are generated by the corresponding scene illustrated on the left side. In all test scenes, we set sound speed to 344m/s, source impulse energy to 1J and air density to 1.21$\text{kg}/\text{m}^3$. We use the Dirichlet boundary condition condition for the first two scenes and Neumann condition for the last scene. The difference between our method and BTM is minimal in the first two cases. Moreover, major diffraction peaks are well-matched in the last one. These results indicate close agreement with BTM on these scenarios. Meanwhile, we also observe improved accuracy of our BEDRF  formulation over UTD in the top example.}
  \label{FIG:DIFFRACTION-IR}
\end{figure*}

We have tested our diffraction BSDF and integrator in various scene configurations and compared the results with those of state-of-the-art diffraction models, notably the BTM and UTD models. The BTM model gives the exact solution for single-edge diffraction for spherical incident waves, and our result matches the BTM solution in both the single-edge and the double-edge cases  (see \autoref{FIG:DIFFRACTION-IR}).

Our time-domain algorithm shows a clear advantage over frequency-domain algorithms like UTD. Specifically, while UTD produces near-perfect results when there are very high number of samples on the frequency domain, it is much less accurate in interactive propagation algorithms~\cite{schissler2014high-order}, where only 4-8 frequency bands are used  and the phase information of the result is discarded. In \autoref{FIG:DIFFRACTION-IR}, the ``8-band UTD'' IR is produced by calculating the UTD result on the central frequency of 8 bands ranging from 62.5-8000Hz, and combining them with crossover filters. The result shows the typical ``ringing'' effect caused by undersampling in the frequency domain. The performance of the resulting propagation algorithm increases linearly with the number of bands.

For more complex scenes where an analytic solution is not known, We compare with the result of Svensson's diffraction toolbox for Matlab, which uses the BTM algorithm for edge diffraction~\cite{SvenssonToolBox, Torres2001ComputationOE}. The scene we used for comparison consists of a floor plane and two boxes of size $\text{12m}\times\text{12m}\times\text{4m}$. We set the highest diffraction order to 2 for our implementation and Svensson's toolbox. Our result matches Svensson's result very well at many diffraction peaks, but there are also some other peaks in Svensson's result not present in ours. This is due to the presence of ``creeping waves'', which is the sound wave that doesn't bounce between faces or edges, but travels along the surface of the geometry. Currently, our algorithm is unable to simulate this phenomenon.

\begin{figure}[t]
  \centering
  \begin{tikzpicture}
    \begin{axis}[
      axis y line = left,
      axis x line = bottom,
      xlabel      = {position},
      ylabel      = {intensity},
      xmin = -10, xmax = 10,
      ymin = 0, ymax = 1,
      legend cell align = left,
      legend pos = north east,
      legend transposed = true,
      legend style = {draw=none, fill=none},
      grid = major
    ]
      \addplot[red-base, unbounded coords=jump] table [x=pos, y=spectrum_theory, col sep=comma] {img/single_slit_spectrum.txt};
      \addlegendentry{ground truth};
      \addplot[blue-base, unbounded coords=jump] table [x=pos, y=spectrum_sampled, col sep=comma] {img/single_slit_spectrum.txt};
      \addlegendentry{Ours};
    \end{axis}
  \end{tikzpicture}
\caption{Stripe pattern produced by single-slit diffraction. The result is calculated using 50000 samples and closely matches the ground truth. }
\label{fig:Diffraction}
\end{figure}
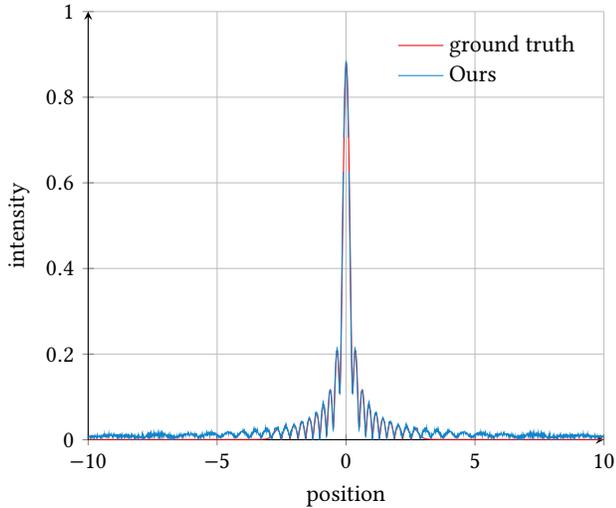

Since our algorithm is amplitude-based, it is also capable of simulating interference of diffraction waves produced by two or more edges. Fig. \ref{fig:Diffraction} shows the result of our algorithm reproducing the single-slit experiment, which involves a plane wave passing through a narrow slit of constant width and infinite length, producing a typical stripe-like pattern. Our result closely matches the precomputed ground truth.

\section{Conclusion, Limitations and Future Work}

In this paper, we have presented BEDRF, a new localized representation of edge diffraction for convex wedges. We also present a path tracer utilizing the BEDRF representation for interactive simulation of sound diffraction in complex models. The flexibility of path tracing allows us to deal with complex geometries with ease, which has been shown in our results.
The BEDRF model and the new path tracer are mutually independent. We can replace BEDRF with BTM or DLSM without too much work. The latter two models may have advantages in some special cases. BEDRF can perform interactive sound propagation on complex models with no preprocessing or model simplification, though its accuracy may not match with BTM in some cases.

The computation cost of ``diffraction path tracing'' is still much higher than traditional path tracing. The formation of BEDRF itself is already more complex than that of common BSDFs for surface reflection. In addition, we use triangle-triangle intersection in our tracing algorithm, which is much more costly than ray-triangle intersection. It seems to be challenging to design a purely ray-based unbiased integrator on geometric edges. However, when we remove the restriction of unbiasedness, we may find an algorithm that relies on ray-triangle intersections only. That is a good topic for future work.

We have evaluated the accuracy on many benchmarks. We observe that the accuracy of our algorithm is nearly-indistinguishable from that of BTM-based solvers in simple scenes. As part of future work, it would be useful to investigate the accuracy of BEDRF with respect to BTM. At the moment, it is unknown whether BEDRF is exact when the incident wave is non-planar. 
It is also worth noting that our path tracer doesn't support the simulation of creeping waves. This phenomenon is especially important for simulating the diffraction effect of convex smooth objects, like spheres and cylinders. Our future simulator will take this phenomenon into consideration.

\bibliographystyle{ACM-Reference-Format}
\bibliography{thesis}


\begin{thebibliography}{54}


\ifx \showCODEN    \undefined \def \showCODEN     #1{\unskip}     \fi
\ifx \showDOI      \undefined \def \showDOI       #1{#1}\fi
\ifx \showISBNx    \undefined \def \showISBNx     #1{\unskip}     \fi
\ifx \showISBNxiii \undefined \def \showISBNxiii  #1{\unskip}     \fi
\ifx \showISSN     \undefined \def \showISSN      #1{\unskip}     \fi
\ifx \showLCCN     \undefined \def \showLCCN      #1{\unskip}     \fi
\ifx \shownote     \undefined \def \shownote      #1{#1}          \fi
\ifx \showarticletitle \undefined \def \showarticletitle #1{#1}   \fi
\ifx \showURL      \undefined \def \showURL       {\relax}        \fi
\providecommand\bibfield[2]{#2}
\providecommand\bibinfo[2]{#2}
\providecommand\natexlab[1]{#1}
\providecommand\showeprint[2][]{arXiv:#2}

\bibitem[Allen and Raghuvanshi(2015)]%
        {Allen2015AerophonesIF}
\bibfield{author}{\bibinfo{person}{Andrew Allen} {and} \bibinfo{person}{Nikunj
  Raghuvanshi}.} \bibinfo{year}{2015}\natexlab{}.
\newblock \showarticletitle{Aerophones in flatland}.
\newblock \bibinfo{journal}{\emph{ACM Transactions on Graphics (TOG)}}
  \bibinfo{volume}{34} (\bibinfo{year}{2015}), \bibinfo{pages}{1 -- 11}.
\newblock


\bibitem[Antani et~al\mbox{.}(2010)]%
        {Antani2010}
\bibfield{author}{\bibinfo{person}{Lakulish Antani}, \bibinfo{person}{Anish
  Chandak}, \bibinfo{person}{Micah Taylor}, {and} \bibinfo{person}{Dinesh
  Manocha}.} \bibinfo{year}{2010}\natexlab{}.
\newblock \showarticletitle{Fast Geometric Sound Propagation with Finite Edge
  Diffraction}.
\newblock \bibinfo{journal}{\emph{Technical Report TR10-011, University of
  North Carolina at Chapel Hill}} (\bibinfo{date}{01} \bibinfo{year}{2010}).
\newblock


\bibitem[Asmail(1991)]%
        {Asmail1991BidirectionalSD}
\bibfield{author}{\bibinfo{person}{Clara~C. Asmail}.}
  \bibinfo{year}{1991}\natexlab{}.
\newblock \showarticletitle{Bidirectional Scattering Distribution Function
  (BSDF): A Systematized Bibliography}.
\newblock \bibinfo{journal}{\emph{Journal of Research of the National Institute
  of Standards and Technology}}  \bibinfo{volume}{96} (\bibinfo{year}{1991}),
  \bibinfo{pages}{215 -- 223}.
\newblock


\bibitem[Bertram et~al\mbox{.}(2005)]%
        {Bertram2005phonon}
\bibfield{author}{\bibinfo{person}{M. Bertram}, \bibinfo{person}{E. Deines},
  \bibinfo{person}{J. Mohring}, \bibinfo{person}{J. Jegorovs}, {and}
  \bibinfo{person}{H. Hagen}.} \bibinfo{year}{2005}\natexlab{}.
\newblock \showarticletitle{Phonon tracing for auralization and visualization
  of sound}. In \bibinfo{booktitle}{\emph{VIS 05. IEEE Visualization, 2005.}}
  \bibinfo{pages}{151--158}.
\newblock
\urldef\tempurl%
\url{https://doi.org/10.1109/VISUAL.2005.1532790}
\showDOI{\tempurl}


\bibitem[Biot and Tolstoy(1957)]%
        {biot1957formulation}
\bibfield{author}{\bibinfo{person}{M~A Biot} {and} \bibinfo{person}{I
  Tolstoy}.} \bibinfo{year}{1957}\natexlab{}.
\newblock \showarticletitle{Formulation of Wave Propagation in Infinite Media
  by Normal Coordinates with an Application to Diffraction}.
\newblock \bibinfo{journal}{\emph{Journal of the Acoustical Society of
  America}} \bibinfo{volume}{29}, \bibinfo{number}{3} (\bibinfo{year}{1957}),
  \bibinfo{pages}{381--391}.
\newblock


\bibitem[Calamia(2009)]%
        {Calamia2009}
\bibfield{author}{\bibinfo{person}{Paul~Thomas Calamia}.}
  \bibinfo{year}{2009}\natexlab{}.
\newblock \emph{\bibinfo{title}{Advances in Edge-Diffraction Modeling for
  Virtual-Acoustic Simulations}}.
\newblock \bibinfo{thesistype}{Ph.\,D. Dissertation}. \bibinfo{address}{USA}.
\newblock Advisor(s) Funkhouser, Thomas.
\newblock
\showISBNx{9781109138252}


\bibitem[Calamia and Svensson(2007)]%
        {calamia2007fast}
\bibfield{author}{\bibinfo{person}{Paul~T. Calamia} {and}
  \bibinfo{person}{U.~Peter Svensson}.} \bibinfo{year}{2007}\natexlab{}.
\newblock \showarticletitle{Fast Time-Domain Edge-Diffraction Calculations for
  Interactive Acoustic Simulations}.
\newblock \bibinfo{journal}{\emph{EURASIP J. Adv. Signal Process}}
  \bibinfo{volume}{2007}, \bibinfo{number}{1} (\bibinfo{date}{Jan.}
  \bibinfo{year}{2007}), \bibinfo{pages}{186}.
\newblock
\showISSN{1110-8657}
\urldef\tempurl%
\url{https://doi.org/10.1155/2007/63560}
\showDOI{\tempurl}


\bibitem[Cao et~al\mbox{.}(2016)]%
        {cao2016interactive}
\bibfield{author}{\bibinfo{person}{Chunxiao Cao}, \bibinfo{person}{Zhong Ren},
  \bibinfo{person}{Carl Schissler}, \bibinfo{person}{Dinesh Manocha}, {and}
  \bibinfo{person}{Kun Zhou}.} \bibinfo{year}{2016}\natexlab{}.
\newblock \showarticletitle{Interactive sound propagation with bidirectional
  path tracing}.
\newblock \bibinfo{journal}{\emph{ACM Transactions on Graphics (TOG)}}
  \bibinfo{volume}{35} (\bibinfo{year}{2016}), \bibinfo{pages}{1 -- 11}.
\newblock


\bibitem[Case(1993)]%
        {Case1993StructuralAA}
\bibfield{author}{\bibinfo{person}{K. Case}.} \bibinfo{year}{1993}\natexlab{}.
\newblock \showarticletitle{Structural Acoustics: A General Form of Reciprocity
  Principles in Acoustics}.
\newblock \bibinfo{journal}{\emph{Mitre Corp. Technical Report JSR-91-193}}.
\newblock


\bibitem[Hamilton(2021)]%
        {Hamilton2021Tutorial}
\bibfield{author}{\bibinfo{person}{Brian Hamilton}.}
  \bibinfo{year}{2021}\natexlab{}.
\newblock \showarticletitle{Tutorial on finite-difference time-domain (FDTD)
  methods for room acoustics simulation}.
\newblock \bibinfo{journal}{\emph{The Journal of the Acoustical Society of
  America}} \bibinfo{volume}{149}, \bibinfo{number}{4} (\bibinfo{year}{2021}),
  \bibinfo{pages}{A92--A93}.
\newblock
\urldef\tempurl%
\url{https://doi.org/10.1121/10.0004614}
\showDOI{\tempurl}
\showeprint{https://doi.org/10.1121/10.0004614}


\bibitem[Jensen(2001)]%
        {Jensen2001RealisticIS}
\bibfield{author}{\bibinfo{person}{Henrik~Wann Jensen}.}
  \bibinfo{year}{2001}\natexlab{}.
\newblock \showarticletitle{Realistic Image Synthesis Using Photon Mapping}.
\newblock


\bibitem[Kajiya(1986)]%
        {kajiya1986the}
\bibfield{author}{\bibinfo{person}{James~T Kajiya}.}
  \bibinfo{year}{1986}\natexlab{}.
\newblock \showarticletitle{The rendering equation}.
\newblock \bibinfo{journal}{\emph{Proceedings of the 13th Annual Conference on
  Computer Graphics and Interactive Techniques}} \bibinfo{volume}{20},
  \bibinfo{number}{4} (\bibinfo{year}{1986}), \bibinfo{pages}{143--150}.
\newblock


\bibitem[Kapralos et~al\mbox{.}(2007)]%
        {Jenkin07acousticalmodeling}
\bibfield{author}{\bibinfo{person}{Bill Kapralos}, \bibinfo{person}{Michael
  R.~M. Jenkin}, {and} \bibinfo{person}{Evangelos~E. Milios}.}
  \bibinfo{year}{2007}\natexlab{}.
\newblock \showarticletitle{Acoustical Modeling with Sonel Mapping}. In
  \bibinfo{booktitle}{\emph{19th International Congress on Acoustics}}.
\newblock


\bibitem[Keller(1997)]%
        {Keller1997InstantR}
\bibfield{author}{\bibinfo{person}{Alexander Keller}.}
  \bibinfo{year}{1997}\natexlab{}.
\newblock \showarticletitle{Instant radiosity}.
\newblock \bibinfo{journal}{\emph{Proceedings of the 24th Annual Conference on
  Computer Graphics and Interactive Techniques}} (\bibinfo{year}{1997}).
\newblock


\bibitem[Keller and Blank(1951)]%
        {keller1951diffraction}
\bibfield{author}{\bibinfo{person}{Joseph~B Keller} {and}
  \bibinfo{person}{Albert Blank}.} \bibinfo{year}{1951}\natexlab{}.
\newblock \showarticletitle{Diffraction and reflection of pulses by wedges and
  corners}.
\newblock \bibinfo{journal}{\emph{Communications on Pure and Applied
  Mathematics}} \bibinfo{volume}{4}, \bibinfo{number}{1}
  (\bibinfo{year}{1951}), \bibinfo{pages}{75--94}.
\newblock


\bibitem[Kirkup(2019)]%
        {Kirkup2019boundary}
\bibfield{author}{\bibinfo{person}{Stephen Kirkup}.}
  \bibinfo{year}{2019}\natexlab{}.
\newblock \showarticletitle{The Boundary Element Method in Acoustics: A
  Survey}.
\newblock \bibinfo{journal}{\emph{Applied Sciences}}  \bibinfo{volume}{9}
  (\bibinfo{date}{04} \bibinfo{year}{2019}), \bibinfo{pages}{1642}.
\newblock
\urldef\tempurl%
\url{https://doi.org/10.3390/app9081642}
\showDOI{\tempurl}


\bibitem[{Kouyoumjian} and {Pathak}(1974)]%
        {UTD1}
\bibfield{author}{\bibinfo{person}{R.~G. {Kouyoumjian}} {and}
  \bibinfo{person}{P.~H. {Pathak}}.} \bibinfo{year}{1974}\natexlab{}.
\newblock \showarticletitle{A uniform geometrical theory of diffraction for an
  edge in a perfectly conducting surface}.
\newblock \bibinfo{journal}{\emph{Proc. IEEE}} \bibinfo{volume}{62},
  \bibinfo{number}{11} (\bibinfo{year}{1974}), \bibinfo{pages}{1448--1461}.
\newblock


\bibitem[Kreyszig(1978)]%
        {kreyszig1978introductory}
\bibfield{author}{\bibinfo{person}{E Kreyszig}.}
  \bibinfo{year}{1978}\natexlab{}.
\newblock \showarticletitle{Introductory Functional Analysis with
  Applications}.
\newblock  (\bibinfo{year}{1978}).
\newblock


\bibitem[Lentz et~al\mbox{.}(2007)]%
        {lentz2007virtual}
\bibfield{author}{\bibinfo{person}{Tobias Lentz}, \bibinfo{person}{Dirk
  Schroder}, \bibinfo{person}{Michael Vorlander}, {and} \bibinfo{person}{Ingo
  Assenmacher}.} \bibinfo{year}{2007}\natexlab{}.
\newblock \showarticletitle{Virtual reality system with integrated sound field
  simulation and reproduction}.
\newblock \bibinfo{journal}{\emph{EURASIP Journal on Advances in Signal
  Processing}} \bibinfo{volume}{2007}, \bibinfo{number}{1}
  (\bibinfo{year}{2007}), \bibinfo{pages}{187--187}.
\newblock


\bibitem[Liu and Manocha(2020)]%
        {SoundSurvey}
\bibfield{author}{\bibinfo{person}{Shiguang Liu} {and} \bibinfo{person}{Dinesh
  Manocha}.} \bibinfo{year}{2020}\natexlab{}.
\newblock \showarticletitle{Sound Synthesis, Propagation, and Rendering: {A}
  Survey}.
\newblock \bibinfo{journal}{\emph{CoRR}}  \bibinfo{volume}{abs/2011.05538}
  (\bibinfo{year}{2020}).
\newblock
\showeprint[arXiv]{2011.05538}
\urldef\tempurl%
\url{https://arxiv.org/abs/2011.05538}
\showURL{%
\tempurl}


\bibitem[Mcnamara et~al\mbox{.}(1990)]%
        {UTD2}
\bibfield{author}{\bibinfo{person}{D. Mcnamara}, \bibinfo{person}{C.
  Pistorius}, {and} \bibinfo{person}{J. Malherbe}.}
  \bibinfo{year}{1990}\natexlab{}.
\newblock \bibinfo{booktitle}{\emph{Introduction to The Uniform Geometrical
  Theory and Diffraction}}.
\newblock


\bibitem[Medwin et~al\mbox{.}(1982)]%
        {medwin1982impulse}
\bibfield{author}{\bibinfo{person}{Herman Medwin}, \bibinfo{person}{Emily
  Childs}, {and} \bibinfo{person}{Gary~M Jebsen}.}
  \bibinfo{year}{1982}\natexlab{}.
\newblock \showarticletitle{Impulse studies of double diffraction: A discrete
  Huygens interpretation}.
\newblock \bibinfo{journal}{\emph{Journal of the Acoustical Society of
  America}} \bibinfo{volume}{72}, \bibinfo{number}{3} (\bibinfo{year}{1982}),
  \bibinfo{pages}{1005--1013}.
\newblock


\bibitem[Mehra et~al\mbox{.}(2013)]%
        {mehra2013wave-based}
\bibfield{author}{\bibinfo{person}{Ravish Mehra}, \bibinfo{person}{Nikunj
  Raghuvanshi}, \bibinfo{person}{Lakulish Antani}, \bibinfo{person}{Anish
  Chandak}, \bibinfo{person}{Sean Curtis}, {and} \bibinfo{person}{Dinesh
  Manocha}.} \bibinfo{year}{2013}\natexlab{}.
\newblock \showarticletitle{Wave-based sound propagation in large open scenes
  using an equivalent source formulation}.
\newblock \bibinfo{journal}{\emph{ACM Transactions on Graphics}}
  \bibinfo{volume}{32}, \bibinfo{number}{2} (\bibinfo{year}{2013}),
  \bibinfo{pages}{19}.
\newblock


\bibitem[Mehra et~al\mbox{.}(2012)]%
        {Mehra2012AnEG}
\bibfield{author}{\bibinfo{person}{Ravish Mehra}, \bibinfo{person}{Nikunj
  Raghuvanshi}, \bibinfo{person}{Lauri Savioja}, \bibinfo{person}{Ming~C. Lin},
  {and} \bibinfo{person}{Dinesh Manocha}.} \bibinfo{year}{2012}\natexlab{}.
\newblock \showarticletitle{An efficient GPU-based time domain solver for the
  acoustic wave equation}.
\newblock \bibinfo{journal}{\emph{Applied Acoustics}}  \bibinfo{volume}{73}
  (\bibinfo{year}{2012}), \bibinfo{pages}{83--94}.
\newblock


\bibitem[Menounou et~al\mbox{.}(2000)]%
        {Menounou2000DLSM}
\bibfield{author}{\bibinfo{person}{Penelope Menounou},
  \bibinfo{person}{Ilene~J. Busch-Vishniac}, {and} \bibinfo{person}{David~T.
  Blackstock}.} \bibinfo{year}{2000}\natexlab{}.
\newblock \showarticletitle{Directive line source model: A new model for sound
  diffraction by half planes and wedges}.
\newblock \bibinfo{journal}{\emph{The Journal of the Acoustical Society of
  America}} \bibinfo{volume}{107}, \bibinfo{number}{6} (\bibinfo{year}{2000}),
  \bibinfo{pages}{2973--2986}.
\newblock
\urldef\tempurl%
\url{https://doi.org/10.1121/1.429327}
\showDOI{\tempurl}
\showeprint{https://doi.org/10.1121/1.429327}


\bibitem[Menounou and Nikolaou(2017)]%
        {Menounou2017DLSM2}
\bibfield{author}{\bibinfo{person}{Penelope Menounou} {and}
  \bibinfo{person}{Petros Nikolaou}.} \bibinfo{year}{2017}\natexlab{}.
\newblock \showarticletitle{Analytical model for predicting edge diffraction in
  the time domain}.
\newblock \bibinfo{journal}{\emph{The Journal of the Acoustical Society of
  America}} \bibinfo{volume}{142}, \bibinfo{number}{6} (\bibinfo{year}{2017}),
  \bibinfo{pages}{3580--3592}.
\newblock
\urldef\tempurl%
\url{https://doi.org/10.1121/1.5014051}
\showDOI{\tempurl}
\showeprint{https://doi.org/10.1121/1.5014051}


\bibitem[{Pind J{\"o}rgensson}(2020)]%
        {Pind2020Wave}
\bibfield{author}{\bibinfo{person}{{Finnur K{\'a}ri} {Pind J{\"o}rgensson}}.}
  \bibinfo{year}{2020}\natexlab{}.
\newblock \emph{\bibinfo{title}{Wave-Based Virtual Acoustics}}.
\newblock \bibinfo{thesistype}{Ph.\,D. Dissertation}.
\newblock


\bibitem[Pisha et~al\mbox{.}(2020)]%
        {Pisha2020VDaT}
\bibfield{author}{\bibinfo{person}{Louis Pisha}, \bibinfo{person}{Siddharth
  Atre}, \bibinfo{person}{John Burnett}, {and} \bibinfo{person}{Shahrokh
  Yadegari}.} \bibinfo{year}{2020}\natexlab{}.
\newblock \showarticletitle{Approximate diffraction modeling for real-time
  sound propagation simulation}.
\newblock \bibinfo{journal}{\emph{The Journal of the Acoustical Society of
  America}}  \bibinfo{volume}{148} (\bibinfo{date}{10} \bibinfo{year}{2020}),
  \bibinfo{pages}{1922--1933}.
\newblock
\urldef\tempurl%
\url{https://doi.org/10.1121/10.0002115}
\showDOI{\tempurl}


\bibitem[Raghuvanshi et~al\mbox{.}(2009)]%
        {Raghuvanshi2009EfficientAA}
\bibfield{author}{\bibinfo{person}{Nikunj Raghuvanshi}, \bibinfo{person}{Rahul
  Narain}, {and} \bibinfo{person}{Ming~C. Lin}.}
  \bibinfo{year}{2009}\natexlab{}.
\newblock \showarticletitle{Efficient and accurate sound propagation using
  adaptive rectangular decomposition.}
\newblock \bibinfo{journal}{\emph{IEEE transactions on visualization and
  computer graphics}}  \bibinfo{volume}{15 5} (\bibinfo{year}{2009}),
  \bibinfo{pages}{789--801}.
\newblock


\bibitem[Ross(1987)]%
        {Ross1987IntroductionTP}
\bibfield{author}{\bibinfo{person}{Sheldon~M. Ross}.}
  \bibinfo{year}{1987}\natexlab{}.
\newblock \showarticletitle{Introduction to Probability and Statistics for
  Engineers and Scientists}.
\newblock


\bibitem[Royden and Fitzpatrick(1988)]%
        {royden1988real}
\bibfield{author}{\bibinfo{person}{Halsey~Lawrence Royden} {and}
  \bibinfo{person}{Patrick Fitzpatrick}.} \bibinfo{year}{1988}\natexlab{}.
\newblock \bibinfo{booktitle}{\emph{Real analysis}}.
  Vol.~\bibinfo{volume}{198}.
\newblock \bibinfo{publisher}{Macmillan New York}.
\newblock


\bibitem[Rungta et~al\mbox{.}(2016)]%
        {Rungta2016Psychoacoustic}
\bibfield{author}{\bibinfo{person}{Atul Rungta}, \bibinfo{person}{Sarah Rust},
  \bibinfo{person}{Nicolas Morales}, \bibinfo{person}{Roberta Klatzky},
  \bibinfo{person}{Ming Lin}, {and} \bibinfo{person}{Dinesh Manocha}.}
  \bibinfo{year}{2016}\natexlab{}.
\newblock \showarticletitle{Psychoacoustic Characterization of Propagation
  Effects in Virtual Environments}.
\newblock \bibinfo{journal}{\emph{ACM Trans. Appl. Percept.}}
  \bibinfo{volume}{13}, \bibinfo{number}{4}, Article \bibinfo{articleno}{21}
  (\bibinfo{date}{jul} \bibinfo{year}{2016}), \bibinfo{numpages}{18}~pages.
\newblock
\showISSN{1544-3558}
\urldef\tempurl%
\url{https://doi.org/10.1145/2947508}
\showDOI{\tempurl}


\bibitem[Rungta et~al\mbox{.}(2018)]%
        {rungta2018diffraction}
\bibfield{author}{\bibinfo{person}{Atul Rungta}, \bibinfo{person}{Carl
  Schissler}, \bibinfo{person}{Nicholas Rewkowski}, \bibinfo{person}{Ravish
  Mehra}, {and} \bibinfo{person}{Dinesh Manocha}.}
  \bibinfo{year}{2018}\natexlab{}.
\newblock \showarticletitle{Diffraction Kernels for Interactive Sound
  Propagation in Dynamic Environments}.
\newblock \bibinfo{journal}{\emph{IEEE Transactions on Visualization and
  Computer Graphics}} \bibinfo{volume}{24}, \bibinfo{number}{4}
  (\bibinfo{year}{2018}), \bibinfo{pages}{1613--1622}.
\newblock


\bibitem[Savioja(2010)]%
        {Savioja2010REALTIME3F}
\bibfield{author}{\bibinfo{person}{Lauri Savioja}.}
  \bibinfo{year}{2010}\natexlab{}.
\newblock \showarticletitle{Real-time 3D finite-difference time-domain
  simulation of low-and mid-frequency room acoustics}. In
  \bibinfo{booktitle}{\emph{13th Int. Conf on Digital Audio Effects}},
  Vol.~\bibinfo{volume}{1}. \bibinfo{pages}{75}.
\newblock


\bibitem[Schissler and Manocha(2016)]%
        {Schissler2016interactive}
\bibfield{author}{\bibinfo{person}{Carl Schissler} {and}
  \bibinfo{person}{Dinesh Manocha}.} \bibinfo{year}{2016}\natexlab{}.
\newblock \showarticletitle{Interactive Sound Propagation and Rendering for
  Large Multi-Source Scenes}.
\newblock \bibinfo{journal}{\emph{ACM Trans. Graph.}} \bibinfo{volume}{36},
  \bibinfo{number}{1}, Article \bibinfo{articleno}{2} (\bibinfo{date}{sep}
  \bibinfo{year}{2016}), \bibinfo{numpages}{12}~pages.
\newblock
\showISSN{0730-0301}
\urldef\tempurl%
\url{https://doi.org/10.1145/2943779}
\showDOI{\tempurl}


\bibitem[Schissler et~al\mbox{.}(2014)]%
        {schissler2014high-order}
\bibfield{author}{\bibinfo{person}{Carl Schissler}, \bibinfo{person}{Ravish
  Mehra}, {and} \bibinfo{person}{Dinesh Manocha}.}
  \bibinfo{year}{2014}\natexlab{}.
\newblock \showarticletitle{High-order diffraction and diffuse reflections for
  interactive sound propagation in large environments}.
\newblock  \bibinfo{volume}{33}, \bibinfo{number}{4} (\bibinfo{year}{2014}),
  \bibinfo{pages}{39}.
\newblock


\bibitem[Schissler et~al\mbox{.}(2021)]%
        {Schissler2021FastDP}
\bibfield{author}{\bibinfo{person}{Carl Schissler}, \bibinfo{person}{Gregor
  M{\"u}ckl}, {and} \bibinfo{person}{Paul~T. Calamia}.}
  \bibinfo{year}{2021}\natexlab{}.
\newblock \showarticletitle{Fast diffraction pathfinding for dynamic sound
  propagation}.
\newblock \bibinfo{journal}{\emph{ACM Transactions on Graphics (TOG)}}
  \bibinfo{volume}{40} (\bibinfo{year}{2021}), \bibinfo{pages}{1 -- 13}.
\newblock


\bibitem[Siltanen et~al\mbox{.}(2007)]%
        {siltanen2007the}
\bibfield{author}{\bibinfo{person}{Samuel Siltanen}, \bibinfo{person}{Tapio
  Lokki}, \bibinfo{person}{Sami Kiminki}, {and} \bibinfo{person}{Lauri
  Savioja}.} \bibinfo{year}{2007}\natexlab{}.
\newblock \showarticletitle{The room acoustic rendering equation}.
\newblock \bibinfo{journal}{\emph{Journal of the Acoustical Society of
  America}} \bibinfo{volume}{122}, \bibinfo{number}{3} (\bibinfo{year}{2007}),
  \bibinfo{pages}{1624--1635}.
\newblock


\bibitem[Stephenson(2010)]%
        {Stephenson2010energetic}
\bibfield{author}{\bibinfo{person}{Uwe~M. Stephenson}.}
  \bibinfo{year}{2010}\natexlab{}.
\newblock \showarticletitle{An Energetic Approach for the Simulation of
  Diffraction within Ray Tracing Based on the Uncertainty Relation}.
\newblock \bibinfo{journal}{\emph{Acta Acustica united with Acustica}}
  \bibinfo{volume}{96}, \bibinfo{number}{3} (\bibinfo{year}{2010}),
  \bibinfo{pages}{516--535}.
\newblock
\showISSN{1610-1928}
\urldef\tempurl%
\url{https://doi.org/doi:10.3813/AAA.918304}
\showDOI{\tempurl}


\bibitem[Stephenson and Svensson(2007)]%
        {Stephenson0improved}
\bibfield{author}{\bibinfo{person}{Uwe~Martin Stephenson} {and}
  \bibinfo{person}{U.~Peter Svensson}.} \bibinfo{year}{2007}\natexlab{}.
\newblock \showarticletitle{An Improved Energetic Approach to Diffraction Based
  on the Uncertainty Principle}. In \bibinfo{booktitle}{\emph{Proceedings of
  the 19th International Congress on Acoustics}} (Madrid, Spain).
\newblock


\bibitem[Svensson(2015)]%
        {SvenssonToolBox}
\bibfield{author}{\bibinfo{person}{U.~Peter Svensson}.}
  \bibinfo{year}{2015}\natexlab{}.
\newblock \bibinfo{title}{Edge diffraction toolbox for Matlab}.
\newblock
\newblock
\urldef\tempurl%
\url{https://folk.ntnu.no/ulfps/software/index.html#EDGE}
\showURL{%
Retrieved May 20, 2022 from \tempurl}


\bibitem[Svensson et~al\mbox{.}(1999)]%
        {svensson1999an}
\bibfield{author}{\bibinfo{person}{U~Peter Svensson}, \bibinfo{person}{Roger~I
  Fred}, {and} \bibinfo{person}{John Vanderkooy}.}
  \bibinfo{year}{1999}\natexlab{}.
\newblock \showarticletitle{An analytic secondary source model of edge
  diffraction impulse responses}.
\newblock \bibinfo{journal}{\emph{Journal of the Acoustical Society of
  America}} \bibinfo{volume}{106}, \bibinfo{number}{5} (\bibinfo{year}{1999}),
  \bibinfo{pages}{2331--2344}.
\newblock


\bibitem[Tang et~al\mbox{.}(2021)]%
        {Tang2021learning}
\bibfield{author}{\bibinfo{person}{Zhenyu Tang}, \bibinfo{person}{Hsien-Yu
  Meng}, {and} \bibinfo{person}{Dinesh Manocha}.}
  \bibinfo{year}{2021}\natexlab{}.
\newblock \showarticletitle{Learning Acoustic Scattering Fields for Dynamic
  Interactive Sound Propagation}. In \bibinfo{booktitle}{\emph{2021 IEEE
  Virtual Reality and 3D User Interfaces (VR)}}. \bibinfo{pages}{835--844}.
\newblock
\urldef\tempurl%
\url{https://doi.org/10.1109/VR50410.2021.00111}
\showDOI{\tempurl}


\bibitem[Tarantola(2008)]%
        {Tarantola08measure}
\bibfield{author}{\bibinfo{person}{Albert Tarantola}.}
  \bibinfo{year}{2008}\natexlab{}.
\newblock \bibinfo{title}{Image and Reciprocal Image of a Measure.
  Compatibility Theorem.}
\newblock
\newblock


\bibitem[Taylor et~al\mbox{.}(2012)]%
        {micah2012guided}
\bibfield{author}{\bibinfo{person}{Micah Taylor}, \bibinfo{person}{Anish
  Chandak}, \bibinfo{person}{Qi Mo}, \bibinfo{person}{Christian Lauterbach},
  \bibinfo{person}{Carl Schissler}, {and} \bibinfo{person}{Dinesh Manocha}.}
  \bibinfo{year}{2012}\natexlab{}.
\newblock \showarticletitle{Guided Multiview Ray Tracing for Fast
  Auralization}.
\newblock \bibinfo{journal}{\emph{IEEE Transactions on Visualization and
  Computer Graphics}} \bibinfo{volume}{18}, \bibinfo{number}{11}
  (\bibinfo{year}{2012}), \bibinfo{pages}{1797--1810}.
\newblock
\urldef\tempurl%
\url{https://doi.org/10.1109/TVCG.2012.27}
\showDOI{\tempurl}


\bibitem[Thompson(2006)]%
        {Thompson2006review}
\bibfield{author}{\bibinfo{person}{Lonny~L. Thompson}.}
  \bibinfo{year}{2006}\natexlab{}.
\newblock \showarticletitle{A review of finite-element methods for
  time-harmonic acoustics}.
\newblock \bibinfo{journal}{\emph{The Journal of the Acoustical Society of
  America}} \bibinfo{volume}{119}, \bibinfo{number}{3} (\bibinfo{year}{2006}),
  \bibinfo{pages}{1315--1330}.
\newblock
\urldef\tempurl%
\url{https://doi.org/10.1121/1.2164987}
\showDOI{\tempurl}
\showeprint{https://doi.org/10.1121/1.2164987}


\bibitem[Torres et~al\mbox{.}(2001)]%
        {Torres2001ComputationOE}
\bibfield{author}{\bibinfo{person}{Rendell~R. Torres},
  \bibinfo{person}{U.~Peter Svensson}, {and} \bibinfo{person}{Mendel Kleiner}.}
  \bibinfo{year}{2001}\natexlab{}.
\newblock \showarticletitle{Computation of edge diffraction for more accurate
  room acoustics auralization.}
\newblock \bibinfo{journal}{\emph{The Journal of the Acoustical Society of
  America}}  \bibinfo{volume}{109 2} (\bibinfo{year}{2001}),
  \bibinfo{pages}{600--10}.
\newblock


\bibitem[Tsingos et~al\mbox{.}(2007)]%
        {tsingos2007instant}
\bibfield{author}{\bibinfo{person}{Nicolas Tsingos}, \bibinfo{person}{Carsten
  Dachsbacher}, \bibinfo{person}{Sylvain Lefebvre}, {and}
  \bibinfo{person}{Matteo Dellepiane}.} \bibinfo{year}{2007}\natexlab{}.
\newblock \showarticletitle{Instant Sound Scattering}. In
  \bibinfo{booktitle}{\emph{Proceedings of the 18th Eurographics Conference on
  Rendering Techniques}} (Grenoble, France) \emph{(\bibinfo{series}{EGSR'07})}.
  \bibinfo{publisher}{Eurographics Association}, \bibinfo{address}{Goslar,
  DEU}, \bibinfo{pages}{111–120}.
\newblock
\showISBNx{9783905673524}


\bibitem[Tsingos et~al\mbox{.}(2001)]%
        {tsingos2001modeling}
\bibfield{author}{\bibinfo{person}{Nicolas Tsingos}, \bibinfo{person}{Thomas
  Funkhouser}, \bibinfo{person}{Addy Ngan}, {and} \bibinfo{person}{Ingrid
  Carlbom}.} \bibinfo{year}{2001}\natexlab{}.
\newblock \showarticletitle{Modeling acoustics in virtual environments using
  the uniform theory of diffraction}.
\newblock  (\bibinfo{year}{2001}), \bibinfo{pages}{545--552}.
\newblock


\bibitem[Veach(1997)]%
        {Guibas1997RobustMC}
\bibfield{author}{\bibinfo{person}{Eric Veach}.}
  \bibinfo{year}{1997}\natexlab{}.
\newblock \emph{\bibinfo{title}{Robust Monte Carlo methods for light transport
  simulation}}.
\newblock \bibinfo{thesistype}{Ph.\,D. Dissertation}.
\newblock


\bibitem[Veach and Guibas(1994)]%
        {Veach1995BidirectionalEF}
\bibfield{author}{\bibinfo{person}{Eric Veach} {and}
  \bibinfo{person}{Leonidas~J. Guibas}.} \bibinfo{year}{1994}\natexlab{}.
\newblock \showarticletitle{Bidirectional Estimators for Light Transport}. In
  \bibinfo{booktitle}{\emph{Proceedings of Eurographics Workshop on
  Rendering'94}}.
\newblock


\bibitem[Veach and Guibas(1997)]%
        {Veach1997MetropolisLT}
\bibfield{author}{\bibinfo{person}{Eric Veach} {and}
  \bibinfo{person}{Leonidas~J. Guibas}.} \bibinfo{year}{1997}\natexlab{}.
\newblock \showarticletitle{Metropolis light transport}. In
  \bibinfo{booktitle}{\emph{Proceedings of the 24th Annual Conference on
  Computer Graphics and Interactive Techniques}}.
\newblock


\bibitem[Vorländer(1989)]%
        {Vorlander1989simulation}
\bibfield{author}{\bibinfo{person}{Michael Vorländer}.}
  \bibinfo{year}{1989}\natexlab{}.
\newblock \showarticletitle{Simulation of the transient and steady‐state
  sound propagation in rooms using a new combined ray‐tracing/image‐source
  algorithm}.
\newblock \bibinfo{journal}{\emph{The Journal of the Acoustical Society of
  America}} \bibinfo{volume}{86}, \bibinfo{number}{1} (\bibinfo{year}{1989}),
  \bibinfo{pages}{172--178}.
\newblock
\urldef\tempurl%
\url{https://doi.org/10.1121/1.398336}
\showDOI{\tempurl}
\showeprint{https://doi.org/10.1121/1.398336}


\bibitem[Yeh et~al\mbox{.}(2013)]%
        {yeh2013wave-ray}
\bibfield{author}{\bibinfo{person}{Hengchin Yeh}, \bibinfo{person}{Ravish
  Mehra}, \bibinfo{person}{Zhimin Ren}, \bibinfo{person}{Lakulish Antani},
  \bibinfo{person}{Dinesh Manocha}, {and} \bibinfo{person}{Ming~C Lin}.}
  \bibinfo{year}{2013}\natexlab{}.
\newblock \showarticletitle{Wave-ray coupling for interactive sound propagation
  in large complex scenes}. In \bibinfo{booktitle}{\emph{International
  Conference on Computer Graphics and Interactive Techniques}},
  Vol.~\bibinfo{volume}{32}. \bibinfo{pages}{165}.
\newblock


\end{thebibliography}



\begin{thebibliography}{4}


\ifx \showCODEN    \undefined \def \showCODEN     #1{\unskip}     \fi
\ifx \showDOI      \undefined \def \showDOI       #1{#1}\fi
\ifx \showISBNx    \undefined \def \showISBNx     #1{\unskip}     \fi
\ifx \showISBNxiii \undefined \def \showISBNxiii  #1{\unskip}     \fi
\ifx \showISSN     \undefined \def \showISSN      #1{\unskip}     \fi
\ifx \showLCCN     \undefined \def \showLCCN      #1{\unskip}     \fi
\ifx \shownote     \undefined \def \shownote      #1{#1}          \fi
\ifx \showarticletitle \undefined \def \showarticletitle #1{#1}   \fi
\ifx \showURL      \undefined \def \showURL       {\relax}        \fi
\providecommand\bibfield[2]{#2}
\providecommand\bibinfo[2]{#2}
\providecommand\natexlab[1]{#1}
\providecommand\showeprint[2][]{arXiv:#2}

\bibitem[Keller and Blank(1951)]%
        {keller1951diffraction}
\bibfield{author}{\bibinfo{person}{Joseph~B Keller} {and}
  \bibinfo{person}{Albert Blank}.} \bibinfo{year}{1951}\natexlab{}.
\newblock \showarticletitle{Diffraction and reflection of pulses by wedges and
  corners}.
\newblock \bibinfo{journal}{\emph{Communications on Pure and Applied
  Mathematics}} \bibinfo{volume}{4}, \bibinfo{number}{1}
  (\bibinfo{year}{1951}), \bibinfo{pages}{75--94}.
\newblock


\bibitem[Medwin et~al\mbox{.}(1982)]%
        {medwin1982impulse}
\bibfield{author}{\bibinfo{person}{Herman Medwin}, \bibinfo{person}{Emily
  Childs}, {and} \bibinfo{person}{Gary~M Jebsen}.}
  \bibinfo{year}{1982}\natexlab{}.
\newblock \showarticletitle{Impulse studies of double diffraction: A discrete
  Huygens interpretation}.
\newblock \bibinfo{journal}{\emph{Journal of the Acoustical Society of
  America}} \bibinfo{volume}{72}, \bibinfo{number}{3} (\bibinfo{year}{1982}),
  \bibinfo{pages}{1005--1013}.
\newblock


\bibitem[Needham(1997)]%
        {Needham1997VisualCA}
\bibfield{author}{\bibinfo{person}{Tristan Needham}.}
  \bibinfo{year}{1997}\natexlab{}.
\newblock \showarticletitle{Visual Complex Analysis}.
\newblock


\bibitem[Tarantola(2008)]%
        {Tarantola08measure}
\bibfield{author}{\bibinfo{person}{Albert Tarantola}.}
  \bibinfo{year}{2008}\natexlab{}.
\newblock \bibinfo{title}{Image and Reciprocal Image of a Measure.
  Compatibility Theorem.}
\newblock
\newblock


\end{thebibliography}

\end{document}


\maketitle

\section{Construction of BEDRF}

Our BEDRF comes from the solution for wedge diffraction of incident planar pulse waves in \cite{keller1951diffraction}. To understand the expression of BEDRF, we need to take a look at the original solution. The following sections require some knowledge on complex analysis. A good reference book on this topic is \cite{Needham1997VisualCA}.

\subsection{Piecewise Constant Function on Complex Unit Circle}

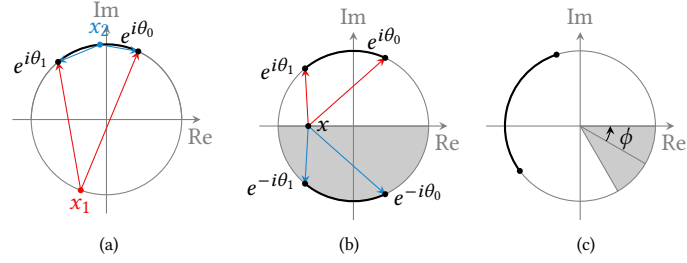
\begin{figure*}[htbp]
    \centering
    \subfigure[]{
        \begin{tikzpicture}
            \coordinate (pos_re) at (1.25,0);
            \coordinate (pos_im) at (0,1.25);
            \coordinate (omega1) at (65:1);
            \coordinate (omega2) at (130:1);
            \coordinate (x_1) at (250:1);
            \coordinate (x_2) at (95:1);
            \draw[gray, -stealth] (-1.25,0)--(pos_re) node[below]{Re};
            \draw[gray, -stealth] (0,-1.25)--(pos_im) node[above]{Im};
            \draw[gray] (0,0)circle(1);
            \draw[gray] (1,0)arc(0:180:1);
            \draw[black, thick] (omega1)arc(65:130:1);
            \draw[red-base,-stealth] (x_1)--(omega1);
            \draw[red-base,-stealth] (x_1)--(omega2);
            \draw[blue-base,-stealth] (x_2)--(omega1);
            \draw[blue-base,-stealth] (x_2)--(omega2);
            \fill[black] (omega1)circle(0.04) node[above]{$e^{i\theta_0}$};
            \fill[black] (omega2)circle(0.04) node[left]{$e^{i\theta_1}$};
            \fill[red-base] (x_1)circle(0.04) node[below]{$x_1$};
            \fill[blue-base] (x_2)circle(0.04) node[above]{$x_2$};
            \label{fig-PWF-1}
        \end{tikzpicture}
    }
    \subfigure[]{
        \begin{tikzpicture}
            \coordinate (pos_re) at (1.25,0);
            \coordinate (pos_im) at (0,1.25);
            \coordinate (omega1) at (65:1);
            \coordinate (omega2) at (130:1);
            \coordinate (omega2a) at (230:1);
            \coordinate (omega1a) at (295:1);
            \coordinate (x) at (-0.6,0);
            \coordinate (x_2) at (95:1);
            \fill[gray!40,draw=gray](0,0)--(-1,0)arc(180:360:1)--cycle;
            \draw[gray, -stealth] (-1.25,0)--(pos_re) node[below]{Re};
            \draw[gray, -stealth] (0,-1.25)--(pos_im) node[above]{Im};
            \draw[gray] (1,0)arc(0:180:1);
            \draw[black, thick] (65:1)arc(65:130:1);
            \draw[black, thick] (230:1)arc(230:295:1);
            \draw[red-base,-stealth] (-0.6,0)--(65:1);
            \draw[red-base,-stealth] (-0.6,0)--(130:1);
            \draw[blue-base,-stealth] (-0.6,0)--(230:1);
            \draw[blue-base,-stealth] (-0.6,0)--(295:1);
            \fill[black] (omega1)circle(0.04) node[above]{$e^{i\theta_0}$};
            \fill[black] (omega2)circle(0.04) node[left]{$e^{i\theta_1}$};
            \fill[black] (omega1a)circle(0.04) node[right]{$e^{-i\theta_0}$};
            \fill[black] (omega2a)circle(0.04) node[left]{$e^{-i\theta_1}$};
            \fill[black] (x)circle(0.04) node[right]{$x$};
        \end{tikzpicture}
    }
    \subfigure[]{
        \begin{tikzpicture}
            \coordinate (pos_re) at (1.25,0);
            \coordinate (pos_im) at (0,1.25);
            \coordinate (omega1) at (108.33:1);
            \coordinate (omega2) at (216.67:1);
            \coordinate (x_1) at (170:1);
            \coordinate (x_2) at (95:1);
            \coordinate (wedge1) at (330:1);
            \coordinate (wedge2) at (360:1);
            \coordinate (center) at (0,0);
            \fill[gray!40,draw=gray](0,0)--(300:1)arc(300:360:1)--cycle;
            \draw[gray, -stealth] (-1.25,0)--(pos_re) node[below]{Re};
            \draw[gray, -stealth] (0,-1.25)--(pos_im) node[above]{Im};
            \draw[gray] (0:0)--(330:1);
            \draw[gray] (0:1)arc(0:300:1);
            \draw[black, thick] (omega1)arc(108.33:216.67:1);
            \fill[black] (omega1)circle(0.04);
            \fill[black] (omega2)circle(0.04);
            \pic ["$\phi$", draw, -stealth, angle eccentricity=1.6, angle radius=0.4cm]{angle=wedge1--center--wedge2};
        \end{tikzpicture}
    }
    \caption{Visualization of the construction process of \autoref{eq-fa1} and \autoref{eq-fb1}. Fig.\ (a) shows the function $f_c(\theta_0,\theta_1)(x)$ whose value is related to the angle between $e^{i\theta_0}-x$ and $e^{i\theta_1}-x$, which is constant on arcs between $e^{i\theta_0}$ and $e^{i\theta_1}$. In Fig.\ (b), we flip $f_c$ vertically and combine with the original $f_c$ to satisfy the boundary condition on the real axis. The upper half of the unit circle is then scaled by $1/\kappa$ to transform the real axis into a wedge, as shown in Fig.\ (c).}
\end{figure*}

The solution for wedge diffraction is related to piecewise constant functions on the unit circle of the complex plane. An example of a piecewise constant function is

\begin{equation}
    f_c(\theta_0,\theta_1)(x)=\frac{1}{\pi}\left(\text{arg}\left(\frac{x-e^{i\theta_1}}{x-e^{i\theta_0}}\right)-\frac{\theta_1-\theta_0}2\right),
\end{equation}

where $\text{arg}(x)$ is the argument of the complex number $x$. This function is analytic within the whole unit circle except at $e^{i\theta_0}$ and $e^{i\theta_1}$. When $0\leq\theta_0<\theta_1\leq2\pi$ and $x$ is on the arc $e^{i\omega},\omega\in (\theta_0,\theta_1)$, we have $f_c(\theta_0,\theta_1)(x)=1$. On the remainder of the unit circle, we have $f_c(\theta_0,\theta_1)(x)=0$. This is because that the angle between $x-e^{i\theta_0}$ and $x-e^{i\theta_1}$ remains constant inside the same arc (See \autoref{fig-PWF-1}). It is obvious that

\begin{equation}
    f_c(\omega,\omega) = 0
    \label{eq:fc_zero}
\end{equation}
and

\begin{equation}
    f_c(\theta_0,\theta_1)+f_c(\theta_1,\theta_2)=f_c(\theta_0,\theta_2).
    \label{eq:fc_add}
\end{equation}

From \autoref{eq:fc_zero} and \autoref{eq:fc_add}, we can derive two useful equations:

\begin{equation}
    f_c(\theta_0,\theta_1)=-f_c(\theta_1,\theta_0),
    \label{eq:fc-r1}
\end{equation}

\begin{equation}
    f_c(\theta_0,\theta_1)-f_c(\theta_2,\theta_3)=f_c(\theta_0,\theta_2)-f_c(\theta_1,\theta_3).
    \label{eq:fc-r2}
\end{equation}

Now we will express $f_c$ as a real value function:

\begin{equation}
    \begin{array}{rl}
        & \text{arg}\left(\frac{re^{i\theta}-e^{i\theta_1}}{re^{i\theta}-e^{i\theta_0}}\right)-\frac{\theta_1-\theta_0}2 \\
        = & \text{arg}\left((re^{i\theta}-e^{i\theta_1})(re^{-i\theta}-e^{-i\theta_0})e^{\frac{\theta_0-\theta_1}2}\right) \\
        = & \text{arg}\left(\left(r^2+e^{i(\theta_1-\theta_0)}-r(e^{i(\theta-\theta_0)}+e^{i(\theta_1-\theta)})\right)e^{\frac{\theta_0-\theta_1}2}\right) \\
        = & \text{arg}\left(r^2e^{\frac{\theta_0-\theta_1}2}+e^{\frac{\theta_1-\theta_0}2}-r(e^{i(\theta-\frac{\theta_0+\theta_1}2)}+e^{i(\frac{\theta_0+\theta_1}2-\theta)})\right) \\
        = & \text{arg}\left(r^2e^{\frac{\theta_0-\theta_1}2}+e^{\frac{\theta_1-\theta_0}2}-2r\cos(\theta-\frac{\theta_0+\theta_1}2)\right) \\
        = & \arctan^*\left(\frac{(1-r^2)\sin(\frac{\theta_1-\theta_0}2)}{(1+r^2)\cos(\frac{\theta_1-\theta_0}2)-2r\cos(\theta-\frac{\theta_0+\theta_1}2)}\right).
    \end{array}
\end{equation}
$\arctan$ is not a univalent function and we use $\arctan^*$ here to avoid ambiguity. We define $\arctan^*$ to be $\arctan$ with the range being $[\frac{\theta_1-\theta_0}2, \pi+\frac{\theta_1-\theta_0}2]$. Now we have

\begin{equation}
    \begin{array}{l}
    f_c(\theta_0,\theta_1)(re^{i\theta})\\
    =\frac1\pi\arctan^*\left(\frac{(1-r^2)\sin(\frac{\theta_1-\theta_0}2)}{(1+r^2)\cos(\frac{\theta_1-\theta_0}2)-2r\cos(\theta-\frac{\theta_0+\theta_1}2)}\right).
    \end{array}
\end{equation}

After this, consider the function

\begin{equation}
    f_a(\theta_0,\theta_1) = f_c(\theta_0,2\pi-\theta_1)-f_c(\theta_1,2\pi-\theta_0),0\leq\theta_0<\theta_1\leq\pi.
    \label{eq-fa}
\end{equation}
On the upper half of the unit circle, using \autoref{eq:fc-r2}, we have $f_a(\theta_0,\theta_1)=f_c(\theta_0,\theta_1)$. On the real axis, we have $f_a=0$ as $f_c(\theta_0,2\pi-\theta_1)$ and $f_c(\theta_1,2\pi-\theta_0)$ is symmetric with respect to the real axis. This is exactly the requirement of the Dirichlet condition. Similarly, consider the function

\begin{equation}
    f_b(\theta_0,\theta_1) = f_c(\theta_0,2\pi-\theta_1)+f_c(\theta_1,2\pi-\theta_0),0\leq\theta_0<\theta_1\leq\pi.
    \label{eq-fb}
\end{equation}
We can see that $f_b$ is symmetric with respect to the real axis, and $f_b$ satisfies the requirement of the Neumann condition on the axis. Notice that there are other ways to satisfy the Neumann condition, like $f_c(\theta_0,\theta_1)+f_c(2\pi-\theta_1,2\pi-\theta_0)$. And we need other restrictions to determine the final expression.

Now we use a conformal transform to bend the real axis into a wedge. We define the following function:

\begin{equation}
    f^*_c(\theta_0,\theta_1,\theta,r,\kappa) = f_c(\kappa\theta_0,\kappa\theta_1)(re^{i\kappa\theta})
    \label{eq-fc}
\end{equation}
where

\begin{equation}
    \kappa=\frac{\pi}{2\pi-2\phi}
\end{equation}

and $\phi$ is the angle of half-wedge. Then we replace $f_c$ in \autoref{eq-fa} and \autoref{eq-fb} with $f^*_c$:

\begin{equation}
    f^*_a(\theta_0,\theta_1) = f^*_c(\theta_0,2\pi-\theta_1)-f^*_c(\theta_1,2\pi-\theta_0)
    \label{eq-fa1}
\end{equation}

\begin{equation}
    f^*_b(\theta_0,\theta_1) = f^*_c(\theta_0,2\pi-\theta_1)+f^*_c(\theta_1,2\pi-\theta_0)
    \label{eq-fb1}
\end{equation}

Since the transform is conformal on the upper half plane, both the Dirichlet and Neumann condition holds on the wedge boundary. It is not hard to imagine the relationship of these functions and the solution for wedge diffraction.

\subsection{Solution for Wedge Diffraction}

In geometric acoustics, if we ignore the diffraction effect, the unoccluded and reflected wave will have discontinuities on certain directions. These discontinuities are generally referred as ``shadow boundaries''. A half-plane occluding a plane wave will generate shadow boundaries on two different directions, one on the boundary of the unoccluded wave, another on the boundary of the reflected wave. A wedge consisting of two half-planes will have 4 possible shadow boundary directions when occluding a plane wave. Assume that the incident direction of the plane wave is $\theta_i$, we list all possible directions below:

\begin{equation}
    \begin{array}{rcl}
        \omega_{lu} & = & \theta_i-\pi \\
        \omega_{lr} & = & -\theta_i-\pi+2\phi \\
        \omega_{ru} & = & \theta_i+\pi \\
        \omega_{rr} & = & -\theta_i+\pi-2\phi
    \end{array}
\end{equation}

\begin{figure*}[htbp]
    \centering
    \subfigure[]{
        \begin{tikzpicture}
            \coordinate (pos_re) at (1.25,0);
            \coordinate (pos_im) at (0,1.25);
            \coordinate (in) at (140:1.2);
            \coordinate (out1) at (320:1.2);
            \coordinate (out2) at (160:1.2);
            \coordinate (x_1) at (170:1);
            \coordinate (x_2) at (95:1);
            \coordinate (wedge1) at (240:1);
            \coordinate (wedge2) at (300:1);
            \coordinate (above) at (90:1);
            \coordinate (bottom) at (270:1);
            \coordinate (center) at (0,0);
            \fill[gray!40,draw=gray](0,0)--(wedge1)arc(240:300:1)--cycle;
            \draw[gray, -stealth] (-1.25,0)--(pos_re) node[below]{$\mathbf{b}$};
            \draw[gray, -stealth] (0,-1.25)--(pos_im) node[above]{$\mathbf{n}$};
            \draw[gray] (300:1)arc(300:360:1);
            \draw[gray] (0:1)arc(0:240:1);
            \draw[red-base,-stealth] (in)--(center);
            \draw[blue-base,-stealth] (center)--(out1);
            \draw[blue-base,-stealth] (center)--(out2);
            \pic ["$\phi$", draw, -stealth, angle eccentricity=1.6, angle radius=0.4cm]{angle=bottom--center--wedge2};
            \pic [red-base,"$-\theta_i$", draw, -stealth, angle eccentricity=1.3, angle radius=0.8cm]{angle=above--center--in};
            \pic [blue-base,"$\omega_{ru}$", draw, stealth-, angle eccentricity=2.2, angle radius=0.2cm]{angle=out1--center--above};
            \pic [blue-base,"$-\omega_{lr}$", draw, -stealth, angle eccentricity=1.8, angle radius=0.3cm]{angle=above--center--out2};
            \label{fig-boundary1}
        \end{tikzpicture}
    }
    \subfigure[]{
        \begin{tikzpicture}
            \coordinate (pos_re) at (1.25,0);
            \coordinate (pos_im) at (0,1.25);
            \coordinate (in) at (100:1.2);
            \coordinate (out1) at (-40:1.2);
            \coordinate (out2) at (200:1.2);
            \coordinate (x_1) at (170:1);
            \coordinate (x_2) at (95:1);
            \coordinate (wedge1) at (240:1);
            \coordinate (wedge2) at (300:1);
            \coordinate (above) at (90:1);
            \coordinate (bottom) at (270:1);
            \coordinate (center) at (0,0);
            \fill[gray!40,draw=gray](0,0)--(wedge1)arc(240:300:1)--cycle;
            \draw[gray, -stealth] (-1.25,0)--(pos_re) node[below]{$\mathbf{b}$};
            \draw[gray, -stealth] (0,-1.25)--(pos_im) node[above]{$\mathbf{n}$};
            \draw[gray] (300:1)arc(300:360:1);
            \draw[gray] (0:1)arc(0:240:1);
            \draw[red-base,-stealth] (in)--(center);
            \draw[blue-base,-stealth] (center)--(out1);
            \draw[blue-base,-stealth] (center)--(out2);
            \pic ["$\phi$", draw, -stealth, angle eccentricity=1.6, angle radius=0.4cm]{angle=bottom--center--wedge2};
            \pic [red-base,"$-\theta_i$", draw, stealth-, angle eccentricity=1.25, angle radius=0.8cm]{angle=above--center--in};
            \pic [blue-base,"$-\omega_{rr}$", draw, stealth-, angle eccentricity=2.4, angle radius=0.2cm]{angle=out1--center--above};
            \pic [blue-base,"$\omega_{lr}$", draw, -stealth, angle eccentricity=1.7, angle radius=0.3cm]{angle=above--center--out2};
            \label{fig-boundary2}
        \end{tikzpicture}
    }
    \subfigure[]{
        \begin{tikzpicture}
            \coordinate (pos_re) at (1.25,0);
            \coordinate (pos_im) at (0,1.25);
            \coordinate (in) at (40:1.2);
            \coordinate (out1) at (220:1.2);
            \coordinate (out2) at (20:1.2);
            \coordinate (x_1) at (170:1);
            \coordinate (x_2) at (95:1);
            \coordinate (wedge1) at (240:1);
            \coordinate (wedge2) at (300:1);
            \coordinate (above) at (90:1);
            \coordinate (bottom) at (270:1);
            \coordinate (center) at (0,0);
            \fill[gray!40,draw=gray](0,0)--(wedge1)arc(240:300:1)--cycle;
            \draw[gray, -stealth] (-1.25,0)--(pos_re) node[below]{$\mathbf{b}$};
            \draw[gray, -stealth] (0,-1.25)--(pos_im) node[above]{$\mathbf{n}$};
            \draw[gray] (300:1)arc(300:360:1);
            \draw[gray] (0:1)arc(0:240:1);
            \draw[red-base,-stealth] (in)--(center);
            \draw[blue-base,-stealth] (center)--(out1);
            \draw[blue-base,-stealth] (center)--(out2);
            \pic ["$\phi$", draw, -stealth, angle eccentricity=1.6, angle radius=0.4cm]{angle=bottom--center--wedge2};
            \pic [red-base,"$\theta_i$", draw, stealth-, angle eccentricity=1.2, angle radius=0.8cm]{angle=in--center--above};
            \pic [blue-base,"$-\omega_{lu}$", draw, -stealth, angle eccentricity=2.6, angle radius=0.2cm]{angle=above--center--out1};
            \pic [blue-base,"$\omega_{rr}$", draw, stealth-, angle eccentricity=1.6, angle radius=0.3cm]{angle=out2--center--above};
            \label{fig-boundary3}
        \end{tikzpicture}
    }
    \subfigure[]{
        \begin{tikzpicture}
            \coordinate (pos_re) at (1.25,0);
            \coordinate (pos_im) at (0,1.25);
            \coordinate (in) at (60:1);
            \coordinate (out1) at (162:1);
            \coordinate (out2) at (48:1);
            \coordinate (out3) at (198:1);
            \coordinate (out4) at (-48:1);
            \coordinate (x_1) at (170:1);
            \coordinate (x_2) at (95:1);
            \coordinate (wedge1) at (180:1);
            \coordinate (wedge2) at (360:1);
            \coordinate (above) at (90:1);
            \coordinate (bottom) at (270:1);
            \coordinate (center) at (0,0);
            \fill[gray!40,draw=gray](1,0)--(-1,0)--(wedge1)arc(180:360:1)--cycle;
            \draw[gray, -stealth] (-1.25,0)--(pos_re);
            \draw[gray, -stealth] (0,-1.25)--(pos_im) node[above]{$0$};
            \draw[gray] (300:1)arc(300:360:1);
            \draw[gray] (0:1)arc(0:240:1);
            \draw[red-base] (in) node[above]{$\kappa\theta_i$}--(center);
            \draw[blue-base] (center)--(out1) node[left]{$\kappa\omega_{lu}$};
            \draw[blue-base] (center)--(out2) node[right]{$\kappa\omega_{rr}$};
            \draw[blue-base] (center)--(out3) node[left]{$\kappa\omega_{lr}$};
            \draw[blue-base] (center)--(out4) node[right]{$\kappa\omega_{ru}$};
            \fill[black] (in)circle(0.04);
            \fill[black] (out1)circle(0.04);
            \fill[black] (out2)circle(0.04);
            \fill[black] (out3)circle(0.04);
            \fill[black] (out4)circle(0.04);
        \end{tikzpicture}
    }
    \caption{Visualization of the shadow boundary directions of wedge diffraction in different cases (Fig.\ (a), (b) and (c)). When multiplying the angles in Fig.\ (c) with the scaling factor $\kappa$, we can see the symmetry of these possible directions with respect to the wedge boundary, as shown in Fig.\ (d). }
    \label{fig-boundary}
\end{figure*}
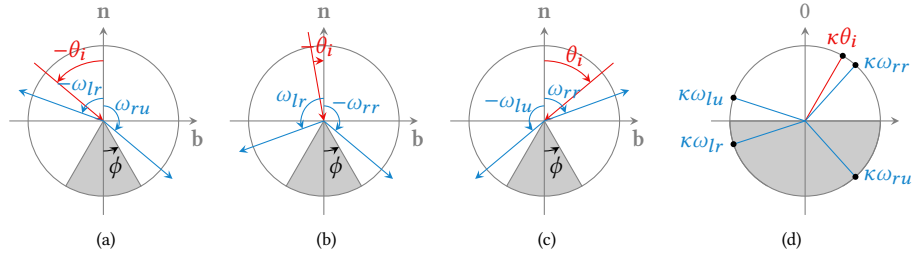

Here we use $l/r$ for the first letter of the subscript to distinguish between left and right half-planes, and $u/r$ for the second letter for unoccluded and reflected waves. The validity of these directions depends on the value of $\theta_i$ and only two directions can be valid in the same time.  When we multiply all the angles with the scaling factor $\kappa$, the shadow boundary directions show an axial symmetry. See \autoref{fig-boundary}.

Now we will take a look at the diffraction solution. We will use the wedge coordinate system in Sect.\ 3.4 of our paper. We set the origin of the wedge coordinate to $(0,0,0)$ where the incident plane impulse touches the wedge at time $t=0$. The incident direction is

\begin{equation}
    \mathbf{v}_i = -\cos\theta_i\cos\varphi_i\mathbf{n}-\sin\theta_i\cos\varphi_i\mathbf{b}-\sin\varphi_i\mathbf{t},
\end{equation}
and the listener $\mathbf{x}$ is at

\begin{equation}
    \mathbf{x} = x_x\mathbf{n} + x_y\mathbf{b} + x_z\mathbf{t} = R^*(\cos\theta_o\cos\varphi_i\mathbf{n}+\sin\theta_o\cos\varphi_i\mathbf{b}-\sin\varphi_i\mathbf{t})
\end{equation}
where
\begin{equation}
    R^*=||\mathbf{x}||_2=\sqrt{x^2_x+x^2_y+x^2_z}.
\end{equation}
Here we put an asterisk over $R$ because it is not the distance from BEDRF to the listener. Without loss of generality, we assume that the outgoing angle is equal to the incident angle on the normal plane of the wedge (which can be achieved by adjusting the origin of the coordinate). One can see from the equations above that the component of $\mathbf{x}$ on the direction $\mathbf{t}$ is not $R^*\sin\varphi_o$, but $-R^*\sin\varphi_i$.

According to \cite{keller1951diffraction}, the solution for wedge diffraction under the Dirichlet boundary is

\begin{equation}
    C_D(\theta_i)+f^*_c(\omega_{ru}, \omega_{lu},\theta_o,r^\kappa,\kappa)-f^*_c(\omega_{lr}, \omega_{rr},\theta_o,r^\kappa,\kappa).
    \label{eq-Dirichlet1}
\end{equation}
The variable $r$ is defined by the following equations:



\begin{equation}
    r=\frac{s-\sqrt{s^2-d^2}}{d},d=\sqrt{x^2_x+x^2_y}=R^*\cos\varphi_i,
    \label{eq-18}
\end{equation}

\begin{equation}
    s=\frac{ct}{\cos\varphi_i}+x_z\tan\varphi_i=\frac{ct-R^*\sin^2\varphi_i}{\cos\varphi_i}.
    \label{eq-19}
\end{equation}
Here $c$ stands for the sound speed. For the Neumann boundary, the solution is

\begin{equation}
    C_N(\theta_i)+f^*_c(\omega_{ru}, \omega_{lu},\theta_o,r^\kappa,\kappa)+f^*_c(\omega_{lr}, \omega_{rr},\theta_o,r^\kappa,\kappa).
    \label{eq-Neumann1}
\end{equation}
In the original paper, the expression of the solution for cases \autoref{fig-boundary1}, \autoref{fig-boundary2} and \autoref{fig-boundary3} are different. We find that the difference can be extracted out as a term $C_D$ or $C_N$, which is irrelated to $t$.

The cross section of the incident wave in the original paper is not a Dirac delta impulse, but the Heaviside step function:

\begin{equation}
    1(t)=\left\{\begin{array}{cc}
        0, & t<0 \\
        1, & t\geq 0
    \end{array}\right.
\end{equation}
And the solution for the delta impulse is the derivative of the original solution over $t$. For a Dirac delta impulse, we can completely ignore $C_D$ and $C_N$ and consider only $df^*_c/dt$. Now we will calculate this derivative. First we have

\begin{equation}
    \begin{array}{rl}
        & \pi f^*_c(\theta_0,\theta_1,\theta,r^\kappa,\kappa) \\
        = & \arctan^*\left(\frac{(1-r^{2\kappa})\sin(\kappa(\frac{\theta_1-\theta_0}2))}{(1+r^{2\kappa})\cos(\kappa(\frac{\theta_1-\theta_0}2))-2r^\kappa\cos(\kappa(\theta-\frac{\theta_0+\theta_1}2))}\right) \\
        = & \arctan^*\left(\frac{\frac{r^{-\kappa}-r^\kappa}{2}\sin(\kappa(\frac{\theta_1-\theta_0}2))}{\frac{r^{-\kappa}+r^\kappa}{2}\cos(\kappa(\frac{\theta_1-\theta_0}2))-\cos(\kappa(\theta-\frac{\theta_0+\theta_1}2))}\right) \\
        = & \arctan^*\left(\frac{\sinh(\kappa\ln r)\sin(\kappa(\frac{\theta_1-\theta_0}2))}{\cos(\kappa(\theta-\frac{\theta_0+\theta_1}2))-\cosh(\kappa\ln r)\cos(\kappa(\frac{\theta_1-\theta_0}2))}\right).
    \end{array}
\end{equation}
Since $d\arctan(x)/dx=d\arctan^*(x)/dx$, we define
\begin{equation}
    \begin{aligned}
        & f(\theta_0,\theta_1,\theta,r,\kappa) \\
        & = \frac{1}{\pi}\arctan\left(\frac{\sinh(\kappa r)\sin(\kappa(\frac{\theta_1-\theta_0}2))}{\cos(\kappa(\theta-\frac{\theta_0+\theta_1}2))-\cosh(\kappa r)\cos(\kappa(\frac{\theta_1-\theta_0}2))}\right)
    \end{aligned}
\end{equation}
and have $f(\ln r)=f^*_c(r^\kappa)$. The expression of $df/dr$ is useful and we'll give it here:

\begin{equation}
    \begin{aligned}
        & \frac{df(\theta_0,\theta_1,\theta,r,\kappa)}{dr} \\
        & = \frac{\kappa}{\pi}\left(\cosh(\kappa r)\sin(\kappa(\frac{\theta_1-\theta_0}2))\cos(\kappa(\theta-\frac{\theta_0+\theta_1}2)) \right. \\
        & \left.-\sin(\kappa(\frac{\theta_1-\theta_0}2))\cos(\kappa(\frac{\theta_1-\theta_0}2))\right) \\
        & / \left((\sinh(\kappa r)\sin(\kappa(\frac{\theta_1-\theta_0}2)))^2\right. \\
        & \left. + (\cosh(\kappa r)\cos(\kappa(\frac{\theta_1-\theta_0}2))-\cos(\kappa(\theta-\frac{\theta_0+\theta_1}2)))^2\right)
    \end{aligned}
\end{equation}

To achieve $df(\ln r)/dt$, we also need the expression of $d\ln r/dt$:

\begin{equation}
    \begin{aligned}
        \frac{d\ln r}{dt}& =\frac{1}{r}\frac{dr}{ds}\frac{ds}{dt} \\
        & =\frac{1}{r}\cdot-\frac{r}{\sqrt{s^2-d^2}}\cdot\frac{c}{\cos\varphi_i} \\
        & =-\frac{c}{\sqrt{s^2-d^2}\cos\varphi_i}.
    \end{aligned}
\end{equation}
Now we can have the diffraction solution for delta impulse incident wave. For the Dirichlet boundary, the solution is

\begin{equation}
    f_D=\frac{d\ln r}{dt}\left(\frac{df}{dr}(\omega_{ru},\omega_{lu},\theta_o,\ln r,\kappa)+\frac{df}{dr}(\omega_{rr},\omega_{lr},\theta_o,\ln r,\kappa)\right).
\end{equation}
And for the Neumann boundary, the solution is
\begin{equation}
    f_N=\frac{d\ln r}{dt}\left(\frac{df}{dr}(\omega_{ru},\omega_{lu},\theta_o,\ln r,\kappa)-\frac{df}{dr}( \omega_{rr},\omega_{lr},\theta_o,\ln r,\kappa)\right).
\end{equation}
The order of 4 shadow boundary directions in the equations above look different from the ones in \autoref{eq-Dirichlet1} and \autoref{eq-Neumann1} because we adjusted them with \autoref{eq:fc-r1} and \autoref{eq:fc-r2}.

\subsection{Breaking the Solution into BEDRF}

In this section, we will break the solutions above into the sum of a series of spherical components centering at different points on the edge, which are our BEDRFs. This is also what Medwin did to the Biot-Tolstoy solution in his assumption \cite{medwin1982impulse}.

In $f_N$ and $f_D$, the wave received at the listener is a function of time $t$. Now this $t$ is explained as the time delay from the edge to the listener. This time delay from a point $\mathbf{x}_e$ on the edge to the listener $x$ consists of two parts: the propagation delay $t_p$, which is the time delay of the wave propagating from $\mathbf{x}_e$ to $\mathbf{x}$, and the wavefront delay $t_w$, which is the time delay of the incident planar wave arriving at $\mathbf{x}_e$. Suppose that $\mathbf{x}_e=x_e\mathbf{t}$, we have

\begin{figure}
    \centering
    \begin{tikzpicture}
        \coordinate (pos_re) at (1.25,0);
        \coordinate (pos_im) at (0,1.25);
        \coordinate (in1) at (-1,1);
        \coordinate (in2) at (0.67544,1);
        \coordinate (out) at (2,1);
        \coordinate (edge1) at (0,0);
        \coordinate (edge2) at (1.67544,0);
        \fill[gray!40] (-3,0)--(-3,-1.5)--(3,-1.5)--(3,0)--cycle;
        \draw[gray, -stealth] (-3,0)--(3,0) node[above]{$\mathbf{t}$};
        \draw[gray, -stealth] (0,-1.25)--(pos_im) node[above]{$\mathbf{n}$};
        \draw[red-base, -stealth] (in1)--(edge1);
        \draw[red-base, -stealth] (edge1)--(out);
        \draw[blue-base, -stealth] (in2)--(edge2);
        \draw[blue-base, -stealth] (edge2)--(out);
        \draw[black, thick] (edge1)--(edge2);
        \draw[black, dashed] (0.5,0.5)--(-1,-1);
        \draw[black, dashed] (1.83772,0.16228)--(0.33772,-1.33772);
        \draw[black, stealth-stealth] (-0.2,-0.2)--(0.63772,-1.03772) node[pos=0.5,below,sloped,font=\footnotesize] {\shortstack{wavefront\\delay}};
        \fill[black] (edge1)circle(0.04) node[above]{$\mathbf{x}_{\text{min}}$};
        \fill[black] (edge2)circle(0.04) node[above]{$\mathbf{x}_{\text{max}}$};
        \fill[black] (out)circle(0.04) node[above]{$\mathbf{x}$};
    \end{tikzpicture}
    \caption{When an incident planer wave interacts with a wedge. The time delay of the wave received at $\mathbf{x}$ is the sum of the wavefront delay and the propagation delay. Given a time point $t_0$, we can find two points $\mathbf{x}_{\text{min}}(t_0)$ and $\mathbf{x}_{\text{max}}(t_0)$ from which the time delay is equal to $t_0$, and the time delay of all the points on the segment between them is less than $t_0$.}
    \label{fig-medwin}
\end{figure}
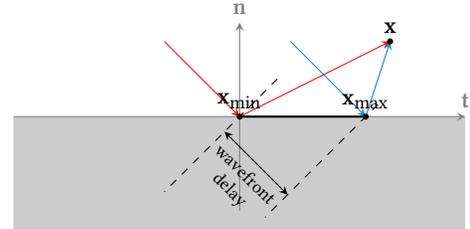

\begin{equation}
    \begin{aligned}
        \mathbf{x}-\mathbf{x}_e & =x_x\mathbf{n} + x_y\mathbf{b} + (x_z-x_e)\mathbf{t} \\
        & =R(\cos\theta_o\cos\varphi_o\mathbf{n}+\sin\theta_o\cos\varphi_o\mathbf{b}+\sin\varphi_o\mathbf{t}), \\
        R& = ||\mathbf{x}-\mathbf{x}_e||_2=\sqrt{x^2_x+x^2_y+(x_z-x_e)^2}.
    \end{aligned}
    \label{eq:listener_pos2}
\end{equation}

\begin{equation}
    ct_p=R
\end{equation}

\begin{equation}
    ct_w=x_e\sin\varphi_i=(x_z-R\sin\varphi_o)\sin\varphi_i
\end{equation}

Combining with \autoref{eq-18} and \autoref{eq-19}, we have
\begin{equation}
    \begin{aligned}
        ct &=c(t_p+t_w) \\
        ct &=R+(x_z-R\sin\varphi_o)\sin\varphi_i \\
        ct &=d(\frac{1}{\cos\varphi_o}-(\tan\varphi_i+\tan\varphi_o)\sin\varphi_i) \\
        ct &=R^*\cos\varphi_i(\frac{1}{\cos\varphi_o}-(\tan\varphi_i+\tan\varphi_o)\sin\varphi_i) \\
        ct &=R^*(\frac{\cos\varphi_i}{\cos\varphi_o}-(\sin\varphi_i+\cos\varphi_i\tan\varphi_o)\sin\varphi_i) \\
        s\cos\varphi_i+R^*\sin^2\varphi_i &= R^*(\frac{\cos\varphi_i}{\cos\varphi_o}-(\sin\varphi_i+\cos\varphi_i\tan\varphi_o)\sin\varphi_i) \\
        s\cos\varphi_i &= R^*(\frac{\cos\varphi_i}{\cos\varphi_o}-\cos\varphi_i\tan\varphi_o\sin\varphi_i) \\
        \frac{s}{d} &= \frac{1-\sin\varphi_i\sin\varphi_o}{\cos\varphi_i\cos\varphi_o} \\
        r &= \frac{1-\sin\varphi_i\sin\varphi_o-|\sin\varphi_i-\sin\varphi_o|}{\cos\varphi_i\cos\varphi_o}
    \end{aligned}
    \label{eq:r}
\end{equation}

Now we'll decompose the original solution into spherical components. We assume that the expression of the spherical component for the Dirichlet case is $\rho_D/R$. This $\rho_D$ is the BEDRF we seek for. Given a time point $t_0$, we can find a segment $\mathbf{x}_{\text{min}}\mathbf{x}_{\text{max}}$ on the edge where all the points have their time delay smaller than $t_0$ (See \autoref{fig-medwin}). Hence

\begin{equation}
    \int_0^{t_0}f_D(t)dt=\int_{x_{\text{min}}(t_0)}^{x_{\text{max}}(t_0)}\frac{\rho_D(\varphi_o(x_e))}{R(x_e)}dx_e.
    \label{eq-33}
\end{equation}
Here $\mathbf{x}_\text{min}\mathbf{x}_\text{max}=[x_\text{min},x_\text{max}]\mathbf{t}$. There are many possible choices for $\rho_D$. And we choose the relationship

\begin{equation}
    f_D(t)=\frac{\rho_D(\varphi_o(x_e))}{R(x_e)}\cdot2\left|\frac{dx_e}{dt}\right|.
\end{equation}
Integrate this equation with respect to $t$ and we'll get \autoref{eq-33}. Combining with \autoref{eq-18} and \autoref{eq-19}, we get

\begin{equation}
    \begin{aligned}
        \rho_D &=\frac{f_DR}{2}\cdot\left|\frac{dt}{dx_e}\right| \\
        &=\frac{f_DR}{2}\cdot\left|\frac{dt}{d\ln r}\cdot\frac{d\ln r}{d\varphi_o}\cdot\frac{d\varphi_o}{dx_e}\right| \\
        &=\frac{f_DR}{2}\cdot\left|\frac{dt}{d\ln r}\cdot\frac{1}{r}\frac{d r}{d\varphi_o}\cdot\frac{d\varphi_o}{dx_e}\right| \\
        &=\frac{f_DR}{2}\cdot\left|\frac{dt}{d\ln r}\cdot\frac{-1}{\sqrt{s^2-d^2}}\frac{d s}{d\varphi_o}\cdot\frac{\cos^2\varphi_o}{d}\right| \\
        &=\frac{f_DR}{2}\cdot\left|\frac{dt}{d\ln r}\cdot\frac{\sin\varphi_i-\sin\varphi_o}{(\frac{s}{d}-r)\cos\varphi_i\cos^2\varphi_o}\cdot\frac{\cos\varphi_o}{R}\right| \\
        &=\frac{f_D}{2}\cdot\left|\frac{dt}{d\ln r}\cdot\frac{1}{\cos\varphi_o}\cdot\cos\varphi_o\right| \\
        &=\frac{f_D}{2}\cdot\left|\frac{dt}{d\ln r}\right|.
    \end{aligned}
\end{equation}
And we obtained the expression of $\rho_D$:

\begin{equation}
    \rho_D=\frac12\left(\frac{df}{dr}(\omega_{lu},\omega_{ru},\theta_o,\ln r,\kappa)+\frac{df}{dr}(\omega_{lr},\omega_{rr},\theta_o,\ln r,\kappa)\right).
\end{equation}

Similarly, the expression of the BEDRF for the Neumann case $\rho_N$ is

\begin{equation}
    \rho_N=\frac12\left(\frac{df}{dr}(\omega_{lu},\omega_{ru},\theta_o,\ln r,\kappa)-\frac{df}{dr}( \omega_{lr},\omega_{rr},\theta_o,\ln r,\kappa)\right).
\end{equation}

\subsubsection{Proof of Reciprocity}

We can see from \autoref{eq:r} that the expression of $r$ is reciprocal. So we only need to deal with $\varphi_i$ and $\varphi_o$. Here we'll only prove that $\frac{df}{dr}(\omega_{lu},\omega_{ru},\theta_o,\ln r,\kappa)$ is reciprocal:

\begin{equation}
    \begin{aligned}
        & \frac{d}{dr}f(\theta_i-\pi,\theta_i+\pi,\theta_o,\ln r,\kappa) \\
        = &\frac{1}{\pi}\left(\frac{d}{dr}\text{arg}\left(\frac{re^{i\kappa\theta_o}-e^{i\kappa(\theta_i+\pi)}}{re^{i\kappa\theta_o}-e^{i\kappa(\theta_i-\pi)}}\right)\right) \\
        = &\frac{1}{\pi}\left(\frac{d}{dr}\text{arg}\left(r^2+e^{i2\kappa\pi}-r(e^{i\kappa(\theta_o-\theta_i+\pi)}+e^{i\kappa(\theta_i-\theta_o+\pi)})\right)\right) \\
        = &\frac{d}{dr}f(\theta_o-\pi,\theta_o+\pi,\theta_i,\ln r,\kappa). \\
    \end{aligned}
\end{equation}
The proof for $\frac{df}{dr}(\omega_{lr},\omega_{rr},\theta_o,\ln r,\kappa)$ is similar.

\subsection{Importance Sampling}

\begin{figure*}[htbp]
    \centering
    \subfigure[BEDRF, absolute value]{
        \includegraphics[width=0.3\linewidth]{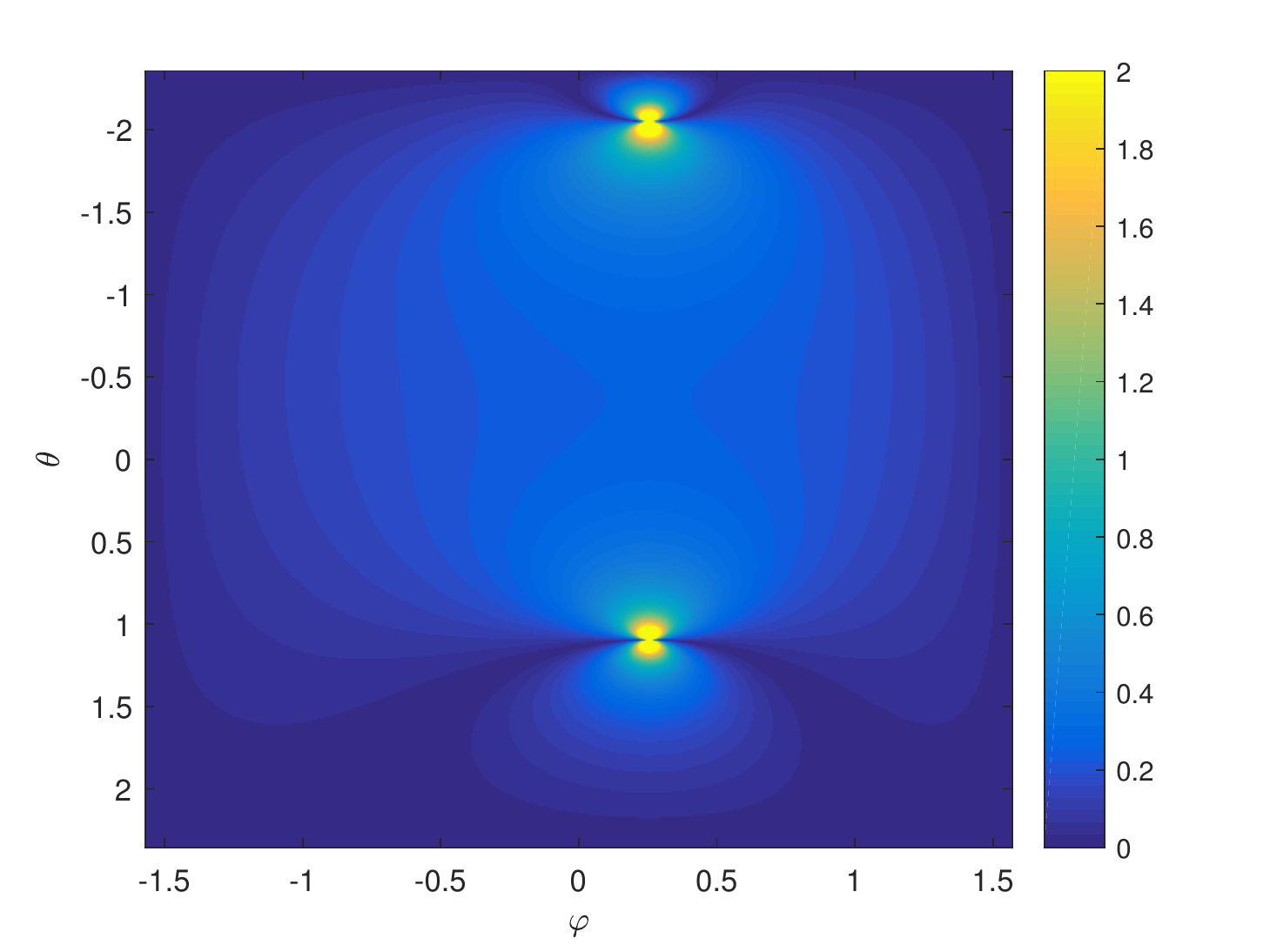}
    }
    \subfigure[PDF, original]{
        \includegraphics[width=0.3\linewidth]{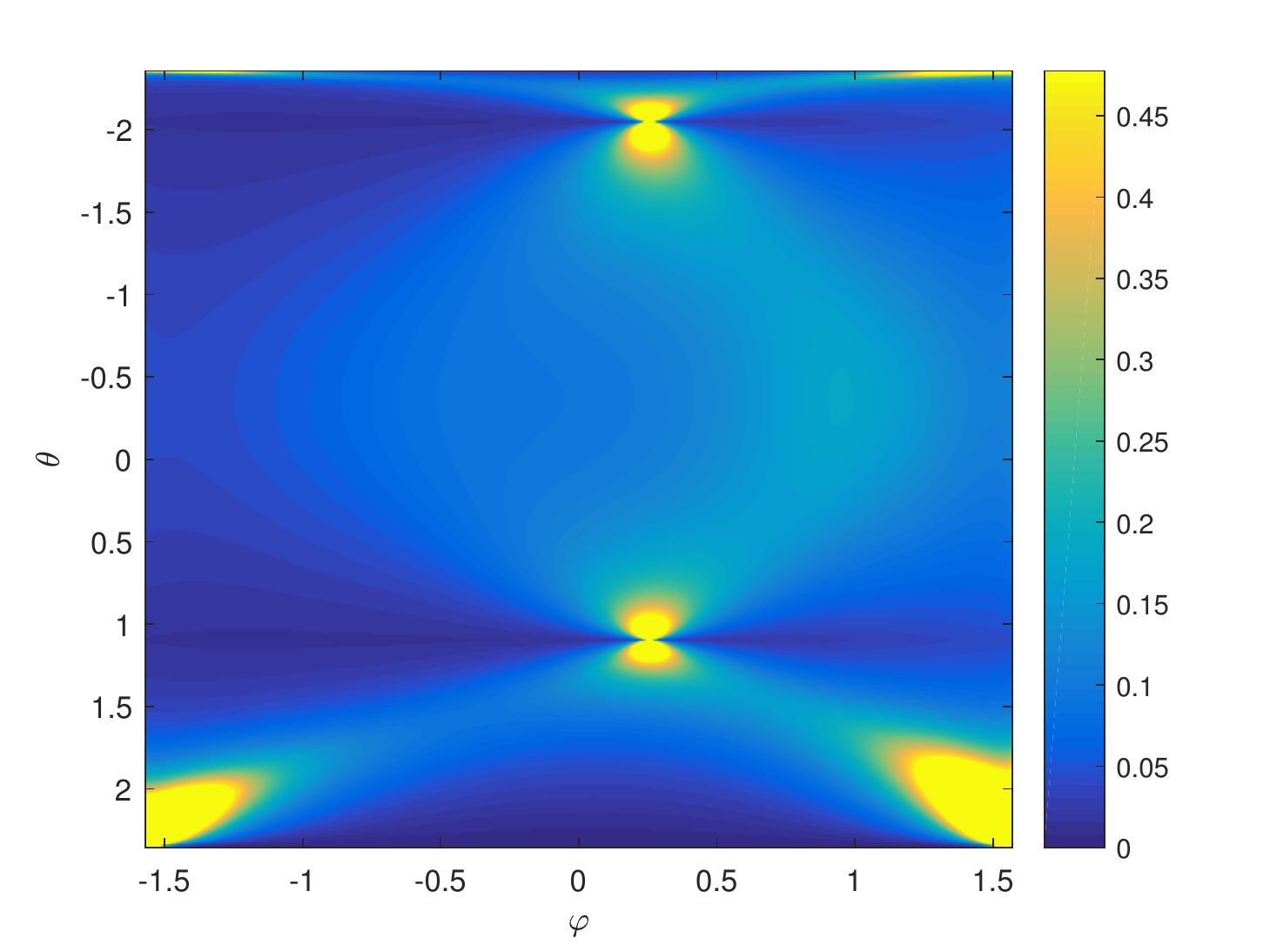}
    }
    \subfigure[PDF, in actual implementation]{
        \includegraphics[width=0.3\linewidth]{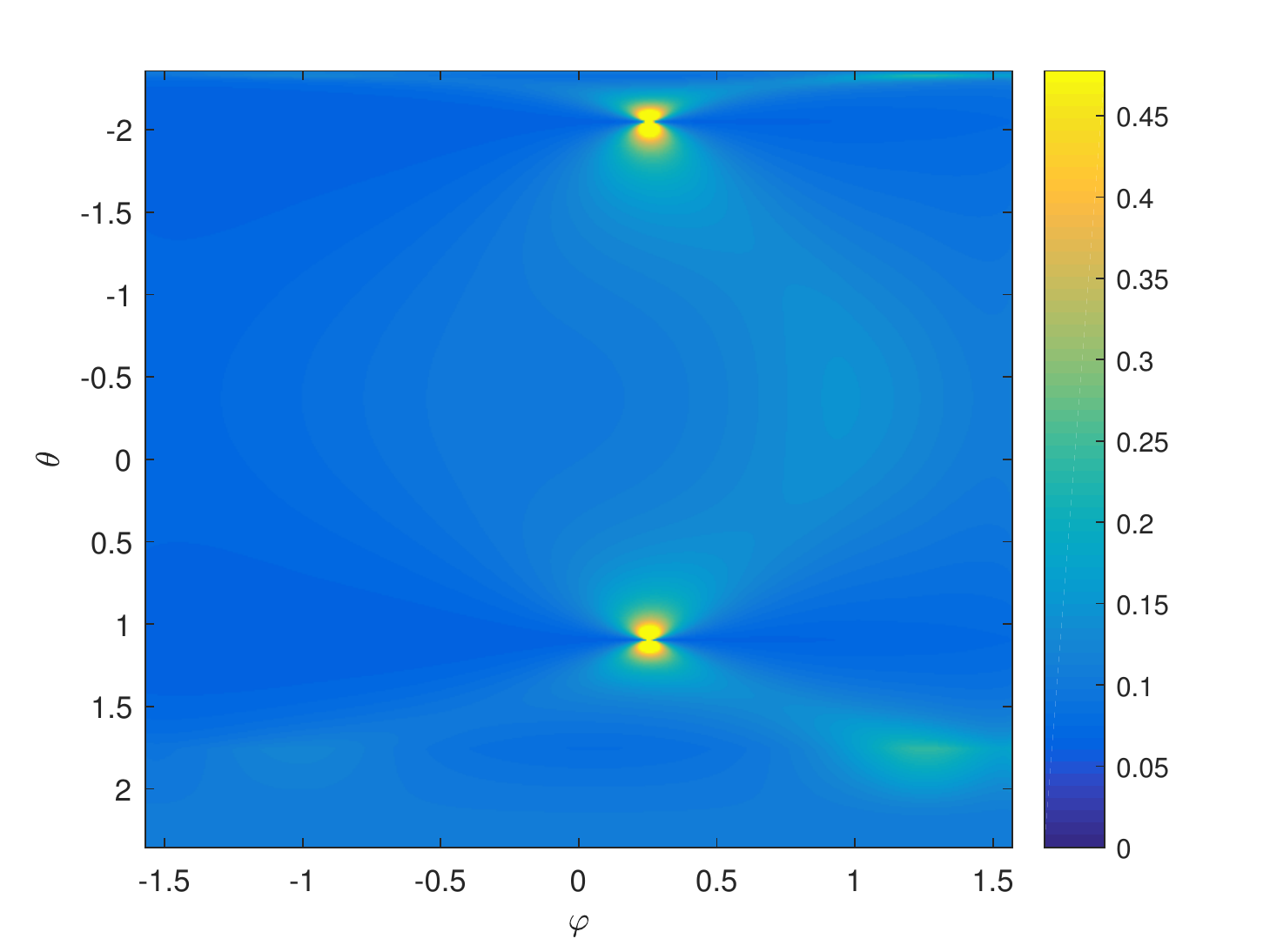}
    }
    \caption{Demonstraction of our importance sampler. Fig. (a) shows a BEDRF for the Dirichlet boundary. Fig. (b) shows the PDF of the corresponding importance sampler according to \autoref{eq:final-sampler}. Fig. (c) shows the actual sampling PDF in our implementation, with a few tweaks to avoid undersampling and suppress the unwanted hotspots in Fig. (b).}
    \label{fig-bedrf}
\end{figure*}

Now we will try to find a sampler whose probability distribution function (PDF) is similar to $\rho_D$ or $\rho_N$. The PDF of our importance sampler $p$ must satisfy the equation
\begin{equation}
    \int_{\phi-\pi}^{\pi-\phi}\int_{-\pi/2}^{\pi/2}p(\theta_o,\varphi_o)\cos\varphi_od\varphi_od\theta_o=1.
\end{equation}
For the convenience of implementation. we would like to sample $\theta_o$ and $\varphi_o$ separately, first $\varphi_o$ and then $\theta_o$. And we need to design two distributions whose PDF is $p_\theta$ and $p_\varphi$. These two PDFs should satisfy the following equations:

\begin{equation}
    p(\theta_o,\varphi_o)\cos\varphi_o=p_\theta(\theta_o)p_\varphi(\theta_o,\varphi_o),
\end{equation}

\begin{equation}
    \int_{\phi-\pi}^{\pi-\phi}p_\theta(\theta_o)d\theta_o=1,
\end{equation}

\begin{equation}
    \int_{-\pi/2}^{\pi/2}p_\varphi(\theta_o,\varphi_o)d\varphi_o=1.
\end{equation}

When implementing a sampler, we would like to know the expression the cumulative distribution function (CDF) instead of PDF. We note the CDF for the distribution of $\theta_o$ as $P_\theta$ and the CDF for $\varphi_o$ as $P_\varphi$. And we have $p_\theta=dP_\theta/d\theta_o$, $p_\varphi=dP_\varphi/d\varphi_o$.

An ideal sampler should have its PDF $p$ proportional to the BEDRF ($p\propto\rho_D$ or $p\propto\rho_N$). Since $p$ is non-negative white $\rho_D$ and $\rho_N$ is not, such a sampler is impossible to achieve. In this section, we will use $\approx$ to represent the ``similiarity'' relationship we look for. We assume that the ``similar'' relationship behaves like the ``proprotional to'' relationship $\propto$:

\begin{equation}
    \begin{aligned}
        f(x)\approx g(x) & \implies h(x)f(x)\approx h(x)g(x), \\
        f(x)\approx g(x) & \implies af(x)\approx g(x), a\text{ is independent from } x.
    \end{aligned}
\end{equation}

Here we try to approach $\rho_D$ first. Starting from $p \approx\rho_D$, we have

\begin{equation}
    \begin{aligned}
        p_\theta p_\varphi & \approx\rho_D\cos\varphi_o \\
        p_\varphi & \approx\frac12\left(\frac{df}{dr}(\omega_{lu},\omega_{ru},\ln r)+\frac{df}{dr}(\omega_{lr},\omega_{rr},\ln r)\right)\cos\varphi_o \\
        \frac{dP_\varphi}{d\varphi_o} & \approx\frac{\kappa r^\kappa}2\left(\frac{df^*_c}{dr}(\omega_{lu},\omega_{ru},r^\kappa)+\frac{df^*_c}{dr}(\omega_{lr},\omega_{rr}, r^\kappa)\right)\cos\varphi_o \\
    \end{aligned}
\end{equation}
If $P_\varphi$ is a function of $r^\kappa$, then
\begin{equation}
    \begin{aligned}
        & \frac{dP_\varphi}{d(r^\kappa)} \\
        \approx & \frac{2d\varphi_o}{d(r^\kappa)}\cdot\frac{\kappa r^\kappa}2\left(\frac{df^*_c}{dr}(\omega_{lu},\omega_{ru},r^\kappa)+\frac{df^*_c}{dr}(\omega_{lr},\omega_{rr}, r^\kappa)\right)\cos\varphi_o \\
        \approx & \left(\frac{df^*_c}{dr}(\omega_{lu},\omega_{ru},r^\kappa)+\frac{df^*_c}{dr}(\omega_{lr},\omega_{rr}, r^\kappa)\right)\cos^2\varphi_o. \\
    \end{aligned}
\end{equation}
From \autoref{eq-18}, we can derive some simple facts about $r$:
\begin{equation}
    \begin{aligned}
        r & \in [0,1], \\
        r=0 &\implies \varphi_o=\pm\frac{\pi}2, \\
        r=1 &\implies \varphi_o=\varphi_i. \\
    \end{aligned}
\end{equation}
From these facts, we have
\begin{equation}
    \label{eq:probability-condition}
    \begin{aligned}
        \frac{dP_\varphi}{d(r^\kappa)}(0) &\approx 0 \\
        \frac{d^2P_\varphi}{d(r^\kappa)^2}(0) &\approx 0 \\
        \frac{d^3P_\varphi}{d(r^\kappa)^3}(0) &\approx -2\left(\frac{df^*_c}{dr}(\omega_{lu},\omega_{ru},0)+\frac{df^*_c}{dr}(\omega_{lr},\omega_{rr}, 0)\right) \\
        \frac{dP_\varphi}{d(r^\kappa)}(1) &\approx \left(\frac{df^*_c}{dr}(\omega_{lu},\omega_{ru},1)+\frac{df^*_c}{dr}(\omega_{lr},\omega_{rr}, 1)\right)\cos^2\varphi_i \\
        \frac{d^2P_\varphi}{d(r^\kappa)^2}(1) &\approx \left(\frac{d^2f^*_c}{dr^2}(\omega_{lu},\omega_{ru},1)+\frac{d^2f^*_c}{dr^2}(\omega_{lr},\omega_{rr}, 1)\right)\cos^2\varphi_i \\
    \end{aligned}
\end{equation}
Similar relationships can be constructed for $\rho_N$.

The equations above is related to the derivative of $f^*_c$ at the endpoint $r^k=0$ and $r^k=1$. The value of the derivatives are given below:
\begin{equation}
    \begin{aligned}
        \frac{df^*_c}{d(r^\kappa)}(0) & = \frac{2}{\pi}\left(\sin(\kappa(\frac{\theta_1-\theta_0}2))(\cos(\kappa(\theta-\frac{\theta_0+\theta_1}2)) \right. \\
        & \left.-\cos(\kappa(\frac{\theta_1-\theta_0}2)))\right) \\
    \end{aligned}
\end{equation}

\begin{equation}
    \frac{df^*_c}{d(r^k)}(1)=\frac{\sin(\kappa(\frac{\theta_1-\theta_0}2))}{\pi(\cos(\kappa(\theta-\frac{\theta_0+\theta_1}2))-\cos(\kappa(\frac{\theta_1-\theta_0}2)))}
    \label{eq-49}
\end{equation}

\begin{equation}
    \frac{d^2f^*_c}{d(r^k)^2}(1)=-\frac{df^*_c}{d(r^k)}(1)
    \label{eq-50}
\end{equation}

We can see from \autoref{eq-49} and \autoref{eq-50} that a good $P_\varphi$ candidate should have $\frac{dP_\varphi}{d(r^\kappa)}(1) = -\frac{d^2P_\varphi}{d(r^\kappa)^2}(1)$. This condition is very hard to satisfy, and it turns out that the best function to achieve this is no other than $f^*_c$ itself. With some modification on $f^*_c$, we get a CDF $P_c$ which satisfies $P'_c(1)=-P''_c(1)$:

\begin{equation}
    \begin{aligned}
        P_c(\theta_a,\theta_b)(x) &= \frac{1}{\pi}\left(\arctan^*\left(\frac{(1-x^2)\sin \theta_a}{(1+x^2)\cos \theta_a-2x\cos \theta_b}\right)\right. \\
        &\left.-\arctan^*\left(\frac{(1-x^2)\sin \theta_a}{(1+x^2)\cos \theta_a+2x\cos \theta_b}\right)\right), \\
        & 0<\theta_b<\theta_a<\frac\pi2.
    \end{aligned}
\end{equation}

This CDF function is monotonic with respect to $x$. Now we calculate its derivative at $x=1$:
\begin{equation}
    \begin{aligned}
        \frac{dP_c(\theta_a,\theta_b)}{dx}(1) &=\frac{1}{\pi}\left(\frac{\sin\theta_a}{\cos\theta_b-\cos\theta_a}-\frac{\sin\theta_a}{-\cos\theta_b-\cos\theta_a}\right) \\
        &=\frac{2\sin\theta_a\cos\theta_b}{\pi(\cos^2\theta_b-\cos^2\theta_a)}. \\
    \end{aligned}
\end{equation}
$\frac{dP_c(\theta_a,\theta_b)}{dx}(1)$ is a monotonically decreasing function with respect to $\theta_b$, and it approaches its lower bound when $\theta_b\to 0$, where $\frac{dP_c(\theta_a,\theta_b)}{dx}(1)\to \frac{2}{\pi\sin\theta_a}$.

Now we'll choose $\theta_a$ and $\theta_b$ for $P_c$, consider the expression of $\frac{dP_\varphi}{d(r^\kappa)}(1)$, we think $\frac{dP_c(\theta_a,\theta_b)}{dx}(1)=-\frac{2}{\pi}\tan\kappa\pi$ is a good choice. Hence we have
\begin{equation}
    \theta_a = \arcsin(\min\{-\frac{\cos^2\varphi_i}{\tan\kappa\pi},1\}).
\end{equation}
The upper threshold $1$ cannot be omitted, or we may have $\theta_a\geq\frac\pi2$. Afterwards, we can calculate the value of $|\frac{dP_\varphi}{d(r^\kappa)}(1)|$. Let $|\frac{dP_\varphi}{d(r^\kappa)}(1)|=x$, we solve $x=\frac{dP_c(\theta_a,\theta_b)}{dx}(1)$ and get

\begin{equation}
    \label{eq:theta-b}
    \theta_b = \arccos\left(\frac{\sin\theta_a}{\pi x}+\sqrt{\left(\frac{\sin\theta_a}{\pi x}\right)^2+\cos^2\theta_a}\right).
\end{equation}
And now $P_c(\theta_a,\theta_b)$ satisfies all the conditions in \autoref{eq:probability-condition} as $P_\varphi$.

If we use $P_c(\theta_a,\theta_b)(r^k)$ for the distribution of $\varphi_o$, we will have
\begin{equation}
    \label{eq:final-sampler}
    p_\varphi = \frac{\kappa r^\kappa}{2\cos\varphi_o}\frac{dP_c(\theta_a,\theta_b)}{dx}(r^k).
\end{equation}
In actual implementation, we use
\begin{equation}
    p_\varphi = a+(1-a)\frac{\kappa r^\kappa}{\cos\varphi_o}\frac{dP_c(\theta_a,\theta_b)}{dx}(r^k)
\end{equation}
where
\begin{equation}
    a = 1-\frac12e^{1 - \frac{2\sin\theta_a}{\pi x}}
\end{equation}
during sampling to prevent the probability being too low, as low BEDRF sampling probability may produce outliers (explained in Sect. 5.5 of our paper). Besides, we also replace $\frac{\sin\theta_a}{\pi x}$ in \autoref{eq:theta-b} with $\min\left\{\frac{\sin\theta_a}{\pi x},\frac12\right\}$ to make the result probability distribution better. See \autoref{fig-bedrf}.

\section{Measure Projection on Triangle}
\label{sect:measure}

Here we will prove the measure conversion formula for edge integration in Sect. 5.2 of our paper. We consider an triangle $\triangle \mathbf{abc}$ and its edge $\mathbf{ab}$. We note the length measure on $\mathbf{ab}$ as $\mu$ and the area measure on $\triangle\mathbf{abc}$ as $\mu^*$. We can project every point in $\triangle \mathbf{abc}$ to $\mathbf{ab}$ by connecting it with $\mathbf{c}$ and intersecting the connecting line with edge $\mathbf{ab}$. This projection is noted as $\Pi$. According to the measure projecting theory in \cite{Tarantola08measure}, the projection induces a measure $\Pi^{-1}(\mu)$ on the triangle.

Now we will calculate the Radon derivative $d\Pi^{-1}(\mu)/d\mu^*$. For a range of length $\Delta x$, all the unoccluded points that projects into the range form a set $S$. Suppose that $S$ is the shaded trapezoid $S$ in \autoref{fig-measure}, then we have

\begin{equation}
    \begin{aligned}
        \int_{\mathbf{x}\in S}d\Pi^{-1}(\mu) &=\Delta x, \\
        \int_{\mathbf{x}\in S}d\mu^* &=(y+\frac12\Delta y)\frac{\Delta x\Delta y}{h}.
    \end{aligned}
\end{equation}
Here $h$ is the half-height of $\triangle \mathbf{abc}$ with respect to the edge $\mathbf{ab}$, which equals to $S\triangle\mathbf{abc}/||\mathbf{b}-\mathbf{a}||_2$. Let $\Delta x$ and $\Delta y$ be infinitely small, and we have $d\mu^*/d\Pi^{-1}(\mu)=\frac{y}{h}\Delta y$.

For other unoccluded areas, we can always regard them as a combination of infinitesimal trapezoids, Suppose that the unoccluded area $S$ is an arbitary set, and a point $\mathbf{x}'\in S$ is projected to $\mathbf{x}\in\mathbf{ab}$. Then

\begin{equation}
      \frac{d\mu^*}{d\Pi^{-1}(\mu)}(\mathbf{x}')=\int_{\mathbf{x}'\in S\cap \mathbf{xc}}\frac{||\mathbf{x}'-\mathbf{c}||_2}{||\mathbf{x}-\mathbf{c}||_2}d\mu_{\mathbf{xc}}.
\end{equation}
Here $\mu_{\mathbf{xc}}$ is the length measure on $\mathbf{xc}$. With a little more modification and we have the equation in Sect. 5.2 of our paper.

\begin{figure}[htbp]
    \begin{tikzpicture}
        \usetikzlibrary{arrows.meta}

        \coordinate (a) at (-2,0);
        \coordinate (b) at (2,0);
        \coordinate (c) at (0.4,3);
        \coordinate (x1) at (-0.6,0);
        \coordinate (x2) at (-0.2,0);
        \coordinate (p1) at (-0.2,1.2);
        \coordinate (p2) at (0.04,1.2);
        \coordinate (q1) at (0,1.8);
        \coordinate (q2) at (0.16,1.8);

        \draw[gray,thick] (a) node[below,black]{$\mathbf{a}$} --(b)  node[below,black]{$\mathbf{b}$} --(c)  node[above,black]{$\mathbf{c}$} --cycle;
        \draw[gray] (x1)--(c)--(x2);
        \draw[gray, fill=gray!20] (p1)--(p2)--(q2)--(q1);
        \draw[gray] (0.04,1.2)--(0.4,1.2);
        \draw[gray] (0.16,1.8)--(0.4,1.8);
        \draw[gray] (0.4,0)--(0.4,3);
        \node[black] at (-0.025,1.5) {$S$};
        \draw[gray,|{Stealth}-{Stealth}|] (-0.6,-0.1)--(-0.2,-0.1)  node[black,midway,below] {$\Delta x$};
        \draw[gray,|{Stealth}-{Stealth}|] (0.45,1.2)--(0.45,1.8)  node[black,midway,right] {$\Delta y$};
        \draw[gray,|{Stealth}-{Stealth}|] (0.45,1.8)--(0.45,3)  node[black,midway,right] {$y$};
    \end{tikzpicture}
    \caption{Illustration of an infinitesimal area $S$ in $\triangle\mathbf{abc}$ in \autoref{sect:measure}.}
    \label{fig-measure}
\end{figure}

\bibliographystyle{ACM-Reference-Format}
\bibliography{thesis}